\documentclass[12pt]{article}
\usepackage{epsfig,rotating,setspace,latexsym,amsmath,epsf,amssymb,bm}
\usepackage{cite}

\newtheorem{Theo}{Theorem}
\newtheorem{Lem}{Lemma}

\begin{document}
\date{}
\title{Dense Gaussian Sensor Networks: Minimum Achievable Distortion
and the Order Optimality of Separation\thanks{This work was
supported by NSF Grants CCR $03$-$11311$, CCF $04$-$47613$ and CCF
$05$-$14846$. It was presented in part at the IEEE International
Conference on Communications, Istanbul, Turkey, June 2006
\cite{Liu_Ulukus:2006} and at the IEEE International Symposium on
Information Theory, Seattle, WA, July 2006
\cite{Liu_Ulukus:2006ISIT}.}}

\author{Nan Liu \qquad Sennur Ulukus \\
\normalsize Department of Electrical and Computer Engineering \\
\normalsize University of Maryland, College Park, MD 20742 \\
\normalsize {\it nkancy@umd.edu} \qquad {\it ulukus@umd.edu} }

\maketitle

\begin{abstract}
We investigate the optimal performance of dense sensor networks by
studying the joint source-channel coding problem. The overall goal
of the sensor network is to take measurements from an underlying
random process, code and transmit those measurement samples to a
collector node in a cooperative multiple access channel with
potential feedback, and reconstruct the entire random process at
the collector node. We provide lower and upper bounds for the
minimum achievable expected distortion when the underlying random
process is Gaussian. When the Gaussian random process satisfies
some general conditions, we evaluate the lower and upper bounds
explicitly, and show that they are of the same order for a wide
range of power constraints. Thus, for these random processes,
under these power constraints, we express the minimum achievable
expected distortion as a function of the power constraint.
Further, we show that the achievability scheme that achieves the
lower bound on the distortion is a separation-based scheme that is
composed of multi-terminal rate-distortion coding and
amplify-and-forward channel coding. Therefore, we conclude that
separation is order-optimal for the dense Gaussian sensor network
scenario under consideration, when the underlying random process
satisfies some general conditions.
\end{abstract}

\newpage

\section{Introduction}
\begin{figure}
\centering
\includegraphics[width=3.5in]{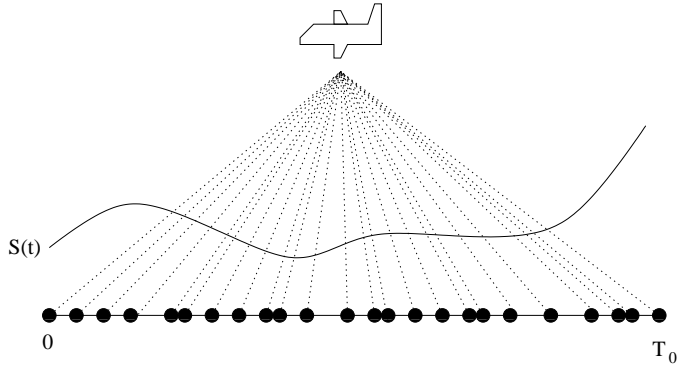}
\caption{Sensor network.} \label{shonda}
\end{figure}

With the recent advances in the hardware technology, small cheap
nodes with sensing, computing and communication capabilities have
become available. In practical applications, it is possible to
deploy a large number of these nodes to sense the environment. In
this paper, we investigate the optimal performance of a dense
sensor network by studying the joint source-channel coding
problem. The sensor network is composed of $N$ sensors, where $N$
is very large, and a single collector node. Each sensor node has
the capability of taking noiseless samples from the underlying
random process, and is equipped with one transmit and one receive
antenna to transmit and receive signals. The overall goal of the
sensor network is to take measurements from an underlying random
process $S(t)$, $0 \leq t \leq T_0$, code and transmit those
measured samples to a collector node, and reconstruct the entire
random process at the collector node, with as little distortion as
possible; see Figure \ref{shonda}. Due to the small distances
between the sensor nodes and the correlation in the measured data,
the underlying sources are correlated, and due to the existence of
receive antennas at the sensor nodes and a transmit antenna at the
collector node, the communication channel is a Gaussian
cooperative multiple access channel with potential feedback. We
investigate the minimum achievable expected distortion and the
corresponding achievability scheme when the underlying random
process is Gaussian.

Following the seminal paper of Gupta and Kumar \cite{Gupta:2000},
which showed that multi-hop wireless ad-hoc networks, where users
transmit independent data and utilize single-user coding, decoding
and forwarding techniques, do not scale up, Scaglione and Servetto
\cite{Servetto:2002} investigated the scalability of the sensor
networks. Sensor networks, where the observed data is correlated,
may scale up for two reasons: first, the correlation among the
sampled data increases with the increasing number of nodes and
hence, the amount of information the network needs to carry does
not increase as fast as in ad-hoc wireless networks; and second,
correlated data facilitates cooperation, and may increase the
information carrying capacity of the network. The goal of the
sensor network in \cite{Servetto:2002} was that each sensor
reconstructs the data measured by all of the sensors using sensor
broadcasting. In this paper, we focus on the case where the
reconstruction is required only at the collector node. Also, in
this paper, the task is not the reconstruction of the data the
sensors measured, but the reconstruction of the underlying random
process.

Gastpar and Vetterli \cite{Gastpar:sensor2005} studied the case
where the sensors observe a noisy version of a linear combination
of $L$ Gaussian random variables which all have the same variance,
code and transmit those observations to a collector node, and the
collector node reconstructs the $L$ random variables. In
\cite{Gastpar:sensor2005}, the expected distortion achieved by
applying separation-based approaches was shown to be exponentially
worse than the lower bound on the minimum achievable expected
distortion. In this paper, we study the case where the data of
interest at the collector node is not a finite number of random
variables, but a random process, which, using Karhunen-Loeve
expansion, can be shown to be equivalent to a set of infinitely
many random variables with varying variances. We assume that the
sensors are able to take noiseless samples, but that each sensor
observes only its own sample. Our upper bound on the minimum
achievable distortion is also developed by using a
separation-based approach, but it is shown to be of the same order
as the lower bound, for a wide range of power constraints, for
random processes that satisfy some general conditions.

El Gamal \cite{ElGamal:2005} studied the capacity of dense sensor
networks and found that all spatially band-limited Gaussian
processes can be estimated at the collector node, subject to any
non-zero constraint on the mean squared distortion. In this paper,
we study the minimum achievable expected distortion for
space-limited, and thus, not band-limited, random processes, and
we show that the minimum achievable expected distortion decreases
to zero as the number of nodes increases, unless the power
constraint is unusually small. Also, in \cite{ElGamal:2005}, it is
assumed that the channel gains between the nodes decrease with the
distance between them, without enforcing any upper bounds. This
implies that, when the sensors are placed very densely, the
channel gains between nearby sensors become unboundedly large.
This physically impossible situation arises because although the
channel model used in \cite{ElGamal:2005} is valid only in the far
field of the transmitter, it is used for all distances. Although
we adopted this channel model in \cite{Liu_Ulukus:2006}, we have
changed the channel model to a more realistic one in this paper
(and \cite{Liu_Ulukus:2006ISIT}), where we assume that the channel
gains decrease with distance, however, they are lower and upper
bounded. The difference in the channel models in
\cite{Liu_Ulukus:2006} and here (and \cite{Liu_Ulukus:2006ISIT}),
does not affect our conclusion, i.e., in both cases, we are able
to find achievable schemes that achieve the lower bound on the
distortion. However, it affects the achievability scheme itself;
in \cite{Liu_Ulukus:2006} the achievability scheme is based on the
basic idea of decode-and-forward as the channel model allows a
significant number of nodes to be able to decode successfully the
signal transmitted by a node, whereas the achievability scheme
here (and \cite{Liu_Ulukus:2006ISIT}) is based on the basic idea
of amplify-and-forward, where due to the lower and upper bounds on
the channel gains a sufficient amount of beamforming effect is
achieved through the amplify-and-forward scheme.

From an information theoretic point of view, our problem is a
joint source-channel coding problem for lossy communication of
correlated sources over a cooperative Gaussian multiple access
channel with potential feedback. The simpler problem of lossless
reconstruction of correlated sources over a multiple access
channel without cooperation or feedback still remains open
\cite{Cover:1980, Dueck:1981, Kang_Ulukus:2005, Kang_Ulukus:2006}.
Therefore, a direct and closed-form expression for the distortion
seems unlikely to be obtained, and consequently, we resort to
developing lower and upper bounds. We first provide lower and
upper bounds for the minimum achievable expected distortion for
arbitrary Gaussian random processes whose Karhunen-Loeve expansion
exists. Then, we focus on the case where the Gaussian random
process also satisfies some general conditions. For these random
processes, we evaluate the lower and upper bounds explicitly, and
show that they are of the same order, for a wide range of power
constraints. Thus, for these random processes, under a wide range
of power constraints, we determine the order-optimal achievability
scheme, and identify the minimum achievable expected distortion.
Our achievability scheme is separation-based: each sensor first
performs multi-terminal source coding \cite{Flynn:1987}, then,
performs channel coding, and utilizes the cooperative nature of
the wireless medium through the amplify-and-forward scheme
\cite{Gastpar:2005}. In multi-user information theory, generally
speaking, the separation principle does not hold. However, in our
case, we have found a scheme which is separation based, and is
order-optimal.

\section{System Model} \label{systemmodel}
The collector node wishes to reconstruct a random process $S(t)$,
for $0 \leq t \leq T_0$, where $t$ denotes the spatial position;
$S(t)$ is assumed to be Gaussian with zero-mean and a continuous
autocorrelation function $K(t,s)$. The $N$ sensor nodes are placed
at positions $0=t_1 \leq t_2 \leq \cdots \leq t_N=T_0$, and
observe samples $\mathbf{S}_N=(S(t_1),S(t_2),\cdots,S(t_N))$. For
simplicity and to avoid irregular cases, we assume that the
sensors are equally spaced, i.e.,
\begin{align}
t_i=\frac{i-1}{N-1}T_0, \qquad  i=1,2, \cdots, N
\end{align}
The distortion measure is the squared error,
\begin{align}
d(s(t),\hat{s}(t)) = \frac{1}{T_0}\int_{0}^{T_0}
(s(t)-\hat{s}(t))^2 dt
\end{align}

Each sensor node and the collector node, denoted as node 0, is
equipped with one transmit and one receive antenna. To simplify
the presentation, from now until Section \ref{sect:discussion}, we
will assume that the collector node does not use its transmit
antenna, and thus, there is no feedback in the system. We will
allow the collector node to use its transmit antenna and provide
feedback to the sensor nodes in Section \ref{sect:discussion}, and
show that the results of the previous sections remain unchanged.
At any time instant, let $X_i$ denote the signal transmitted by
node $i$, and $Y_j$ denote the signal received at node $j$. Let
$h_{ij}$ denote the channel gain from node $i$ to node $j$. Then,
the received signal at node $j$ can be written as,
\begin{align}
Y_j=\sum_{i=1,i \neq j}^N h_{ij} X_i+Z_j, \qquad j=0,1,2,\cdots,N
\end{align}
where $\{Z_j\}_{j=0}^N$ is a vector of $N+1$ independent and
identically distributed, zero-mean, unit-variance Gaussian random
variables. Therefore, the channel model of the network is such
that all nodes hear a linear combination of the signals
transmitted by all other nodes at that time instant. We assume
that the channel gain $h_{ij}$ is bounded, i.e.,
\begin{align}
\bar{h}_l \leq h_{ij} \leq \bar{h}_u, \qquad i=1,\cdots,N, \quad
j=0,1,\cdots,N \label{channelbounded}
\end{align}
where $\bar{h}_u$ and $\bar{h}_l$ are positive constants
independent of $N$. This model is very general and should be
satisfied very easily. By the conservation of energy, $h_{ij}^2
\leq 1$, and since all nodes are within finite distances of each
other, the channel gains should be lower bounded as well.

We assume that all sensors have the same individual power
constraint $P(N)/N$, where $P(N)$, which we will call the total
power, is the sum of the individual power constraints, and it is a
function of $N$. The two most interesting cases for $P(N)$ are
$P(N)=N P_{\text{ind}}$ where each sensor has its individual power
constraint $P_{\text{ind}}$, and $P(N)=P_{\text{tot}}$ where the
total power is a constant $P_{\text{tot}}$ and does not depend on
the number of sensors. In the latter case, when more and more
sensor nodes are deployed, the individual power of each sensor
node decreases as $P_{\text{tot}}/N$. Our goal is to determine the
scheme that achieves the minimum achievable expected distortion
$D^N$ at the collector node for a given total power $P(N)$, and
also to determine the rate at which this distortion goes to zero
as a function of the number of sensor nodes and the total power.

Next, we give a more precise definition of our problem. Each
sensor node observes a sample of a sequence of spatial random
processes $\{S^{(l)}(t)\}_{l=1}^ n$ i.i.d. in time, where index
$l$ denotes time, $t$ denotes the spatial position, and $n$ is the
block length of the sequence of random processes, and also the
delay parameter. For now, we assume that $n$ channel uses are
allowed for $n$ realizations of the random process; the case where
we allow the number of channel uses and the number of observations
to differ will be treated in Section \ref{sect:discussion}. At
time instant $m$, sensor node $j$ transmits
\begin{align}
X_j(m)=F_j^{(m)}(\{S^{(l)}(t_j)\}_{l=1}^n,
\{Y_j^{(l)}\}_{l=1}^{m-1}), \quad m=1,2,\cdots,n, \quad
j=1,2,\cdots,N
\end{align}
i.e., it transmits a signal that is a function of its observations
of the entire block of random process samples and also the signal
it received before time $m$. We are interested in the performance
in the information-theoretic sense and hence, we allow the delay
$n$ to be arbitrarily large. By the assumption of identical
individual power constraints, we have
\begin{align}
\frac{1}{n} \sum_{m=1}^{n}  E[X_j^2(m)] \leq \frac{P(N)}{N}, \quad
j=1,2,\cdots,N
\end{align}
The collector node reconstructs the random process as
\begin{align}
\{\hat{S}^{(l)}(t), t \in [0,
T_0]\}_{l=1}^n=G(Y_0^{(1)},Y_0^{(2)},\cdots,Y_0^{(n)})
\end{align}
For fixed encoding functions of the nodes
$\{F_j^{(m)}\}_{m=1,j=1}^{m=n,j=N}$ and the decoding function of
the collector node $G$, the achieved expected distortion is
\begin{align}
\frac{1}{n} \sum_{l=1}^n E\left[d\left(S^{(l)}(t),
\hat{S}^{(l)}(t)\right)\right]
\end{align}
and we are interested in the smallest achievable expected
distortion over all encoding and decoding functions where $n$ is
allowed to be arbitrarily large.

In this paper, our purpose is to understand the behavior of the
minimum achievable expected distortion when the number of sensor
nodes is very large. We introduce the big-O and big-$\Theta$
notations. We say that $f$ is O($g$), if there exist constants $c$
and $k$, such that $|f(N)| \leq c|g(N)|$ for all $N>k$; we say
that $f$ is $\Theta(g)$, if there exist constants $c_1$, $c_2$ and
$k$ such that $c_1|g(N)| \leq |f(N)| \leq c_2|g(N)|$ for all
$N>k$. All logarithms are defined with respect to base $e$, and
$\left\lfloor x \right\rfloor$ denotes the largest integer smaller
than or equal to $x$.

\section{A Class of Gaussian Random Processes}\label{defineA}
For a Gaussian random process $S(t)$ with a continuous
autocorrelation function, we perform the Karhunen-Loeve expansion
\cite{Papoulis:book},
\begin{align}
S(t)=\sum_{k=0}^\infty S_k \phi_k(t)
\end{align}
to obtain the ordered eigenvalues $\{\lambda_k\}_{k=0}^\infty$,
and the corresponding eigenfunctions $\{\phi_k(t), t \in
[0,T_0]\}_{k=0}^\infty$.

Let $\mathcal{A}$ be the set of Gaussian random processes on
$[0,T_0]$ with continuous autocorrelation functions, that satisfy
the following conditions:

\begin{enumerate}
\item There exist nonnegative constant $d$ and nonnegative
integers $c_l$, $c_u$, $K_0 \geq c_u+1$ and two sequences of
numbers $\{\lambda_k'\}_{k=0}^\infty$ and
$\{\lambda_k''\}_{k=0}^\infty$ defined as
\begin{align}
\lambda_k'=\left\{
\begin{array}{ll}
\lambda_k, & k \leq K_0 \\
\frac{d}{(k+c_l)^x}, & k > K_0 \label{definelambdak1}
\end{array}
\right.
\end{align}
and
\begin{align}
\lambda_k''=\left\{
\begin{array}{ll}
\lambda_k, & k \leq K_0 \\
\frac{d}{(k-c_u)^x} , & k > K_0 \label{definelambdak2}
\end{array}
\right.
\end{align}
for some constant $x>1$, such that
\begin{align}
\lambda_k' \leq \lambda_k \leq \lambda_k'' \label{orderorder}
\end{align}
The condition that $x>1$ is without loss of generality, because
for all continuous autocorrelations, the eigenvalues decrease
faster than $k^{-1}$.

\item In addition to continuity, $K(t,s)$ satisfies the Lipschitz
condition of order $1/2 < \alpha \leq 1$, i.e., there exists a
constant $B>0$ such that
\begin{align}
|K(t_1,s_1)-K(t_2,s_2)| \leq B
\left(\sqrt{(t_1-t_2)^2+(s_1-s_2)^2}\right)^\alpha
\end{align}
for all $t_1,s_1,t_2,s_2 \in [0,T_0]$.

\item For $k=0,1,\cdots$, the function $\phi_k(s)$ and the
function $K(t,s)\phi_k(s)$ as a function of $s$ satisfy the
following condition: there exist positive constants $B_1$, $B_2$,
$B_3$, $B_4$, $\beta \leq 1$, $\gamma \leq 1$, and nonnegative
constant $\tau$, independent of $k$, such that
\begin{align}
|\phi_k(s_1)-\phi_k(s_2)|  \leq B_3(k+B_4)^\tau
|s_1-s_2|^\gamma \label{worried4}
\end{align}
and
\begin{align}
|K(t,s_1)\phi_k(s_1)-K(t,s_2)\phi_k(s_2)|  \leq B_2(k+B_1)^\tau
|s_1-s_2|^\beta
\end{align}
for all $t, s_1,s_2 \in [0,T_0]$.
\end{enumerate}

The reasons why these conditions are needed for the explicit
evaluation of the lower and upper bounds on the minimum achievable
expected distortion will be clear from the proofs. Here, we
provide some intuition as to why they are needed. Condition 1
states that we consider random processes that have eigenvalues
$\lambda_k$ which decrease at a rate of approximately $k^{-x}$.
The rate of decrease in the eigenvalues is an indication of the
amount of randomness the random process contains. Thus, the
minimum achievable expected distortion depends crucially on the
rate of decrease parameter $x$. The lower (upper) bound on the
eigenvalues in (\ref{orderorder}) will be used to calculate the
lower (upper) bound on the minimum achievable expected distortion.
Conditions 2 and 3 are needed because instead of the random
process itself that is of interest to the collector node, the
collector node, at best, can know only the sampled values of the
random process. How well the the entire process can be
approximated from its samples is of great importance in obtaining
quantitative results. Lipschitz conditions describe the quality of
this approximation well. By condition 3, we require the variation
in the eigenfunction $\phi_k$ to be no faster than $k^\tau$. We
note that the well-known trigonometric basis satisfies this
condition.

We also note that our conditions are quite general. Many random
processes satisfy these conditions, including the Gauss-Markov
process, Brownian motion process, centered Brownian bridge, etc.
For example, a Gauss-Markov process, also known as the
Ornstein-Uhlenbeck process \cite{Uhlenbeck:1930,Wang:1945}, is
defined as a random process that is stationary, Gaussian,
Markovian, and continuous in probability. It is known that the
autocorrelation function of this process is
\cite{Doob:1942,Breiman:book,Karatzas:book}
\begin{align}
K(t,s)=\frac{\sigma^2}{2 \eta} e^{-\eta |t-s|} \label{nine}
\end{align}
The Karhunen-Loeve expansion of the Gauss-Markov process yields
the eigenfunctions \break $\{\phi_k(t)\}_{k=0}^\infty$
\begin{align}
\phi_k(t)=b_k \left(\cos \sqrt{\frac{\sigma^2}{\lambda_k}-\eta^2}
t+ \frac{\eta}{\sqrt{\frac{\sigma^2}{\lambda_k}-\eta^2}} \sin
\sqrt{\frac{\sigma^2}{\lambda_k}-\eta^2} t \right)
\end{align}
where $\{\lambda_k\}_{k=0}^\infty$ are the corresponding
eigenvalues and $b_k$ are positive constants chosen such that the
eigenfunctions $\phi_k(t)$ have unit energy. It can be shown that
$\{\lambda_k\}_{k=0}^\infty$ may be bounded as
\begin{align}
\lambda_k' \leq \lambda_k \leq \lambda_k'' \label{bound}
\end{align}
where $\{\lambda_k'\}_{k=1}^\infty$ is defined as
\begin{align}
\lambda_k'=\left\{
\begin{array}{ll}
\lambda_k, & k \leq K_0 \\
\frac{\sigma^2 T_0^2}{\left(k+1\right)^2 \pi^2}, & k > K_0
\label{definelambd1}
\end{array}
\right.
\end{align}
with $K_0=\max\left(2,\left\lfloor \frac{\eta^2
T_0^2}{\pi^2}-\frac{3}{4}\right\rfloor\right)$, and
$\{\lambda_k''\}_{k=1}^\infty$ is defined as
\begin{align}
\lambda_k''=\left\{
\begin{array}{ll}
\lambda_k, & k \leq K_0 \\
\frac{\sigma^2T_0^2}{\left(k-1\right)^2 \pi^2}, & k > K_0
\end{array}
\right.
\end{align}

Thus, we observe that the Gauss-Markov process satisfies the
conditions defined in this section with $x=2$ and
$\alpha=\beta=\tau=\gamma=1$. In fact, in a preliminary conference
version of our work \cite{Liu_Ulukus:2006}, we focused
specifically on the Gauss-Markov process and presented results
similar to those here. We also note, as discussed in the
Introduction section, that the channel model in
\cite{Liu_Ulukus:2006} is somewhat different than here, and
therefore the order-optimal achievability schemes in
\cite{Liu_Ulukus:2006} and here are different.

The lower and upper bounds on the minimum achievable expected
distortion will be calculated using $\{\lambda_k'\}_{k=0}^\infty$
and $\{\lambda_k''\}_{k=0}^\infty$, respectively. Some properties
of $\{\lambda_k'\}_{k=0}^\infty$ and
$\{\lambda_k''\}_{k=0}^\infty$ which will be used in later proofs
are stated in Lemmas \ref{lambdak1} and \ref{lambdak2} and proved
in Appendix \ref{property}.

\section{A Lower Bound on the Achievable Distortion} \label{seclowerbound}
\subsection{Arbitrary Gaussian Random Processes}
A lower bound is obtained by assuming that all of the sensor nodes
know the random process exactly, and, the sensor network forms an
$N$-transmit 1-receive antenna point-to-point system to transmit
the random process to the collector node. Let $C_u^N$ be the
capacity of this point-to-point system and $D_p(R)$ be the
distortion-rate function of the random process $S(t)$
\cite{Berger:book}. In this point-to-point system, the separation
principle holds, and therefore
\begin{align}
D^N \geq D_p(C_u^N)
\end{align}
To evaluate $D_p(C_u^N)$, we first find the distortion-rate
function, $D_p(R)$, of $S(t)$ \cite[Section 4.5]{Berger:book} as,
\begin{align}
R(\theta)= \sum_{k=0}^\infty \max \left(0, \frac{1}{2} \log
\left(\frac{\lambda_k}{\theta} \right) \right) \label{rate}
\end{align}
and
\begin{align}
D(\theta)=T_0^{-1} \sum_{k=0}^\infty \min (\theta,
\lambda_k)\label{distortion}
\end{align}
Next, we find $C_u^N$, the capacity of the $N$-transmit 1-receive
antenna point-to-point system \cite{Telatar:1999} as,
\begin{align}
C_u^N=\frac{1}{2} \log \left(1+\sum_{i=1}^N h^2 _{i0} P(N) \right)
\label{Cupper}
\end{align}
To see how $C_u^N$ changes with $N$, using (\ref{Cupper}) and
(\ref{channelbounded}), we can lower and upper bound $C_u^N$ as
\begin{align}
\frac{1}{2} \log \left(1+\bar{h}_l^2 NP(N) \right) \leq C_u^N \leq
\frac{1}{2} \log \left(1+\bar{h}_u^2 NP(N) \right) \label{Barbara}
\end{align}
For arbitrary Gaussian random processes, a lower bound on the
minimum achievable expected distortion is
\begin{align}
D_l^N=D_p(C_u^N) \label{ulukus1}
\end{align}

\subsection{The Class of Gaussian Random Processes in $\mathcal{A}$}
Next, we evaluate $D_p(C_u^N)$ for the class of Gaussian
random processes in $\mathcal{A}$. Based on the structure of the
eigenvalues in (\ref{definelambdak1}) and (\ref{orderorder}), and
the properties of $\{\lambda_k'\}_{k=0}^\infty$ in Lemma
\ref{lambdak1} in Appendix \ref{property}, we have the following
lemma.

\begin{Lem} \label{cut1}
For Gaussian random processes in $\mathcal{A}$, for any constant
$0 < \kappa <1$, we have
\begin{align}
R(\theta) &\geq  \frac{\kappa x d^{\frac{1}{x}}}{2}
\theta^{-{\frac{1}{x}}} \label{uselaterboundb1} \\
D(\theta) & \geq  \kappa \left(1+\frac{\kappa}{x-1}
\right)\frac{d^{\frac{1}{x}}}{T_0} \theta^{1-\frac{1}{x}}
\label{newtheorem2}
\end{align}
when $\theta$ is small enough .
\end{Lem}
A proof of Lemma \ref{cut1} is provided in Appendix \ref{Rlower}.

Using Lemma \ref{cut1}, we present in the next theorem a lower
bound for the distortion-rate function of the random process.
\begin{Theo} \label{Pat}
For Gaussian random processes in $\mathcal{A}$, for any constant
$0 < \kappa <1$, we have
\begin{align}
D_p(R) \geq \kappa \left(1+\frac{\kappa}{x-1}
\right)\left(\frac{\kappa x}{2}\right)^{x-1} \frac{d}{ T_0}
R^{1-x} \label{siggy2}
\end{align}
when $R$ is large enough.
\end{Theo}
A proof of Theorem \ref{Pat} is provided in Appendix \ref{happy}.

We will divide our discussion into two separate cases based on the
total power, $P(N)$. For the first case, $P(N)$ is such that
\begin{align}
\lim_{N \rightarrow \infty} \frac{1}{NP(N)}=0
\label{powerconstraint}
\end{align}
The cases $P(N)=N P_{\text{ind}}$ and $P(N)=P_{\text{tot}}$ are
included in $P(N)$ satisfying (\ref{powerconstraint}).
\begin{Theo} \label{laibuji1}
For Gaussian random processes in $\mathcal{A}$, when $P(N)$ is
such that (\ref{powerconstraint}) is satisfied,
for any constant $0 < \kappa < 1$, a lower bound on the minimum
achievable expected distortion is
\begin{align}
D_l^N=D_p(C_u^N)
 \geq \kappa^2 \left(1+\frac{\kappa}{x-1} \right)\left(\kappa
x\right)^{x-1} \frac{d}{ T_0} \left(\frac{1}{\log \left( N P(N)
\right)}\right)^{x-1} \label{singgy4}
\end{align}
when $N$ is large enough.
\end{Theo}
A proof of Theorem \ref{laibuji1} is provided in Appendix
\ref{laibuji2}.

Hence, when total power $P(N)$ satisfies (\ref{powerconstraint}),
a lower bound on the achievable distortion is
\begin{align}
\Theta \left(\left(\frac{1}{\log(NP(N))}\right)^{x-1} \right)
\end{align}

For the second case, $P(N)$ is such that (\ref{powerconstraint})
is not satisfied. In this case, $C_u^N$ is either a constant
independent of $N$ or goes to zero as $N$ goes to infinity. The
minimum achievable distortion does not go to zero with increasing
$N$.

Therefore, for all possible total power $P(N)$, a lower bound on
the distortion is
\begin{align}
\Theta \left(  \min \left(\left(\frac{1}{\log(NP(N))}
\right)^{x-1}, 1 \right)  \right) \label{barcelona}
\end{align}

When the total power $P(N)$ grows almost exponentially with the
number of nodes, the lower bound on the minimum achievable
expected distortion in (\ref{barcelona}) decreases inverse
polynomially with $N$. Even though this provides excellent
distortion performance, it is impractical since sensor nodes are
low energy devices and it is often difficult, if not impossible,
to replenish their batteries. When the total power $P(N)$ is such
that (\ref{powerconstraint}) is not satisfied, the transmission
power is so low that the communication channels between the
sensors and the collector node are as if they do not exist. From
(\ref{barcelona}), the lower bound on the estimation error in this
case is on the order of 1, which is equivalent to the collector
node blindly estimating $S(t)=0$ for all $t \in [0, T_0]$. Even
though the consumed total power $P(N)$ is very low in this case,
the performance of the sensor network is unacceptable; even the
lower bound on the minimum achievable expected distortion does not
decrease to zero with the increasing number of nodes. For
practically meaningful total power values, including the cases of
$P(N)=N P_{\text{ind}}$ and $P(N)=P_{\text{tot}}$, the lower bound
on the minimum achievable expected distortion in (\ref{barcelona})
decays to zero at the rate of
\begin{align}
\frac{1}{\left(\log N\right)^{x-1}}
\end{align}

\section{An Upper Bound on the Achievable Distortion} \label{secupperbound}
\subsection{Arbitrary Gaussian Random Processes}
Any distortion found by using any achievability scheme will serve
as an upper bound for the minimum achievable expected distortion.
We consider the following separation-based achievable scheme.
First, we perform multi-terminal rate-distortion coding at all
sensor nodes using \cite[Theorem 1]{Flynn:1987}. After obtaining
the indices of the rate-distortion codes, we transmit the indices
as independent messages using the amplify-and-forward method
introduced in \cite{Gastpar:2005}. The distortion obtained using
this scheme will be denoted as $D_u^N$.

We apply \cite[Theorem 1]{Flynn:1987}, generalized to $N$ sensor
nodes in \cite[Theorem 1]{Chen:2004}, to obtain an achievable
rate-distortion point.
\begin{Theo} \label{achratedis}
For all Gaussian random processes, if the individual rates are
equal, the following sum rate and distortion are achievable,
\begin{align}
D_a^N(\theta') &   = \frac{1}{T_0}\int_{0}^{T_0}
\left(K(t,t)-\frac{T_0}{N-1}\bm{\rho}_N^T(t)
\left(\Sigma_N'+\theta'
I\right)^{-1} \bm{\rho}_N(t) \right) dt \\
R_a^N(\theta')  & =\sum_{k=0}^{N-1} \frac{1}{2} \log
\left(1+\frac{\mu_k^{(N)'}}{\theta'} \right)
\end{align}
where
\begin{align}
\bm{\rho}_N(t)=\begin{bmatrix} K\left(t,0\right) &
K\left(t,\frac{T_0}{N-1}\right) & K\left(t,\frac{2
T_0}{N-1}\right) \cdots & K\left(t,T_0\right)
\end{bmatrix}^T
\end{align}
and
\begin{align}
\Sigma_N & =E[\mathbf{S}_N \mathbf{S}_N^T] \nonumber \\
&= \left[
\begin{array}{cccc}
K(0,0) & K\left(0,\frac{T_0}{N-1}\right) & \cdots & K\left(0,T_0\right)\\
K\left(\frac{T_0}{N-1},0\right) & K\left(\frac{T_0}{N-1},\frac{T_0}{N-1}\right) & \cdots & K\left(\frac{T_0}{N-1}, T_0\right)\\
\vdots & \vdots & \vdots & \vdots\\
K(T_0,0)& K\left(T_0,\frac{T_0}{N-1}\right) & \cdots & K(T_0,T_0)
\end{array} \right]
\end{align}
and $\Sigma_N'=\frac{T_0}{N-1}\Sigma_N$ and
$\mu_0^{(N)'},\mu_1^{(N)'},\cdots,\mu_{N-1}^{(N)'}$ are the
eigenvalues of $\Sigma_N'$.
\end{Theo}
A proof of Theorem \ref{achratedis} is provided in Appendix
\ref{achievableratedistortion}.

We further evaluate $D_a^N(\theta')$ in the next lemma.
\begin{Lem} \label{whatnoname}
For all Gaussian random processes, we have
\begin{align}
D_a^N(\theta') & \leq 2A^{(N)}+B^{(N)}+D_b^N(\theta')
\label{houjia}
\end{align}
where $A^{(N)}$, $B^{(N)}$ and $D_b^N(\theta') $ are defined as
\begin{align}
A^{(N)}=& \frac{1}{T_0}\sum_{i=1}^{N-1}
\int_{\frac{i-1}{N-1}T_0}^{\frac{i}{N-1}T_0}
\left(K(t,t)-K \left(\frac{i-1}{N-1}T_0,\frac{i-1}{N-1}T_0\right) \right) dt \nonumber \\
&+\frac{2}{T_0}\sum_{i=1}^{N-1}
\int_{\frac{i-1}{N-1}T_0}^{\frac{i}{N-1}T_0}
\left(\bm{\rho}_N\left(\frac{i-1}{N-1}T_0 \right)-\bm{\rho}_N(t)
\right)_i dt
\end{align}
and
\begin{align}
B^{(N)} =\frac{2}{T_0}\sum_{i=1}^{N-1}
\int_{\frac{i-1}{N-1}T_0}^{\frac{i}{N-1}T_0}
\left|\left|\bm{\rho}_N\left(\frac{i-1}{N-1}T_0
\right)-\bm{\rho}_N(t) \right|\right| dt
\end{align}
and
\begin{align}
D_b^N(\theta') = \frac{1}{T_0} \sum_{k=0}^{N-1}
\left(\frac{1}{\theta'}+\frac{1}{\mu_k^{(N)'}} \right)^{-1}
\end{align}
respectively.
\end{Lem}
A proof of Lemma \ref{whatnoname} is provided in Appendix
\ref{breakdistortion}. Lemma \ref{whatnoname} tells us that the
expected distortion achieved by using the separation-based scheme
is upper bounded by the sum of three types of distortion. The
first two types of distortion, $A^{(N)}$ and $B^{(N)}$, have
nothing to do with the rate and only depend on how well the
samples approximate the entire random process. The third
distortion, $D_b^N(\theta')$, depends on the rate through variable
$\theta'$.

Now, we determine an achievable rate for the communication channel
from the sensor nodes to the collector node. The channel in its
nature is a multiple access channel with potential cooperation
between the transmitters. The capacity region for this channel is
not known. We get an achievable sum rate, with identical
individual rates, for this channel by using the idea presented in
\cite{Gastpar:2005}.
\begin{Theo} \label{generalelgamal}
When the total power $P(N)$ is such that there exists an $\epsilon
> 0$ where
\begin{align}
\lim_{N \rightarrow \infty} P(N) N^{\frac{1}{2}-\epsilon}> 1
\label{powersad1}
\end{align}
for any constant $0 < \kappa < 1$, the following sum rate is
achievable,
\begin{align}
C_a^N= \kappa \nu \log (NP(N)) \label{goli}
\end{align}
where $\nu$ is a positive constant independent of $N$,
\begin{align}
\label{beta_def} \nu=  \min \left(\frac{ \epsilon}{1+2\epsilon},
\frac{1}{4} \right)
\end{align}
when $N$ is large enough. The individual rates of the sensor nodes
are the same. Otherwise, the sum rate approaches a positive
constant or zero as $N \rightarrow \infty$.
\end{Theo}
A proof of Theorem \ref{generalelgamal} is provided in Appendix
\ref{achievablerateamplify}. Theorem \ref{generalelgamal} shows
that when the total power is such that (\ref{powersad1}) is
satisfied, the achievable rate increases with $N$. Furthermore,
the achievable rate is the same as the upper bound on the
achievable sum rate in (\ref{Barbara}) order-wise. Otherwise, the
achievable rate is either a positive constant or decreases to
zero, which will result in poor estimation performance at the
collector node.

The function $R_a^N\left(\theta'\right)$ is a strictly decreasing
function of $\theta'$, thus, the inverse function exists, which we
will denote as $\theta_a^N(R)$. Let us define $D_a(R)$ as the
composition of the two functions $D_a^N(\theta')$ and
$\theta_a^N(R)$, i.e.,
\begin{align}
D_a(R)=D_a^N(\theta_a^N(R)) \label{lab}
\end{align}
An upper bound on the minimum achievable distortion, i.e., the
achievable distortion by the separation-based scheme described
above, is
\begin{align}
D_u^N & = D_a\left(C_a^N\right) \label{sigmaDeqn2}
\end{align}
We will perform this calculation when the underlying random
process is in $\mathcal{A}$.

\subsection{The Class of Gaussian Random Processes in $\mathcal{A}$}
We analyze the three types of distortion in (\ref{houjia}) for
Gaussian random processes in $\mathcal{A}$. We will focus on
$A^{(N)}$ and $B^{(N)}$ in Lemma \ref{addgeneral}, and on
$D_b^N(\theta')$ in Lemma \ref{nothingleft}.

\begin{Lem} \label{addgeneral}
For Gaussian random processes in $\mathcal{A}$, we have
\begin{align}
A^{(N)}&= O \left(N^{-\alpha} \right) \label{AA}\\
B^{(N)} & = O \left(N^{\frac{1}{2}-\alpha} \right) \label{BB}
\end{align}
\end{Lem}
A proof of Lemma \ref{addgeneral} is provided in Appendix
\ref{lipliplip}. The result depends crucially on condition 2 in
the definition of $\mathcal{A}$ in Section \ref{defineA}.  Note
that since $1/2 < \alpha \leq 1$, both $A^{(N)}$ and $B^{(N)}$
decrease to zero inverse polynomially as $N$ goes to infinity.

It remains to calculate the functions $R_a^N(\theta')$ and
$D_b^N(\theta')$ for random processes in $\mathcal{A}$. To do so,
we need some properties of $\{\mu_k^{(N)'}\}_{k=0}^{N-1}$ which
are stated in Lemmas \ref{appro} and \ref{divergence} and proved
in Appendix \ref{preproof}. Lemma \ref{appro} is of great
importance, as it serves as a tool to link
$\{\mu_k^{(N)'}\}_{k=0}^{N-1}$ to $\{\lambda_k\}_{k=0}^\infty,$
which is used in the derivation of the lower bound in Section
\ref{seclowerbound}, through the lower and upper bounds
$\{\lambda_k'\}_{k=0}^\infty$ and $\{\lambda_k''\}_{k=0}^\infty$.
Armed with the properties of $\mu_k^{(N)'}$, $\lambda_k'$ and
$\lambda_k''$ in Lemmas \ref{lambdak1}, \ref{lambdak2},
\ref{appro} and \ref{divergence} in Appendices \ref{property} and
\ref{preproof}, we can show the following lemma. First, we define
two sequences $\vartheta_L^N$ and $\vartheta_U^N$, which are
functions of $N$, that satisfy
\begin{align}
\lim_{N \rightarrow \infty} \frac{1}{\vartheta_L^N N^{\min
\left(\frac{x \gamma}{2\tau}, \frac{\alpha x}{x-1}, \frac{\beta
x}{x+\tau+1} \right)}} =0, \qquad \lim_{N \rightarrow \infty}
\vartheta_U^N =0 \label{increasing}
\end{align}

\begin{Lem} \label{nothingleft}
For Gaussian random processes in $\mathcal{A}$, for any constant
$0 < \kappa <1$, lower and upper bounds for the function
$R_a^N(\theta')$ are
\begin{align}
\frac{\kappa x d^{\frac{1}{x}}}{4} \theta'^{-\frac{1}{x}} \leq
R_a^N(\theta') \leq \frac{ d^{\frac{1}{x}}\left(x^2-(1-\log
2)x+(1-\log 2)\right)}{2(x-1) \kappa^2} \theta'^{-\frac{1}{x}}
\label{singapore1}
\end{align}
and an upper bound for the function $D_b^N(\theta')$ is
\begin{align}
D_b^N(\theta') \leq \frac{ d^{\frac{1}{x}} \left(1+\kappa^2(x-1)
\right)}{\kappa^3 (x-1) T_0} \theta'^{1-\frac{1}{x}}
\label{nuoconclude1}
\end{align}
for $\theta' \in [\vartheta_L^N, \vartheta_U^N]$ and $N$ large
enough.
\end{Lem}
A proof of Lemma \ref{nothingleft} is provided in Appendix
\ref{Rlowerupper}. The proof of Lemma \ref{nothingleft} uses
conditions 1, 2 and 3 in Section \ref{defineA}. Let us define a
sequence $\vartheta_{LL}^N$, which is a function of $N$, that
satisfies
\begin{align}
\lim_{N \rightarrow \infty} \frac{1}{\vartheta_{LL}^N N^{\min
\left(\frac{x \gamma}{2 \tau}, \frac{(\alpha-1/2) x}{x-1},
\frac{\beta x}{x+\tau+1} \right)}} =0 \label{bestbestluck}
\end{align}
Combining (\ref{houjia}), (\ref{AA}), (\ref{BB}),
(\ref{singapore1}) and (\ref{nuoconclude1}), we have the following
theorem.
\begin{Theo} \label{singhope}
For Gaussian random processes in $\mathcal{A}$, for any constant
$0 < \kappa <1$, the achievable distortion-rate function,
$D_a(R)$, is upper bounded as
\begin{align}
D_a(R) \leq \frac{ d(1+\kappa^2(x-1))\left( x^2-(1-\log 2) x +
(1-\log 2)\right)^{x-1}}{T_0 \kappa^{2x+2}2^{x-1}(x-1)^x} R^{1-x}
\label{siggy1}
\end{align}
for $R$ in the interval of
\begin{align}
\left[\frac{ d^{\frac{1}{x}}\left(x^2-(1-\log 2)x+(1-\log
2)\right)}{2(x-1) \kappa^2}
 \left(\vartheta_U^N \right)^{-\frac{1}{x}}, \frac{\kappa x
d^{\frac{1}{x}}}{4} \left(\vartheta_{LL}^N
\right)^{-\frac{1}{x}}\right] \label{partialinterval}
\end{align}
when $N$ is large enough.
\end{Theo}
A proof of Theorem \ref{singhope} is provided in Appendix
\ref{hopehopehope}. This theorem shows that when $R$ is in the
interval (\ref{partialinterval}), the achievable distortion-rate
function is the same as the lower bound on the distortion-rate
function in (\ref{siggy2}) order-wise.

\begin{Theo} \label{bestluck}
For Gaussian random processes in $\mathcal{A}$, when the sum power
constraint satisfies (\ref{powersad1}) and
\begin{align}
\lim_{N \rightarrow \infty} \frac{NP(N)}{e^{N^{\min
 \left(\frac{ \gamma}{2 \tau}, \frac{2 \alpha-1}{2(x-1)} ,\frac{\beta }{x+\tau+1} \right)}}}=0
\label{powerfinalhope}
\end{align}
an upper bound on the minimum achievable expected distortion, or
equivalently, the achievable rate in the separation-based scheme,
is
\begin{align}
D_u^N &= D_a\left(C_a^N\right) \\
&\leq \frac{ d(1+\kappa^2(x-1))\left( x^2-(1-\log 2) x + (1-\log
2)\right)^{x-1}}{T_0 \kappa^{3x+1}2^{x-1}(x-1)^x \nu^{x-1}}
\left(\frac{1}{\log (NP(N))}\right)^{x-1} \label{singgy5}
\end{align}
when $N$ is large enough.
\end{Theo}
A proof of Theorem \ref{bestluck} is provided in Appendix
\ref{bestluck1}. Theorem \ref{bestluck} implies that, when the sum
power constraint satisfies (\ref{powersad1}) and
(\ref{powerfinalhope}), an upper bound on the minimum achievable
expected distortion is
\begin{align}
\Theta \left( \left( \frac{1}{\log (NP(N))} \right)^{x-1} \right)
\label{upperboundrepeat}
\end{align}

For the interesting cases of $P(N)=N P_{\text{ind}}$ and
$P(N)=P_{\text{tot}}$, the upper bound on the minimum achievable
expected distortion decays to zero at the rate of
\begin{align}
\frac{1}{\left(\log N\right)^{x-1}}
\end{align}
When the sum power constraint is such that (\ref{powersad1}) is
not satisfied, an upper bound on the minimum achievable expected
distortion is $\Theta(1)$.

\section{Comparison of the Lower and Upper Bounds for Gaussian Random
Processes in $\mathcal{A}$} \label{ddd}

\subsection{Order-wise Comparison of Lower and Upper Bounds}
In this section, we compare the lower bound in (\ref{barcelona})
and the upper bound in (\ref{upperboundrepeat}). When the total
power is large, i.e., $P(N)$ is so large that
(\ref{powerfinalhope}) is not satisfied, our methods in finding
the upper bound do not apply. Even though our lower bound in
(\ref{barcelona}) is valid, we have not shown whether the lower
and upper bounds meet. However, in this case, $P(N)$ is larger
than $\frac{e^{N^{\min
 \left(\frac{ \gamma}{2 \tau}, \frac{2 \alpha-1}{2(x-1)} ,\frac{\beta }{x+\tau+1} \right)}}}{N}$, and this
region of total power is not of practical interest.

When the total power is medium, i.e., $P(N)$ is in the wide range
of $N^{-1/2+\epsilon}$ to \break $\frac{e^{N^{\min
 \left(\frac{ \gamma}{2 \tau}, \frac{2 \alpha-1}{2(x-1)} ,\frac{\beta }{x+\tau+1} \right)}}}{N}$, our lower and upper bounds do meet and the minimum
achievable expected distortion is
\begin{align}
D^N=\Theta \left(\frac{1}{\left(\log (NP(N)) \right)^{x-1}}
\right)
\end{align}
The order-optimal achievability scheme is a separation-based
scheme, which uses distributed rate-distortion coding as described
in \cite{Flynn:1987} and optimal single-user channel coding with
amplify-and-forward method as described in \cite{Gastpar:2005}. In
fact, when the total power is medium, as shown in (\ref{siggy2})
and (\ref{siggy1}), lower and upper bounds on the distortion-rate
function, $D_p(R)$ and $D_a(R)$ coincide order-wise. In addition,
as shown in (\ref{Barbara}) and (\ref{goli}), the lower and upper
bounds on the achievable sum rate, $C_a^N$ and $C_u^N$, coincide
order-wise as well. The practically interesting cases of $P(N)=N
P_{\text{ind}}$ and $P(N)=P_{\text{tot}}$ fall into this region of
medium total power. In both of these cases, the minimum achievable
expected distortion decreases to zero at the rate of
\begin{align}
\frac{1}{\left(\log N \right)^{x-1}} \label{boring}
\end{align}
Hence, the total power $P(N)=P_{\text{tot}}$ performs as well as
$P(N)=N P_{\text{ind}}$ ``order-wise'', and therefore, in practice
we may prefer to choose $P(N)=P_{\text{tot}}$. In fact, we can
decrease the total power to $P(N)=N^{-1/3}$ and the minimum
achievable distortion will still decrease to zero at the rate in
(\ref{boring}).

When the total power is small, i.e., $P(N)$ ranges from $N^{-1}$
to $N^{-1/2}$, our lower and upper bounds do not meet. Our lower
bound in (\ref{barcelona}) decreases to zero as
$\frac{1}{\left(\log N \right)^{x-1}}$ but our upper bound is a
non-zero constant. The main discrepancy between our lower and
upper bounds comes from the gap between the lower and upper bounds
on the sum capacities, $C_a^N$ and $C_u^N$, for a cooperative
multiple access channel. In fact, when the total power is small,
as shown in (\ref{siggy2}) and (\ref{siggy1}), lower and upper
bounds on the distortion-rate function, $D_p(R)$ and $D_a(R)$
still coincide order-wise. This total power region should be of
practical interest, because in this region, the sum power
constraint is quite low, and yet the lower bound on the distortion
is of the same order as one would obtain with any $P(N)$ which
increases polynomially with $N$.  Hence, from the lower bound, it
seems that this region potentially has good performance. However,
our separation-based upper bound does not meet the lower bound,
and whether the lower bound can be achieved remains an open
problem.

When the total power is very small, i.e., $P(N)$ is less than
$N^{-1}$, our lower and upper bounds meet and the minimum
achievable expected distortion is a constant that does not
decrease to zero with increasing $N$. This case is not of
practical interest because of the unacceptable distortion.

In the case of Gauss-Markov random process, we have $x=2$ and
$\alpha=\beta=\tau=\gamma=1$. Inserting these values into the
above results, we see that in the medium total power region, i.e.,
$P(N)$ is in the wide range of $N^{-1/2+\epsilon}$ to
$\frac{e^{N^{1/4}}}{N}$, the minimum achievable expected
distortion is
\begin{align}
D^N=\Theta \left(\frac{1}{\log (NP(N)) } \right)
\label{zuihouzuihou1}
\end{align}
For the Gauss-Markov random process, in the cases of $P(N)=N
P_{\text{ind}}$ and $P(N)=P_{\text{tot}}$, the minimum achievable
expected distortion decreases to zero at the rate of
\begin{align}
\frac{1}{\log N } \label{zuihouzuihou2}
\end{align}
The conclusions in (\ref{zuihouzuihou1}) and (\ref{zuihouzuihou2})
were derived in \cite{Liu_Ulukus:2006} under a different channel
assumption. For the channel assumption in \cite{Liu_Ulukus:2006},
the order-optimal achievability scheme was determined to be a
decode-and-forward based scheme. The range of medium power constraints was shown to be slightly larger in \cite{Liu_Ulukus:2006},
i.e., $P(N)$ in the range
of $N^{-1/2+\epsilon}$ to
$\frac{e^{N^{1/3}}}{N}$, and this is because it was specifically derived for the Gauss-Markov process,
instead of general Gaussian random processes as in this work.

\subsection{Comparison of the Constants in the Lower and Upper Bounds}
Though the lower and upper bounds meet order-wise in a wide range
of total power constraints, the constants in front of them are
different and we aim to compare these constants for various total
power constraints in this section.

Combining (\ref{singgy4}) and (\ref{singgy5}), when $P(N)$
satisfies (\ref{powersad1}) and (\ref{powerfinalhope}), the
minimum distortion $D^N$ satisfies

\begin{align}
\kappa^2 & \left(1+\frac{\kappa}{x-1} \right)\left(\kappa
x\right)^{x-1} \frac{d}{ T_0}
\left(\frac{1}{\log \left( N P(N) \right)}\right)^{x-1} \leq D^N \nonumber \\
& \leq \frac{ d(1+\kappa^2(x-1))\left( x^2-(1-\log 2) x + (1-\log
2)\right)^{x-1}}{T_0 \kappa^{3x+1}2^{x-1}(x-1)^x \nu^{x-1}}
\left(\frac{1}{\log (NP(N))}\right)^{x-1}
\end{align}
Note that $\kappa$ can be made as close to 1 as possible for large
enough $N$. Let $\pi(x,\nu)$ be the ratio of the constant in the
lower bound and the constant in the upper bound when $N$ is large
enough. Then,
\begin{align}
\pi(x,\nu)=\left(2\nu\right)^{x-1} \left(\frac{x^2-x}{x^2-(1-\log
2) x + (1-\log 2)}\right)^{x-1}
\end{align}
Here, $x$ is a parameter of the underlying Gaussian random process
and $\nu$ depends on the total power constraint of the sensor
nodes, $P(N)$. It is straightforward to see that since from
(\ref{beta_def}), $\nu \leq 1/4$, $\pi(x,\nu)$ is a monotonically
decreasing function of $x$ for a fixed $\nu$. Hence, we conclude
that the constants in front of the lower and upper bounds differ
more as $x$ gets large. Since $x$ is an indication of how much
randomness the random process contains, this means that the more
random the random process, the more the constants in the lower and
upper bounds meet. For a fixed underlying random process, i.e.,
for a fixed $x$, $\pi(x,\nu)$ is a decreasing function of $\nu$.
This means that the less the total power we have, the more
different the constants will be.

In the Gauss-Markov random process, $x=2$. When $P(N)=N
P_{\text{ind}}$ and $P(N)=P_{\text{tot}}$, the ratio of the two
constants is
\begin{align}
\pi(2,1/4)= \frac{1}{3+\log 2} \simeq 0.2708
\end{align}
When $P(N)=N^{-\omega}$, $0 < \omega < \frac{1}{2}$, the ratio of
the two constants is
\begin{align}
\pi\left(2,\frac{1}{2}-\frac{1}{4}\frac{1}{1-\omega}\right)=\left(\frac{1}{2}-\frac{1}{4}\frac{1}{1-\omega}\right)\frac{4}{3+\log
2}
\end{align}
For example, when $P(N)=N^{-1/3}$, the ratio of the constants is
\begin{align}
\pi\left(2,1/8\right)= \frac{1}{2}\pi(2,1/4) \simeq 0.1354
\end{align}

\section{Further Remarks} \label{sect:discussion}
We have shown that the minimum achievable expected distortion
behaves order-wise as
\begin{align}
\Theta \left( \left( \frac{1}{\log (NP(N))} \right)^{x-1} \right)
\label{everything}
\end{align}
Due to the order-optimality of separation, this result can be
generalized straightforwardly to several other scenarios.

The result in (\ref{everything}) still holds when we allow the
collector node to use its transmit antenna with an arbitrary power
constraint. The collector node, using its transmit antenna, can
send some form of feedback to the sensor nodes. However, the lower
bound on the minimum distortion remains unchanged in this case,
because in deriving our lower bound, we assumed that all sensor
nodes know the entire random process, thus, forming a
point-to-point system. In a point-to-point system, feedback,
perfect or not, does not change the capacity. Meanwhile, our upper
bound is still valid, as in this achievable scheme, we choose not
to utilize the feedback link. Hence, our result in
(\ref{everything}) remains valid. For similar reasons, our result
in (\ref{everything}) remains valid, when we consider a sum power
constraint $P(N)$, instead of individual identical power
constraints of $P(N)/N$ for all sensors.

The result in (\ref{everything}) still holds when we allow $K$
channel uses per realization of the random process, where $K$ is a
constant independent of $N$. This is because both lower and upper
bounds are derived using separation-based schemes. The minimum
achievable distortion still behaves as (\ref{everything}), and the
number $K$ will only effect the constant in front. Due to the same
reasoning, the minimum achievable distortion behaves as
(\ref{everything}) when we allow multiple transmit and receive
antennas at each node, as long as the number of antennas on each
node is a constant, independent of $N$.

\section{Conclusions}
In this paper, we investigated the performance of dense sensor
networks by studying the joint source-channel coding problem. We
provided lower and upper bounds for the minimum achievable
expected distortion when the underlying random process is
Gaussian. When the random process satisfies some general
conditions, we evaluated the lower and upper bounds explicitly,
and showed that they are both of order $\frac{1}{\left(\log (N
P(N))\right)^{x-1}}$ for a wide range of total power ranging from
$N^{-\frac{1}{2}+\epsilon}$ to $\frac{e^{N^{\min
 \left(\frac{ \gamma}{2 \tau}, \frac{2 \alpha-1}{2(x-1)} ,\frac{\beta }{x+\tau+1} \right)}}}{N}$. In the most interesting cases when the total power
is a constant or grows linearly with $N$, the minimum achievable
expected distortion decreases to zero at the rate of
$\frac{1}{\left(\log N \right)^{x-1}}$. For random processes that
satisfy these general conditions, under these power constraints,
we have found that an order-optimal scheme is a separation-based
scheme, that is composed of distributed rate-distortion coding
\cite{Flynn:1987} and amplify-and-forward channel coding
\cite{Gastpar:2005}.

\section{Appendix}
\subsection{Some properties of $\lambda_k'$ and $\lambda_k''$} \label{property}
In this subsection, we provide two lemmas which characterize some
properties of $\{\lambda_k'\}_{k=0}^\infty$ and
$\{\lambda_k''\}_{k=0}^\infty$, defined in (\ref{definelambdak1})
and (\ref{definelambdak2}), which will be useful in deriving our
main results.
\begin{Lem} \label{lambdak1}
For any constant $0 < \kappa <1$, we have
\begin{align}
\sum_{k=\left\lfloor\frac{d^{\frac{1}{x}}}{\theta^{\frac{1}{x}}}
-c_l+1\right\rfloor}^\infty \lambda_k' \geq \frac{\kappa
d^{\frac{1}{x}}}{(x-1)} \theta^{1-\frac{1}{x}} \label{needlower1}
 \end{align}
and
\begin{align}
\sum_{k=0}^{\left\lfloor\frac{d^{\frac{1}{x}}}{\theta^{\frac{1}{x}}}
-c_l\right\rfloor}\frac{1}{2} \log\left(\frac{\lambda_k'}{\theta}
\right) \geq  \frac{\kappa x d^{\frac{1}{x}}}{2}
\theta^{-{\frac{1}{x}}} \label{needlower2}
\end{align}
when $\theta$ is small enough.
\end{Lem}

\begin{Lem} \label{lambdak2}
For any constant $0 < \kappa <1$, we have
\begin{align}
 \sum_{k=\left\lfloor \left(\frac{d}{\theta}
\right)^{\frac{1}{x}}+c_u \right\rfloor+1}^\infty
 \lambda_k''  \leq \frac{ d^\frac{1}{x}}{(x-1) \kappa} \theta^{1-\frac{1}{x}}  \label{needupper}
\end{align}
and
\begin{align}
\sum_{k=0}^{\left\lfloor \left(\frac{d}{\theta}
\right)^{\frac{1}{x}}+c_u \right\rfloor}& \frac{1}{2} \log
\left(1+\frac{\lambda_k''}{\theta} \right) \leq \left(\frac{\log
2+x}{2 \kappa}  \right) d^{\frac{1}{x}} \theta^{-\frac{1}{x}}
\label{needupper2}
\end{align}
when $\theta$ is small enough.
\end{Lem}

\subsubsection{Proof of Lemma \ref{lambdak1}}
We will first prove (\ref{needlower1}).
\begin{align}
\sum_{k=\left\lfloor\frac{d^{\frac{1}{x}}}{\theta^{\frac{1}{x}}}
-c_l+1\right\rfloor}^\infty \lambda_k' &=
\sum_{k=\left\lfloor\frac{d^{\frac{1}{x}}}{\theta^{\frac{1}{x}}}
-c_l+1\right\rfloor}^\infty \frac{d}{(k+c_l)^x} \label{need1}\\
&
=d\sum_{k=\left\lfloor\frac{d^{\frac{1}{x}}}{\theta^{\frac{1}{x}}}
+1\right\rfloor}^ \infty \frac{1}{k^x} \\
 & \geq \frac{d}{x-1} \frac{1}{\left\lfloor\frac{d^{\frac{1}{x}}}{\theta^{\frac{1}{x}}}
+1\right\rfloor^{x-1}} \label{need2} \\
& \geq  \frac{\kappa d^{\frac{1}{x}}}{(x-1)}
\theta^{1-\frac{1}{x}} \label{needum}
 \end{align}
 where (\ref{need1}) is true when $\theta$ is small enough,
 more specifically, when $\left\lfloor\frac{d^{\frac{1}{x}}}{\theta^{\frac{1}{x}}}
-c_l+1\right\rfloor>K_0$. We have (\ref{need2}) because of the
inequality
\begin{align}
\sum_{k=n}^\infty \frac{1}{k^x} \geq \int_n^\infty \frac{1}{y^x}
dy=  \frac{1}{(x-1)n^{x-1}}
\end{align}
and (\ref{needum}) is true when $\theta$ is small enough, i.e.,
for any $0 < \kappa < 1$, there exists a $\theta_0(\kappa)>0$ such
that when $0 < \theta \leq \theta_0(\kappa)$, (\ref{needum}) is
true.

Next, we will prove (\ref{needlower2}).
\begin{align}
&\sum_{k=0}^{\left\lfloor\frac{d^{\frac{1}{x}}}{\theta^{\frac{1}{x}}}
-c_l\right\rfloor}\frac{1}{2} \log\left(\frac{\lambda_k'}{\theta}
\right) \nonumber \\
 =& \sum_{k=0}^{K_0} \frac{1}{2} \log
\left(\frac{\lambda_k}{\theta} \right) +
\sum_{k=K_0+1}^{\left\lfloor\frac{d^{\frac{1}{x}}}{\theta^{\frac{1}{x}}}\right\rfloor
-c_l} \frac{1}{2} \log
\left(\frac{d}{\left(k+c_l\right)^x \theta} \right) \label{smalltheta17}\\
=&\sum_{k=0}^{K_0} \frac{1}{2} \log \left(\frac{\lambda_k}{\theta}
\right)+
\sum_{k=K_0+c_l+1}^{\left\lfloor\frac{d^{\frac{1}{x}}}{\theta^{\frac{1}{x}}}\right\rfloor
} \frac{1}{2} \log
\left(\frac{d}{k^x \theta} \right)\\
 = & \sum_{k=0}^{K_0} \frac{1}{2} \log
\left(\frac{\lambda_k}{d} \right)+\sum_{k=0}^{K_0} \frac{1}{2}
\log \left(\frac{d}{\theta} \right) +
\frac{1}{2}\left(\left\lfloor\frac{d^{\frac{1}{x}}}{\theta^{\frac{1}{x}}}\right\rfloor
-c_l-K_0 \right) \log \left(\frac{d}{\theta} \right) -
\frac{x}{2}\log
\prod_{k=K_0+c_l+1}^{\left\lfloor\frac{d^{\frac{1}{x}}}{\theta^{\frac{1}{x}}}\right\rfloor
} k \\
 = & \sum_{k=0}^{K_0} \frac{1}{2} \log
\left(\frac{\lambda_k}{d} \right) +
\frac{1}{2}\left(\left\lfloor\frac{d^{\frac{1}{x}}}{\theta^{\frac{1}{x}}}\right\rfloor
-c_l+1 \right) \log \left(\frac{d}{\theta} \right) -
\frac{x}{2}\log \left(
\left\lfloor\frac{d^{\frac{1}{x}}}{\theta^{\frac{1}{x}}}\right\rfloor!
\right)
+\frac{x}{2} \log \left((K_0+c_l)! \right) \\
\geq &
\frac{1}{2}\left(\left\lfloor\frac{d^{\frac{1}{x}}}{\theta^{\frac{1}{x}}}\right\rfloor
-c_l+1 \right) \log \left(\frac{d}{\theta} \right)- \frac{x}{2}
\left(\left\lfloor\frac{d^{\frac{1}{x}}}{\theta^{\frac{1}{x}}}\right\rfloor+\frac{1}{2}
\right) \log
\left\lfloor\frac{d^{\frac{1}{x}}}{\theta^{\frac{1}{x}}}\right\rfloor
+ \frac{x}{2}
\left\lfloor\frac{d^{\frac{1}{x}}}{\theta^{\frac{1}{x}}}\right\rfloor
-
\frac{x}{24 \left\lfloor\frac{d^{\frac{1}{x}}}{\theta^{\frac{1}{x}}}\right\rfloor} \nonumber \\
& +\sum_{k=0}^{K_0} \frac{1}{2} \log
\left(\frac{\lambda_k}{d} \right) +\frac{x}{2} \log \left((K_0+c_l)! \right) -\frac{x}{4} \log(2 \pi)\label{stirlingagain}\\
\geq & \frac{x}{2}
\left\lfloor\frac{d^{\frac{1}{x}}}{\theta^{\frac{1}{x}}}\right\rfloor
+ \frac{x}{2} \left(-c_l+\frac{1}{2} \right) \log
\left\lfloor\frac{d^{\frac{1}{x}}}{\theta^{\frac{1}{x}}}\right\rfloor
-
\frac{x}{24 \left\lfloor\frac{d^{\frac{1}{x}}}{\theta^{\frac{1}{x}}}\right\rfloor} +c_3 \label{definec3}\\
\geq & \frac{\kappa x d^{\frac{1}{x}}}{2} \theta^{-{\frac{1}{x}}}
\label{thetasmallenough1}
\end{align}
where (\ref{smalltheta17}) is true when $\theta$ is small enough,
more specifically, when
$\left\lfloor\frac{d^{\frac{1}{x}}}{\theta^{\frac{1}{x}}}
-c_l\right\rfloor > K_0$. (\ref{stirlingagain}) follows by using
Stirling's approximation,
\begin{align}
n! < \sqrt{2 \pi} n^{n+\frac{1}{2}}e^{-n+\frac{1}{12n}}
\end{align}
(\ref{definec3}) follows because $c_3$ is a constant, independent
of $\theta$, defined as
\begin{align}
c_3 \overset{\triangle} = \sum_{k=0}^{K_0} \frac{1}{2} \log
\left(\frac{\lambda_k}{d} \right) +\frac{x}{2} \log
\left((K_0+c_l)! \right) -\frac{x}{4} \log(2 \pi)
\end{align}
and (\ref{thetasmallenough1}) is true when $\theta$ is small
enough, i.e., for any $0 < \kappa < 1$, there exists a
$\theta_1(\kappa)>0$ such that when $0 < \theta \leq
\theta_1(\kappa)$, (\ref{thetasmallenough1}) is true.

Therefore, for any $0 < \kappa <1$, (\ref{needlower1}) and
(\ref{needlower2}) hold when $\theta$ is small enough.


\subsubsection{Proof of Lemma \ref{lambdak2}}
We will first prove (\ref{needupper}).
\begin{align}
 \sum_{k=\left\lfloor \left(\frac{d}{\theta} \right)^{\frac{1}{x}}+c_u \right\rfloor+1}^\infty
 \lambda_k''& =  \sum_{k=\left\lfloor \left(\frac{d}{\theta} \right)^{\frac{1}{x}}+c_u \right\rfloor+1}^\infty
 \frac{d}{(k-c_u)^x} \label{yexuyao}\\
 & = \sum_{k=\left\lfloor \left(\frac{d}{\theta} \right)^{\frac{1}{x}} \right\rfloor+1}^\infty
 \frac{d}{k^x}\\
 & = \frac{d}{(x-1)\left(\left\lfloor \left(\frac{d}{\theta} \right)^{\frac{1}{x}} \right\rfloor \right)^{x-1}}
 \label{anotherint} \\
 & \leq \frac{ d^\frac{1}{x}}{(x-1) \kappa} \theta^{1-\frac{1}{x}} \label{latersmall3}
\end{align}
where (\ref{yexuyao}) follows when $\theta$ is small enough, more
specifically, when $\left\lfloor \left(\frac{d}{\theta}
\right)^{\frac{1}{x}}+c_u \right\rfloor+1>K_0$. In obtaining
(\ref{anotherint}) we used
\begin{align}
\sum_{k=n}^\infty \frac{1}{k^x} \leq \int_{n-1}^\infty
\frac{1}{y^x} dy = \frac{1}{(x-1)(n-1)^{x-1}}
\end{align}
and (\ref{latersmall3}) follows when $\theta$ is small enough,
i.e., for any $0 < \kappa < 1$, there exists a
$\theta_2(\kappa)>0$ such that when $0 < \theta \leq
\theta_2(\kappa)$, (\ref{latersmall3}) is true.

Next, we will prove (\ref{needupper2}).
\begin{align}
\sum_{k=0}^{\left\lfloor \left(\frac{d}{\theta}
\right)^{\frac{1}{x}}+c_u \right\rfloor}& \frac{1}{2} \log
\left(1+\frac{\lambda_k''}{\theta} \right)
\nonumber \\
= & \sum_{k=0}^{K_0} \frac{1}{2} \log
\left(1+\frac{\lambda_k}{\theta} \right)+
\sum_{k=K_0+1}^{\left\lfloor \left(\frac{d}{\theta}
\right)^{\frac{1}{x}}+c_u \right\rfloor} \frac{1}{2} \log
\left(1+\frac{d}{(k-c_u)^x\theta} \right)\label{worry}\\
\leq & \sum_{k=0}^{K_0} \frac{1}{2} \log
\left(\frac{2\lambda_k}{\theta} \right)+
\sum_{k=K_0+1}^{\left\lfloor \left(\frac{d}{\theta}
\right)^{\frac{1}{x}}+c_u \right\rfloor} \frac{1}{2} \log
\left(\frac{2d}{(k-c_u)^x\theta} \right) \label{trueforanyk}\\
= & \sum_{k=0}^{K_0} \frac{1}{2} \log
\left(\frac{2\lambda_k}{\theta} \right)+ \sum_{k=c_u+1}^
{\left\lfloor \left(\frac{d}{\theta} \right)^{\frac{1}{x}}+c_u
\right\rfloor} \frac{1}{2} \log \left(\frac{2d}{(k-c_u)^x\theta}
\right)-\sum_{k=c_u+1}^{K_0} \frac{1}{2} \log
\left(\frac{2d}{(k-c_u)^x\theta} \right)\\
= & \left\lfloor \left(\frac{d}{\theta}
\right)^{\frac{1}{x}}\right\rfloor \frac{1}{2} \log 2 -\frac{x}{2}
\log \left(\left\lfloor \left(\frac{d}{\theta}
\right)^{\frac{1}{x}} \right\rfloor ! \right) + \frac{1}{2}
\left(\left\lfloor \left(\frac{d}{\theta} \right)^{\frac{1}{x}}
\right\rfloor \right) \log \frac{d}{\theta}
\nonumber \\
& +\frac{c_u+1}{2} \log \frac{d}{\theta}  + c_1 \label{laterconstantc1}\\
\leq & \left\lfloor \left(\frac{d}{\theta}
\right)^{\frac{1}{x}}\right\rfloor \frac{1}{2} \log 2 -\frac{x}{4}
\log (2 \pi) -\frac{x}{2} \left(\left\lfloor
\left(\frac{d}{\theta} \right)^{\frac{1}{x}}\right\rfloor +
\frac{1}{2} \right) \log \left\lfloor \left(\frac{d}{\theta}
\right)^{\frac{1}{x}}\right\rfloor + \frac{x}{2}
\left\lfloor \left(\frac{d}{\theta} \right)^{\frac{1}{x}}\right\rfloor \nonumber \\
& -\frac{x}{24 \left\lfloor \left(\frac{d}{\theta}
\right)^{\frac{1}{x}}\right\rfloor+2} + \frac{1}{2}
\left(\left\lfloor \left(\frac{d}{\theta} \right)^{\frac{1}{x}}
\right\rfloor \right) \log \frac{d}{\theta}
 +\frac{c_u+1}{2} \log \frac{d}{\theta} + c_1 \label{stirlinguse}\\
 = & \left\lfloor \left(\frac{d}{\theta} \right)^{\frac{1}{x}}\right\rfloor \left(\frac{\log 2 + x}{2} \right)
+ \frac{x}{2} \left\lfloor \left(\frac{d}{\theta}
\right)^{\frac{1}{x}}\right\rfloor \log
\frac{\left(\frac{d}{\theta} \right)^{\frac{1}{x}}} {\left\lfloor
\left(\frac{d}{\theta} \right)^{\frac{1}{x}}\right\rfloor}
-\frac{x}{4} \log \left\lfloor \left(\frac{d}{\theta} \right)^{\frac{1}{x}}\right\rfloor \nonumber \\
& +\frac{c_u+1}{2} \log \frac{d}{\theta} -\frac{x}{24 \left\lfloor
\left(\frac{d}{\theta} \right)^{\frac{1}{x}}\right\rfloor+2}
-\frac{x}{4} \log (2 \pi)
 + c_1 \\
\leq & \left\lfloor \left(\frac{d}{\theta}
\right)^{\frac{1}{x}}\right\rfloor \left(\frac{\log 2 + x}{2}
\right) + \frac{x}{2} \left\lfloor \left(\frac{d}{\theta}
\right)^{\frac{1}{x}}\right\rfloor \log \left( 1+ \frac{1}
{\left\lfloor \left(\frac{d}{\theta}
\right)^{\frac{1}{x}}\right\rfloor} \right)
-\frac{x}{4} \log \left\lfloor \left(\frac{d}{\theta} \right)^{\frac{1}{x}}\right\rfloor \nonumber \\
& +\frac{c_u+1}{2} \log \frac{d}{\theta} -\frac{x}{24 \left\lfloor
\left(\frac{d}{\theta} \right)^{\frac{1}{x}}\right\rfloor+2}
-\frac{x}{4} \log (2 \pi)
 + c_1 \\
 \leq & \left\lfloor \left(\frac{d}{\theta} \right)^{\frac{1}{x}}\right\rfloor \left(\frac{\log 2+x}{2}  \right)
-\frac{x}{4} \log \left\lfloor \left(\frac{d}{\theta}
\right)^{\frac{1}{x}}\right\rfloor +\frac{c_u+1}{2} \log
\frac{d}{\theta}
-\frac{x}{24\left\lfloor \left(\frac{d}{\theta} \right)^{\frac{1}{x}}\right\rfloor+2} \nonumber \\
& -\frac{x}{4} \log (2 \pi)+\frac{x}{2}
 + c_1
 \label{explain}\\
 \leq & \left(\frac{\log 2+x}{2 \kappa }  \right) d^{\frac{1}{x}} \theta^{-\frac{1}{x}} \label{thetasmall2}
 \end{align}
 where (\ref{worry}) is true when $\theta$ is small enough, more specifically, when
 $\left\lfloor \left(\frac{d}{\theta}
\right)^{\frac{1}{x}}+c_u \right\rfloor > K_0$. We have
(\ref{trueforanyk}) because
\begin{align}
\frac{d}{(k-c_u)^x \theta}>1
\end{align}
for all $k$ between $K_0+1$ and $\left\lfloor
\left(\frac{d}{\theta} \right)^{\frac{1}{x}}+c_u \right\rfloor$,
and when $\theta$ is small enough such that
\begin{align}
\theta \leq \lambda_k, \quad k=1,2,\cdots, K_0
\end{align}
We have (\ref{laterconstantc1}) because we defined
\begin{align}
c_1 \overset{\triangle}{=} \sum_{k=1}^{K_0} \frac{1}{2} \log
\frac{2 \lambda_k}{d} -\sum_{k=c_u+1}^{K_0} \frac{1}{2} \log
\frac{2}{(k-c_u)^x}
\end{align}
We used Stirling's approximation,
\begin{align}
n! >\sqrt{2 \pi} n^{n+\frac{1}{2}}e^{-n+\frac{1}{12n+1}}
\end{align}
to obtain (\ref{stirlinguse}), and (\ref{explain}) follows by
using
\begin{align}
\log (1+x) \leq x, \qquad  x \geq 0 \label{loglog}
\end{align}
(\ref{thetasmall2}) follows when $\theta$ is small enough, i.e.,
for any $0 < \kappa < 1$, there exists a $\theta_3(\kappa)>0$ such
that when $0 < \theta \leq \theta_3(\kappa)$, (\ref{thetasmall2})
is true.

Therefore, for any $0 < \kappa <1$, (\ref{needupper}) and
(\ref{needupper2}) hold when $\theta$ is small enough.



\subsection{Proof of Lemma \ref{cut1}} \label{Rlower}
For any $0 < \kappa <1$, when $\theta$ is small enough, the
results of Lemma \ref{lambdak1} hold.

From (\ref{rate}), we have
\begin{align}
R(\theta) &= \sum_{k=0}^\infty \max \left(0, \frac{1}{2} \log \left(\frac{\lambda_k}{\theta} \right) \right)\\
& \geq \sum_{k=0}^\infty \max \left(0, \frac{1}{2} \log
\left(\frac{\lambda_k'}{\theta} \right) \right)
\label{introduceprime}\\
& =
\sum_{k=0}^{\left\lfloor\frac{d^{\frac{1}{x}}}{\theta^{\frac{1}{x}}}
-c_l\right\rfloor}\frac{1}{2} \log\left(\frac{\lambda_k'}{\theta}
\right) \label{Atlantis}\\
& \geq  \frac{\kappa x d^{\frac{1}{x}}}{2} \theta^{-{\frac{1}{x}}}
\label{uselaterbound1}
\end{align}
where in (\ref{introduceprime}) we have used the definition of
sequence $\lambda_k'$ in (\ref{definelambdak1}) and the
observation in (\ref{orderorder}). (\ref{Atlantis}) follows when
$\theta$ is small enough, more specifically, when
$\theta<\lambda_{K_0}$ and
$\left\lfloor\frac{d^{\frac{1}{x}}}{\theta^{\frac{1}{x}}}
-c_l\right\rfloor> K_0$. (\ref{uselaterbound1}) follows from
(\ref{needlower2}) in Lemma \ref{lambdak1}.

From (\ref{distortion}), we have
\begin{align}
D(\theta) & = T_0^{-1} \sum_{k=0}^\infty \min (\theta, \lambda_k)\\
& \geq T_0^{-1} \sum_{k=0}^\infty \min (\theta, \lambda_k') \label{introduceprime2}\\
&= T_0^{-1}
\sum_{k=0}^{\left\lfloor\frac{d^{\frac{1}{x}}}{\theta^{\frac{1}{x}}}
-c_l\right\rfloor} \theta +T_0^{-1}
\sum_{\left\lfloor\frac{d^{\frac{1}{x}}}{\theta^{\frac{1}{x}}}
-c_l+1\right\rfloor}^\infty \lambda_k' \label{Dallas}\\
&\geq T_0^{-1}
\left(\left\lfloor\frac{d^{\frac{1}{x}}}{\theta^{\frac{1}{x}}}
\right\rfloor-c_l+1\right) \theta +T_0^{-1} \frac{\kappa d^{\frac{1}{x}}}{(x-1)} \theta^{1-\frac{1}{x}} \label{uselemma2}\\
& \geq \kappa \left(1+\frac{\kappa}{x-1}
\right)\frac{d^{\frac{1}{x}}}{T_0} \theta^{1-\frac{1}{x}}
\label{largeN2}
\end{align}
where in (\ref{introduceprime2}) we have used the definition of
sequence $\lambda_k'$ in (\ref{definelambdak1}) and the
observation in (\ref{orderorder}). (\ref{Dallas}) follows when
$\theta$ is small enough, more specifically, when
$\theta<\lambda_{K_0}$ and
$\left\lfloor\frac{d^{\frac{1}{x}}}{\theta^{\frac{1}{x}}}
-c_l+1\right\rfloor> K_0$. (\ref{uselemma2}) follows from
(\ref{needlower1}) in Lemma \ref{lambdak1}. (\ref{largeN2}) is
true for small enough $\theta$, i.e., for any $0 < \kappa < 1$,
there exists a $\theta_4(\kappa)>0$ such that when $0 < \theta
\leq \theta_4(\kappa)$, (\ref{largeN2}) is true.

Therefore, for any $0 < \kappa <1$, (\ref{uselaterboundb1}) and
(\ref{newtheorem2}) hold when $\theta$ is small enough.


\subsection{Proof of Theorem \ref{Pat}} \label{happy}
$R(\theta)$ is a strictly decreasing function when
$\theta<\lambda_1$. Hence, when $\theta< \lambda_1$, the inverse
function $\theta(R)$ exists. For any $0< \kappa <1$, when $\theta$
is small enough, or equivalently, when $R$ is large enough, from
(\ref{uselaterboundb1}) in Lemma \ref{cut1},
we have
\begin{align}
\theta(R) \geq d \left(\frac{\kappa x}{2} \right)^x R^{-x}
\label{newtheorem1}
\end{align}
Using (\ref{newtheorem1}) and (\ref{newtheorem2}), for any $0 <
\kappa <1$,  (\ref{siggy2}) holds when $R$ is large enough, since $D(\theta)$ is a nondecreasing function of $\theta$.

\subsection{Proof of Theorem \ref{laibuji1}} \label{laibuji2}
When $P(N)$ is such that (\ref{powerconstraint}) is satisfied,
from (\ref{Barbara}), we see that in this case $C_u^N$ increases
monotonically in $N$. Hence, when $N$ is large enough, $C_u^N$
will be large enough such that Theorem \ref{Pat} holds.
Hence, for any constant $0 < \kappa < 1$, a lower bound on the
minimum achievable expected distortion is
\begin{align}
D_l^N&=D_p(C_u^N) \\
& \geq \kappa \left(1+\frac{\kappa}{x-1} \right)\left(\frac{\kappa x}{2}\right)^{x-1} \frac{d}{ T_0}(C_u^N)^{1-x} \\
& \geq \kappa \left(1+\frac{\kappa}{x-1} \right)\left(\kappa
x\right)^{x-1} \frac{d}{ T_0}
\left(\frac{1}{\log \left( 1+ \bar{h}_u^2 N P(N) \right)}\right)^{x-1} \label{Cancun}\\
& \geq \kappa^2 \left(1+\frac{\kappa}{x-1} \right)\left(\kappa
x\right)^{x-1} \frac{d}{ T_0} \left(\frac{1}{\log \left( N P(N)
\right)}\right)^{x-1} \label{laugh}
\end{align}
where (\ref{Cancun}) follows from (\ref{Barbara}), and the last
step follows when $N$ is large enough, i.e., there exists an
$N_{0}(\kappa)>0$, such that when $N > N_{0}(\kappa)$,
(\ref{laugh}) is true.

Therefore, when $P(N)$ is such that (\ref{powerconstraint}) is
satisfied, for any $0 < \kappa < 1$, (\ref{singgy4}) is true when
$N$ is large enough.

\subsection{Proof of Theorem \ref{achratedis}} \label{achievableratedistortion}
We restate the generalization of \cite[Theorem 1]{Flynn:1987},
which appeared in \cite[Theorem 1]{Chen:2004} for $N$ sensor nodes
below. This provides us with an achievable rate-distortion point.
\begin{Theo}\cite{Flynn:1987, Chen:2004}
If the individual rates are equal, a rate-distortion sum rate
$R_c$ and distortion $D_c$ are achievable if there exist random
variables $T_1,T_2,\cdots,T_N$ with
\begin{align}
(S(t), S_{\{i\}^c}, T_{\{i\}^c}) \rightarrow S_i \rightarrow T_i,
\qquad i=1,2,\cdots, N
\end{align}
and an estimator function
\begin{align}
\hat{S}(t)=g(T_1,T_2,\cdots,T_N)
\end{align}
such that
\begin{align}
R_c &\geq I(S_1,S_2,\cdots,S_N;T_1,T_2,\cdots,T_N)\\
D_c &\geq E[d(S(t),g(T_1,T_2,\cdots,T_N))]
\end{align}
\end{Theo}
We obtain an achievable rate-distortion point when we specify the
relationship between $\left(S(t), \{S_i\}_{i=1}^\infty,
\{T_i\}_{i=1}^\infty \right)$ as
\begin{align}
T_i=S_i+W_i, \qquad i=1,2,\cdots,N
\end{align}
where $W_i, i=1,2,\cdots,N$, are i.i.d. Gaussian random variables
with zero-mean and variance $\sigma_D^2$ and independent of
everything else. Here, we can adjust $\sigma_D^2$ to achieve
various feasible rate-distortion points \cite{Flynn:1987}.

We choose the MMSE estimator to estimate $S(t)$ from observations
$\{T_k\}_{k=1}^N$. Hence, the achieved distortion is
\begin{align}
D_c^N(\sigma_D^2)=\frac{1}{T_0}\int_{0}^{T_0}
\left(K(t,t)-\bm{\rho}_N^T(t) \left(\Sigma_N+\sigma_D ^2
I\right)^{-1} \bm{\rho}_N(t) \right) dt
\end{align}
The sum rate required to achieve this distortion is
\begin{align}
R_c^N(\sigma_D^2)& = I(S_1,S_2,\cdots,S_N;T_1,T_2,\cdots,T_N)\nonumber \\
 & = \frac{1}{2} \log \det \left(I + \frac{1}{\sigma_D^2} \Sigma_N \right)\\
&=\sum_{k=0}^{N-1} \frac{1}{2} \log
\left(1+\frac{\mu_k^{(N)}}{\sigma_D^2} \right)
\end{align}
where $\mu_0^{(N)},\mu_1^{(N)},\cdots,\mu_{N-1}^{(N)}$ are the
eigenvalues of $\Sigma_N$.

Next, let $\theta'=\frac{T_0}{N-1}\sigma_D^2$,
$\Sigma_N'=\frac{T_0}{N-1}\Sigma_N$ and
$\mu_k^{(N)'}=\frac{T_0}{N-1} \mu_k^{(N)}$. We define two
functions of $\theta'$ as
\begin{align}
R_a^N(\theta') \overset{\triangle}=R_c(\sigma_D^2) &=
\sum_{k=0}^{N-1} \frac{1}{2} \log
\left(1+\frac{\mu_k^{(N)'}}{\theta'} \right)
\end{align}
and
\begin{align}
D_a^N(\theta') \overset{\triangle}=D_c^N(\sigma_D^2)  =
\frac{1}{T_0}\int_{0}^{T_0}
\left(K(t,t)-\frac{T_0}{N-1}\bm{\rho}_N^T(t)
\left(\Sigma_N'+\theta' I\right)^{-1} \bm{\rho}_N(t) \right) dt
\end{align}
and by definition, sum rate $R_a^N(\theta')$ and distortion
$D_a^N(\theta')$ are achievable for an arbitrary Gaussian random
processes.

\subsection{Proof of Lemma \ref{whatnoname}} \label{breakdistortion}
Using the matrix inversion lemma \cite{Horn:book},
\begin{align}
\left(\Sigma_N'+\theta' I\right)^{-1}= \Sigma_N^{'-1}-
\Sigma_N^{'-1}\left(\frac{1}{\theta'}I+\Sigma_N^{'-1}\right)^{-1}
\Sigma_N^{'-1}
\end{align}
we have
\begin{align}
D_a^N  (\theta') = &\frac{1}{T_0}\int_{0}^{T_0}
\left(K(t,t)-\frac{T_0}{N-1}\bm{\rho}_N^T(t)
\Sigma_N^{'-1} \bm{\rho}_N(t) \right) dt \nonumber \\
&+
 \frac{1}{N-1}\int_{0}^{T_0}
 \bm{\rho}_N^T(t)\Sigma_N^{'-1}\left(\frac{1}{\theta'}I +\Sigma_N^{'-1}\right)^{-1} \Sigma_N^{'-1} \bm{\rho}_N(t) dt\\
= & D_s^{(N)}+D^{(N)}(\theta') \label{Little1}
\end{align}
where we have defined
\begin{align}
D_s^{(N)} \overset{\triangle}{=} & \frac{1}{T_0}\int_{0}^{T_0}
\left(K(t,t)-\frac{T_0}{N-1}\bm{\rho}_N^T(t)
\Sigma_N^{'-1} \bm{\rho}_N(t) \right) dt \\
D^{(N)}(\theta') \overset{\triangle}{=} &
\frac{1}{N-1}\int_{0}^{T_0}
 \bm{\rho}_N^T(t)\Sigma_N^{'-1}\left(\frac{1}{\theta'}I +\Sigma_N^{'-1}\right)^{-1} \Sigma_N^{'-1} \bm{\rho}_N(t) dt
\end{align}
We continue evaluating $D^{(N)}(\theta')$,
\begin{align}
D^{(N)}&(\theta') \nonumber\\
 =& \frac{1}{N-1}\sum_{i=1}^{N-1}
\int_{\frac{i-1}{N-1}T_0}^{\frac{i}{N-1}T_0}  \left(\bm{\rho}_N^T
\left(\frac{i-1}{N-1}T_0 \right)-\bm{\Delta}_i^T(t)\right)
\Sigma_N^{'-1}
\left(\frac{1}{\theta'}I+\Sigma_N^{'-1}\right)^{-1}  \nonumber\\
& \hspace{1.5in}\Sigma_N^{'-1}
\left(\bm{\rho}_N\left(\frac{i-1}{N-1}T_0 \right)-\bm{\Delta}_i(t)\right) dt\\
= & \frac{1}{T_0}\sum_{i=1}^{N-1}
 \left(\left(\frac{1}{\theta'}I+\Sigma_N^{'-1}\right)^{-1} \right)_{(i,i)}
- 2\frac{1}{T_0}\sum_{i=1}^{N-1}
\int_{\frac{i-1}{N-1}T_0}^{\frac{i}{N-1}T_0}
\left(\left(\frac{1}{\theta'}I+\Sigma_N^{'-1}\right)^{-1} \Sigma_N^{'-1}\bm{\Delta}_i(t) \right)_i dt \nonumber \\
& + \frac{1}{N-1}\sum_{i=1}^{N-1}
\int_{\frac{i-1}{N-1}T_0}^{\frac{i}{N-1}T_0} \bm{\Delta}_i^T(t)
\Sigma_N^{'-1} \left(\frac{1}{\theta'}I+\Sigma_N^{'-1}\right)^{-1}
\Sigma_N^{'-1} \bm{\Delta}_i(t) dt \label{errorterm}
\end{align}
where $\bm{\Delta}_i(t)$ is defined as
\begin{align}
\bm{\Delta}_i(t)=\bm{\rho}_N\left(\frac{i-1}{N-1}T_0
\right)-\bm{\rho}_N(t)
\end{align}
for $\frac{i-1}{N-1}T_0 \leq t \leq \frac{i}{N-1}T_0$, and
(\ref{errorterm}) follows based on the fact that
\begin{align}
\bm{\rho}_N^T \left(\frac{i-1}{N-1}T_0 \right)
\Sigma_N^{'-1}=\frac{N-1}{T_0} \mathbf{e}_i
\end{align}
where $\mathbf{e}_i$ is the row vector whose $i$-th entry is 1 and
all other entries are 0.

The eigenvalues  of $\Sigma_N^{'-1}
\left(\frac{1}{\theta'}I+\Sigma_N^{'-1}\right)^{-1}
\Sigma_N^{'-1}$ are
\begin{align}
\frac{\theta'}{\mu_k^{(N)'}+\theta'}\frac{1}{\mu_k^{(N)'}}, \qquad
k=0,1,\cdots,N-1
\end{align}
which are smaller than the corresponding eigenvalues of
$\Sigma_N^{'-1}$, i.e., $\frac{1}{\mu_k^{(N)'}}$. Thus, the third
term in (\ref{errorterm}) is bounded by
\begin{align}
\frac{1}{N-1}\sum_{i=1}^{N-1}
\int_{\frac{i-1}{N-1}T_0}^{\frac{i}{N-1}T_0}& \bm{\Delta}_i^T(t)
\Sigma_N^{'-1}\left(\frac{1}{\theta'}I+\Sigma_N^{'-1}\right)^{-1}
\Sigma_N^{'-1}
\bm{\Delta}_i(t) dt \nonumber \\
& \leq \frac{1}{N-1}\sum_{i=1}^{N-1}
\int_{\frac{i-1}{N-1}T_0}^{\frac{i}{N-1}T_0} \bm{\Delta}_i^T(t)
\Sigma_N^{'-1} \bm{\Delta}_i(t) dt \label{ulukus2}
\end{align}
To further upper bound the third term in (\ref{errorterm}), we
write
\begin{align}
D_s^{(N)} & = \frac{1}{T_0}\int_{0}^{T_0} \left(K(t,t)-\frac{T_0}{N-1}\bm{\rho}_N^T(t) \Sigma_N^{'-1} \bm{\rho}_N(t) \right) dt\\
& = \frac{1}{T_0}\sum_{i=1}^{N-1}
\int_{\frac{i-1}{N-1}T_0}^{\frac{i}{N-1}T_0}\left( K(t,t)
-\frac{T_0}{N-1}\left(\bm{\rho}_N^T \left(\frac{i-1}{N-1}T_0
\right)-\bm{\Delta}_i(t)^T\right) \right.\nonumber \\
& \hspace{1.5in}\left.
\Sigma_N^{'-1} \left(\bm{\rho}_N\left(\frac{i-1}{N-1}T_0\right)-\bm{\Delta}_i(t)\right) \right) dt\\
& = \frac{1}{T_0}\sum_{i=1}^{N-1}
\int_{\frac{i-1}{N-1}T_0}^{\frac{i}{N-1}T_0} \left(K(t,t)-K
\left(\frac{i-1}{N-1}T_0,\frac{i-1}{N-1}T_0\right) \right) dt  \nonumber \\
& \hspace{0.2in} +\frac{2}{T_0}\sum_{i=1}^{N-1}
\int_{\frac{i-1}{N-1}T_0}^{\frac{i}{N-1}T_0}
\left(\bm{\Delta}_i(t) \right)_i dt- \frac{1}{N-1}\sum_{i=1}^{N-1}
\int_{\frac{i-1}{N-1}T_0}^{\frac{i}{N-1}T_0}\bm{\Delta}_i(t)^T
\Sigma_N^{'-1} \bm{\Delta}_i(t) dt \\
&= A^{(N)} - \frac{1}{N-1}\sum_{i=1}^{N-1}
\int_{\frac{i-1}{N-1}T_0}^{\frac{i}{N-1}T_0}\bm{\Delta}_i(t)^T
\Sigma_N^{'-1} \bm{\Delta}_i(t) dt \label{changeindex}
\end{align}
where we have defined
\begin{align}
A^{(N)}=&\frac{1}{T_0}\sum_{i=1}^{N-1}
\int_{\frac{i-1}{N-1}T_0}^{\frac{i}{N-1}T_0} \left(K(t,t)-K
\left(\frac{i-1}{N-1}T_0,\frac{i-1}{N-1}T_0\right) \right) dt
+\frac{2}{T_0}\sum_{i=1}^{N-1}
\int_{\frac{i-1}{N-1}T_0}^{\frac{i}{N-1}T_0}
\left(\bm{\Delta}_i(t) \right)_i dt \\
 =&\frac{1}{T_0}\sum_{i=1}^{N-1}
\int_{\frac{i-1}{N-1}T_0}^{\frac{i}{N-1}T_0}
\left(K(t,t)-K \left(\frac{i-1}{N-1}T_0,\frac{i-1}{N-1}T_0\right) \right) dt \nonumber \\
&+\frac{2}{T_0}\sum_{i=1}^{N-1}
\int_{\frac{i-1}{N-1}T_0}^{\frac{i}{N-1}T_0}
\left(\bm{\rho}_N\left(\frac{i-1}{N-1}T_0 \right)-\bm{\rho}_N(t)
\right)_i dt
\end{align}
Then, we have the third term in (\ref{errorterm}) upper bounded by
$A^{(N)}$ because of (\ref{ulukus2}), (\ref{changeindex}) and the
fact
 that $D_s^{(N)}$ is non-negative, i.e.,
\begin{align}
\frac{1}{N-1}\sum_{i=1}^{N-1}
\int_{\frac{i-1}{N-1}T_0}^{\frac{i}{N-1}T_0}& \bm{\Delta}_i^T(t)
\Sigma_N^{'-1}\left(\frac{1}{\theta'}I+\Sigma_N^{'-1}\right)^{-1}
\Sigma_N^{'-1} \bm{\Delta}_i(t) dt \leq A^{(N)} \label{Little2}
\end{align}
Furthermore, we can see from (\ref{changeindex}) that
\begin{align}
D_s^{(N)} \leq A^{(N)} \label{Little3}
\end{align}

Now, we evaluate the second term in (\ref{errorterm}). Since,
\begin{align}
\left|\left(\left(\frac{1}{\theta'}I+\Sigma_N^{'-1}\right)^{-1}
\Sigma_N^{'-1}\bm{\Delta}_i(t) \right)_i\right| & \leq
\left|\left|\left(\frac{1}{\theta'}I+\Sigma_N^{'-1}\right)^{-1}
\Sigma_N^{'-1}\bm{\Delta}_i(t) \right| \right|\\
& \leq \left|\left|
\left(\frac{1}{\theta'}I+\Sigma_N^{'-1}\right)^{-1}
\Sigma_N^{'-1} \right| \right|_2 \cdot ||\bm{\Delta}_i(t) ||\\
& = \max_{0 \leq k \leq N-1} \left(\mu_k^{(N)'}\right)^{-1}
\left(\frac{1}{\theta'}+\frac{1}{\mu_k^{(N)'}} \right)^{-1}
||\bm{\Delta}_i(t) ||\\
& \leq ||\bm{\Delta}_i(t) ||
\end{align}
where $||\cdot||_2$ denotes the spectral norm of a matrix, which
is defined as the largest eigenvalue of a matrix \cite{Horn:book}.
Therefore, the second term in (\ref{errorterm}) is bounded by
\begin{align}
\left|\frac{2}{T_0}\sum_{i=1}^{N-1}
\int_{\frac{i-1}{N-1}T_0}^{\frac{i}{N-1}T_0} \right. & \left.
\left(\left(\frac{1}{\theta'}I+\Sigma_N^{'-1}\right)^{-1}
\Sigma_N^{'-1}\bm{\Delta}_i(t) \right)_i dt \right|
\nonumber\\
& \leq \frac{2}{T_0}\sum_{i=1}^{N-1}
\int_{\frac{i-1}{N-1}T_0}^{\frac{i}{N-1}T_0}
\left|\left(\left(\frac{1}{\theta'}I+\Sigma_N^{'-1}\right)^{-1} \Sigma_N^{'-1}\bm{\Delta}_i(t) \right)_i\right| dt \\
& \leq \frac{2}{T_0}\sum_{i=1}^{N-1}
\int_{\frac{i-1}{N-1}T_0}^{\frac{i}{N-1}T_0}
||\bm{\Delta}_i(t) || dt \\
& = B^{(N)} \label{Little4}
\end{align}
where we have defined $B^{(N)}$ as
\begin{align}
B^{(N)} &=\frac{2}{T_0}\sum_{i=1}^{N-1}
\int_{\frac{i-1}{N-1}T_0}^{\frac{i}{N-1}T_0} ||\bm{\Delta}_i(t)
|| dt \\
& =\frac{2}{T_0}\sum_{i=1}^{N-1}
\int_{\frac{i-1}{N-1}T_0}^{\frac{i}{N-1}T_0}
\left|\left|\bm{\rho}_N\left(\frac{i-1}{N-1}T_0
\right)-\bm{\rho}_N(t) \right|\right| dt
\end{align}
Finally, the first term in (\ref{errorterm}) is bounded by
\begin{align}
\frac{1}{T_0}\sum_{i=1}^{N-1}
 \left(\left(\frac{1}{\theta'}I+\Sigma_N^{'-1}\right)^{-1} \right)_{(i,i)} &\leq
 \frac{1}{T_0}\sum_{i=1}^{N}
 \left(\left(\frac{1}{\theta'}I+\Sigma_N^{'-1}\right)^{-1} \right)_{(i,i)} \\
 & = \frac{1}{T_0} \text{tr} \left(\left(\frac{1}{\theta'}I+\Sigma_N^{'-1}\right)^{-1}  \right) \\
 & = \frac{1}{T_0} \sum_{k=0}^{N-1}
\left(\frac{1}{\theta'}+\frac{1}{\mu_k'^{(N)}} \right)^{-1} \label{errorfirstterm} \\
& \overset{\triangle}{=} D_b^N(\theta') \label{Little5}
\end{align}
where the last step is by the definition of $D_b^N(\theta')$.
Hence, for an arbitrary Gaussian random process, by
(\ref{Little1}), (\ref{errorterm}), (\ref{Little2}),
(\ref{Little3}), (\ref{Little4}) and (\ref{Little5}), we have
shown that
\begin{align}
D_a^N(\theta') \leq 2A^{(N)}+B^{(N)}+D_b^N(\theta')
\end{align}

\subsection{Proof of Theorem \ref{generalelgamal}}
\label{achievablerateamplify} Each round of communication will
take $2Nn$ time slots. In the $[2n(i-1)]$-th to $[2ni]$-th time
slots, node $i$ transmits at rate $R_i$, while all other nodes act
as relay nodes and transmit no data of their own. In the end, the
achievable sum rate is $\frac{\sum_{i=1}^N R_i}{N}$. We will show
that each node can achieve $R_i=C_a^N$, and thus, all nodes can
achieve the sum rate of $C_a^N$ with identical individual rates.

We will consider the transmission of the data of node $i$. Node
$i$ codes its message using capacity achieving single-user coding
techniques with codeword length $n$. Each codeword symbol requires
two time slots. In the first time slot, node $i$ transmits its
codeword symbol using power $P(N)$. All other nodes remain silent,
and receive a noisy version of node $i$'s transmitted signal. The
collector node ignores its received signal, which is suboptimal
but eases calculation and does not affect the scaling law of the
achievable rate. Identical individual power constraints of
$P(N)/N$ for the nodes are satisfied, since all nodes take turns
and node $i$ will do this only $1/N$-th of the time, therefore its
transmit power in $1/N$-th of the time is $P(N)$. In the second
time slot, all sensor nodes, except node $i$, amplify and forward
what they have received in the previous time slot to the collector
node using an individual power constraint $P(N)/N$. The collector
node, after $2n$ time slots, decodes using capacity achieving
single-user decoding techniques. Now, we calculate the rate
achievable with this scheme. In the first time slot, sensor node
$j$ receives
\begin{align}
Y_j=h_{ij}X_i+Z_j, \qquad i,j=1,2,\cdots, N, \quad j \neq i
\end{align}
and in the second time slot, sensor node $j$ transmits
\begin{align}
X_j&=\beta_{ij} Y_j \\
&= \beta_{ij} h_{ij}X_i+\beta_{ij}Z_j, \qquad i,j=1,2,\cdots,N,
\quad j \neq i
\end{align}
where $\beta_{ij}$ is the scaling coefficient of node $j$ when it
amplifies the signal it received from node $i$. In order to
satisfy the identical individual power constraints,
$\{\beta_{ij}\}_{j=1, j \neq i}^N$ have to satisfy
\begin{align}
\beta_{ij}^2 \left(h_{ij}^2P(N)+1 \right) \leq \frac{P(N)}{N},
\qquad \forall i,j=1,2,\cdots,N
\end{align}
The collector node receives
\begin{align}
Y_0& =\sum_{j=1, j \neq i}^N h_{j0} X_j+Z_0 \\
& = \left(\sum_{j=1, j \neq i}^N \beta_{ij}
h_{ij}h_{j0}\right)X_i+\left(\sum_{j=1, j \neq i}^N
h_{j0}\beta_{ij} Z_j \right)+Z_0
\end{align}
Therefore, the achievable rate is,
\begin{align}
\frac{1}{4} \log \left(1+ \frac{ \left(\sum_{j=1, j \neq i}^N
\beta_{ij}  h_{ij}h_{j0}\right)^2 P(N)} {\sum_{j=1, j \neq i}^N
\left(\beta_{ij} h_{j0} \right)^2+1} \right) \label{whatnow}
\end{align}
where we have $\frac{1}{4}$ because we used two time slots to
transmit one codeword symbol. We choose
\begin{align}
\beta_{ij}=\zeta h_{ij} h_{j0}
\end{align}
where, in order to satisfy the power constraint, the constant
$\zeta$ must satisfy
\begin{align}
\zeta^2 \leq \frac{P(N)}{ h_{ij}^4 h_{j0}^2 NP(N)+  h_{ij}^2
h_{j0}^2 N}, \quad \forall i,j \label{directlyabove2}
\end{align}
We can choose $\zeta$ as
\begin{align}
\zeta^2=\frac{P(N)}{\bar{h}_u^6NP(N) +\bar{h}_u^4N}
\end{align}
Thus, from (\ref{whatnow}), an lower bound on the achievable rate
is
\begin{align}
\frac{1}{4} \log \left(1+ \frac{ \zeta^2 \left(\sum_{j=1, j \neq
i}^N h_{ij}^2 h_{j0}^2 \right)^2 P(N)} { \zeta^2 \left(\sum_{j=1,
j \neq i}^N h_{ij}^2h_{j0}^4\right)+1} \right)
 \geq \frac{1}{4} \log \left(1+\frac{\bar{h}_l^8 \zeta^2 (N-1)^2
P(N)}{\bar{h}_u^6 \zeta^2 N +1} \right) \overset{\triangle}{=}
C_b^N
\end{align}
Clearly, rate $C_b^N$ can be achievable by any node $i$. We have
\begin{align}
C_b^N & = \frac{1}{4} \log \left(1+\frac{\bar{h}_l^8
\left(P(N)\right)^2 \frac{(N-1)^2}{N}
}{2\bar{h}_u^6  P(N) +\bar{h}_u^4} \right) \\
& \geq \frac{1}{4} \log \left(1+\frac{\bar{h}_l^8
\left(P(N)\right)^2 N }{4\bar{h}_u^6
P(N) +2\bar{h}_u^4} \right) \label{Singgy}
\end{align}
where the last step follows when $N$ is large enough such that $
\frac{(N-1)^2}{N} > \frac{N}{2} $.

When $P(N)$ is such that
\begin{align}
\lim_{N \rightarrow \infty} \frac{1}{P(N)}=0
\end{align}
for any $0 < \kappa < 1$, we have,
\begin{align}
C_b^N & \geq \frac{1}{4} \log \left(1+\frac{\bar{h}_l^8
\left(P(N)\right)^2 N
}{8\bar{h}_u^6 P(N) } \right) \label{gone2}\\
& = \frac{1}{4} \log \left(1+\frac{\bar{h}_l^8
}{8\bar{h}_u^6  }NP(N) \right)\\
& \geq \frac{\kappa}{4} \log \left(NP(N) \right) \label{gone1}
\end{align}
for $N$ large enough, i.e., there exists $N_{1}(\kappa)>0$, such
that when $N > N_{1}(\kappa)$, (\ref{gone2}) and (\ref{gone1}) are
true.

When $P(N)$ is such that
\begin{align}
\lim_{N \rightarrow \infty} P(N)=l
\end{align}
and $l$ is a number that satisfies $ 0 < l < \infty$, fix some
small $l_0>0$, there exists an $N_{2}(l_0)>0$ such that when $N >
N_{2}(l_0)$, we have,
\begin{align}
l-l_0 < P(N) < l+l_0
\end{align}
Hence, when $N > N_{2}(l_0)$, for any $0 < \kappa < 1$,
\begin{align}
C_b^N & \geq \frac{1}{4} \log \left(1+\frac{\bar{h}_l^8 (l-l_0)}{4\bar{h}_u^6 (l+l_0)+2 \bar{h}_u^4} P(N) N \right)\\
& \geq \frac{\kappa}{4} \log \left(NP(N) \right) \label{gone3}
\end{align}
where the last step follows when $N$ is large enough, i.e., when
there exists an $N_{3}(\kappa)>0$, such that when $N > \max
\left(N_{2}(l_0),
 N_{3}(\kappa)\right)$, (\ref{gone3}) is true.

When $P(N)$ is such that
\begin{align}
\lim_{N \rightarrow \infty} P(N)=0
\end{align}
and there exists a constant $0 < \epsilon < \frac{1}{2}$, such
that
\begin{align}
\lim_{N \rightarrow \infty} P(N) N^{\frac{1}{2}-\epsilon} > 1
\end{align}
we have, for $0 < \kappa < 1$,
\begin{align}
C_b^N & \geq \frac{1}{4} \log \left(1+\frac{\bar{h}_l^8}{4 \bar{h}_u^4}(P(N))^2 N \right) \label{gone4} \\
& \geq \frac{\kappa}{4} \log \left((P(N))^2 N \right) \label{gone5}\\
& = \frac{\kappa}{4} \log (N P(N)) + \frac{\kappa}{4} \log (P(N)) \\
& \geq \frac{\kappa}{4}\frac{4 \epsilon}{1+2 \epsilon} \log
(NP(N)) \label{gone6}
\end{align}
where the last step follows from
\begin{align}
\frac{\kappa}{4}\left(1-\frac{4 \epsilon}{1+2 \epsilon}\right)
\log (NP(N))+\frac{\kappa}{4} \log (P(N)) = \frac{\kappa}{4}
\frac{2}{1+2\epsilon} \log (P(N) N^{\frac{1}{2}-\epsilon}) \geq 0
\label{gone7}
\end{align}
when $N$ is large enough, i.e., there exists an $N_{4}(\kappa)>0$,
such that when $N > N_{4}(\kappa)$, (\ref{gone4}), (\ref{gone5})
and (\ref{gone7}) are true, and therefore, (\ref{gone6}) is true.

Thus, combining all possible cases of $P(N)$, we see that when
$P(N)$ is such that there exists a constant $\epsilon>0$, such
that
\begin{align}
\lim_{N \rightarrow \infty} P(N) N^{\frac{1}{2}-\epsilon} > 1
\label{Singapore}
\end{align}
for any $0 < \kappa < 1$, the following rate $C_a^N$ is
achievable,
\begin{align}
C_a^N =\kappa \nu \log(NP(N)) \label{indrate68}
\end{align}
where constant $\nu$ is
\begin{align}
\nu=  \min \left(\frac{ \epsilon}{1+2\epsilon}, \frac{1}{4}
\right)
\end{align}\
when $N$ is large enough.

Since all nodes take turns applying the same scheme, the
individual rates of all sensors are the same, and the achievable
sum rate is (\ref{indrate68}).

For all other $P(N)$, from (\ref{Singgy}), we see that the
achievable sum rate approaches a positive constant or zero as $N$
goes to infinity.

\subsection{Proof of Lemma \ref{addgeneral}} \label{lipliplip}
We first consider $A^{(N)}$.
\begin{align}
A^{(N)}=& \frac{1}{T_0}\sum_{i=1}^{N-1}
\int_{\frac{i-1}{N-1}T_0}^{\frac{i}{N-1}T_0}
\left(K(t,t)-K\left(\frac{i-1}{N-1}T_0,\frac{i-1}{N-1}T_0\right) \right) dt \nonumber \\
& +\frac{2}{T_0}\sum_{i=1}^{N-1}
\int_{\frac{i-1}{N-1}T_0}^{\frac{i}{N-1}T_0}
\left(\bm{\rho}_N \left(\frac{i-1}{N-1}T_0 \right)-\bm{\rho}_N(t) \right)_i dt \\
\leq & \frac{1}{T_0}\sum_{i=1}^{N-1}
\int_{\frac{i-1}{N-1}T_0}^{\frac{i}{N-1}T_0}
\left|K(t,t)-K\left(\frac{i-1}{N-1}T_0,\frac{i-1}{N-1}T_0 \right) \right| dt \nonumber \\
 &+\frac{2}{T_0}\sum_{i=1}^{N-1}
\int_{\frac{i-1}{N-1}T_0}^{\frac{i}{N-1}T_0} \left|
\left(\bm{\rho}_N \left(\frac{i-1}{N-1}T_0\right)-\bm{\rho}_N(t) \right)_i \right| dt \\
\leq & \frac{1}{T_0}\sum_{i=1}^{N-1}
\int_{\frac{i-1}{N-1}T_0}^{\frac{i}{N-1}T_0} B
\left(\frac{\sqrt{2}}{N-1}T_0\right)^\alpha dt
 +\frac{2}{T_0}\sum_{i=1}^{N-1}
\int_{\frac{i-1}{N-1}T_0}^{\frac{i}{N-1}T_0} B \left(\frac{1}{N-1}T_0\right)^\alpha dt \label{lip2}\\
=& B \left(2^{\frac{\alpha}{2}}+2 \right)T_0^\alpha \frac{1}{(N-1)^\alpha} \\
=& \Theta \left(N^{-\alpha}\right)
\end{align}
where (\ref{lip2}) follows from condition 2 in Section
\ref{defineA}. Using similar ideas, we have
\begin{align}
B^{(N)} & =\frac{2}{T_0}\sum_{i=1}^{N-1}
\int_{\frac{i-1}{N-1}T_0}^{\frac{i}{N-1}T_0}
\left|\left|\bm{\rho}_N\left(\frac{i-1}{N-1}T_0 \right)-\bm{\rho}_N(t) \right|\right| dt \\
& = \frac{2}{T_0}\sum_{i=1}^{N-1}
\int_{\frac{i-1}{N-1}T_0}^{\frac{i}{N-1}T_0}
\left(\sum_{m=0}^{N-1} \left|K\left(\frac{i-1}{N-1}T_0,\frac{mT_0}{N-1}\right)-K\left(t,\frac{mT_0}{N-1}\right) \right|^2\right)^{\frac{1}{2}} dt \\
& \leq \frac{2}{T_0}\sum_{i=1}^{N-1}
\int_{\frac{i-1}{N-1}T_0}^{\frac{i}{N-1}T_0}
\left(\sum_{m=0}^{N-1} \left(B \left(\frac{T_0}{N-1} \right)^\alpha \right)^2 \right)^{\frac{1}{2}} dt\\
& = 2B T_0^{ \alpha} \frac{N^{\frac{1}{2}}}{(N-1)^{ \alpha}} \\
& = \Theta \left(N^{\frac{1}{2}- \alpha} \right)
\end{align}

\subsection{Some properties of $\mu_k^{(N)'}$} \label{preproof}

\begin{Lem} \label{appro}
For all Gaussian random processes in $\mathcal{A}$, let $K_1(N)$
be a sequence of numbers that satisfies
\begin{align}
\lim_{N \rightarrow \infty} & \frac{1}{K_1(N)}=0  \label{conwor1}\\
\lim_{N \rightarrow \infty} &
\frac{\left(K_1(N)+B_4\right)^{2\tau}}{(N-1)^\gamma}=0 \label{conwor2}\\
\lim_{N \rightarrow \infty} &
\frac{K_1(N)^{x+1+\tau}}{(N-1)^\beta}=0 \label{worried2}
\end{align}
Then, for each $k$ such that $k \leq K_1(N)$, there exists an
eigenvalue $\mu'^{(N)}$, different for each $k$, of $\Sigma_N'$ such that
\begin{align}
\left| \mu'^{(N)} -\lambda_k \right|   \leq d_1
\frac{\left(k+B_7\right)^\tau}{(N-1)^\beta} \label{lemmacon}
\end{align}
for some $d_1 >0$ and some positive integer $B_7$, both
independent of $k$ and $N$, when $N$ is large enough.
\end{Lem}
Lemma \ref{appro} shows that the convergence of $\mu_k^{(N)}$ to
$\lambda_k$ is not uniform, and the approximation of $\mu_k^{(N)}$
using $\lambda_k$ is accurate only when $k
<<N^{\frac{\gamma}{\tau}}$ and $\lambda_k >> d_1
\frac{\left(k+B_7\right)^\tau}{(N-1)^\beta} $. When the conditions of Lemma \ref{appro} are satisfied, we label the $\mu'^{(N)}$ that satisfies
(\ref{lemmacon}) to be $\mu_k^{(N)'}$ for $k \leq K_1(N)$. The remaining $N-K_1(N)$ eigenvalues of $\mu'{(N)}$ will be labelled according to
the order from large to small.


\begin{Lem} \label{divergence}
For all Gaussian random processes in $\mathcal{A}$, define two
sequences $\vartheta_L^N$ and $\vartheta_U^N$ as
\begin{align}
\lim_{N \rightarrow \infty} \frac{1}{\vartheta_L^N N^{\min
\left(\frac{x \gamma}{2\tau}, \frac{\alpha x}{x-1}, \frac{\beta
x}{x+\tau+1} \right)}} =0, \qquad \lim_{N \rightarrow \infty}
\vartheta_U^N =0
\end{align}
For any constant $0 < \kappa <1$, we have
\begin{align}
\sum_{k=\left\lfloor \left(\frac{d}{\theta'}
\right)^{\frac{1}{x}}+c_u \right\rfloor +1}^{N-1} \mu_k^{(N)'}
\leq \frac{d^\frac{1}{x}}{(x-1) \kappa^2} \theta'^{1-\frac{1}{x}}
\label{nomorename}
\end{align}
when $\theta' \in [\vartheta_L^N, \vartheta_U^N]$ and $N$ is large
enough.
\end{Lem}
Lemma \ref{divergence} shows that the sum of the eigenvalues that
do not converge to $\lambda_k$ for $k=0,1, \cdots, \left\lfloor
\left(\frac{d}{\theta'} \right)^{\frac{1}{x}}+c_u \right\rfloor$
is quite small.

\subsubsection{Proof of Lemma \ref{appro}}

By definition, $\lambda_k$ for any $k$
satisfies
\begin{align}
\lambda_k \phi_k\left(\frac{l-1}{N-1}T_0 \right)=\int_0^{T_0}
K\left(\frac{l-1}{N-1}T_0 ,s \right) \phi_k(s) ds, \quad \forall
l=1,2,\cdots,N \label{eigenfunction1}
\end{align}
We rewrite the right hand side of (\ref{eigenfunction1}) by
\begin{align}
\frac{T_0}{N-1}\sum_{i=1}^{N}
K\left(\frac{l-1}{N-1}T_0,\frac{i-1}{N-1}T_0 \right) \phi_k
\left(\frac{i-1}{N-1}T_0 \right) +
\epsilon_N^k\left(\frac{l-1}{N-1}T_0 \right)
 \quad \forall l=1,2,\cdots,N
\label{eigenfunction2}
\end{align}
where $\epsilon_N^k\left(\frac{l-1}{N-1}T_0 \right)$ is defined as
\begin{align}
&\sum_{i=1}^{N-1} \int_{\frac{i-1}{N-1}T_0}^{\frac{i}{N-1}T_0}
\left(K\left(\frac{l-1}{N-1}T_0,s\right) \phi_k(s)
-K\left(\frac{l-1}{N-1}T_0,\frac{i-1}{N-1}T_0
\right) \phi_k \left(\frac{i-1}{N-1}T_0 \right) \right) ds \nonumber \\
&-\frac{T_0}{N-1} K \left(\frac{l-1}{N-1}T_0, T_0 \right)
\phi_k(T_0) \label{Dora}
\end{align}
Using (\ref{eigenfunction1}) and (\ref{eigenfunction2}), we have
for any $l=1,2,\cdots,N$,
\begin{align}
\lambda_k \phi_k\left(\frac{l-1}{N-1}T_0
\right)= \frac{T_0}{N-1}\sum_{i=1}^{N} K\left(\frac{l-1}{N-1}T_0
,\frac{i-1}{N-1}T_0 \right) \phi_k \left(\frac{i-1}{N-1}T_0
\right) + \epsilon_N^k\left(\frac{l-1}{N-1}T_0 \right)
\end{align}
i.e., we have
\begin{align}
\lambda_k   \sqrt{\frac{T_0}{N-1}} \phi_k\left(\frac{l-1}{N-1}T_0
\right)= &\frac{T_0}{N-1}\sum_{i=1}^{N} K\left(\frac{l-1}{N-1}T_0
,\frac{i-1}{N-1}T_0 \right)  \sqrt{\frac{T_0}{N-1}} \phi_k \left(\frac{i-1}{N-1}T_0
\right) \\
& +  \sqrt{\frac{T_0}{N-1}} \epsilon_N^k\left(\frac{l-1}{N-1}T_0 \right)
\end{align}
Let us define vector $\mathbf{a}_k^{(N)}$ of length of $N$  by defining
its $l$-th element to be $\sqrt{\frac{T_0}{N-1}} \epsilon_N^k \left(\frac{l-1}{N-1}T_0
\right) $ and vector $\mathbf{b}_k^{(N)}$ of length of $N$  by defining
its $l$-th element to be $ \sqrt{\frac{T_0}{N-1}}\phi_k\left(\frac{l-1}{N-1}T_0 \right)
$, we have in matrix form
\begin{align}
\lambda_k \mathbf{b}_k^{(N)}=\Sigma_N' \mathbf{b}_k^{(N)}+\mathbf{a}_k^{(N)} \label{worried3}
\end{align}
The links between the eigenvalues of $\Sigma_N'$ and the eigenvalues of $K(t,s)$, i.e., the $\lambda_k$s, will be determined using
(\ref{worried3}).
To do this, we first bound three quantities, $\left|\left|\mathbf{a}_k^{(N)} \right| \right|$, $\left| \left| \mathbf{b}_k^{(N)}
\right| \right|$, $\left| \mathbf{b}_m^{(N)^T} \mathbf{b}_l^{(N)} \right|$ for $k,m,l \leq K_1(N)$ and $m \neq l$.

We first upper bound $\left| \phi_k(T_0) \right|$. Let $F_k(s)$ be
defined as
\begin{align}
\int_{0}^s \phi_k^2(t) dt
\end{align}
Then, by the mean value theorem on interval $[0,T_0]$, we have
that there exists a $T' \in [0,T_0]$, such that
\begin{align}
1=F_k(T_0)-F_k(0)=\phi_k^2(T')
\end{align}
Hence, using condition 3 in Section \ref{defineA}, we have
\begin{align}
\left| \phi_k(t)-\phi_k(T')\right| \leq B_3(k+B_4)^\tau
T_0^\gamma, \quad t \in [0,T_0]
\end{align}
Thus,
\begin{align}
\left|\phi_k(t) \right| \leq B_3(k+B_4)^\tau T_0^\gamma+1, \quad t \in [0,T_0]  \label{Dora4}
\end{align}

Now, we analyze the norm of $\mathbf{a}_k^{(N)}$. From the
definition of $\epsilon_N^k\left(\frac{l-1}{N-1}T_0 \right)$ in
(\ref{Dora}), we have
\begin{align}
&\left|\epsilon_N^k\left(\frac{l-1}{N-1}T_0 \right) \right| \nonumber \\
& \leq  \sum_{i=1}^{N-1}
\int_{\frac{i-1}{N-1}T_0}^{\frac{i}{N-1}T_0}
\left|K\left(\frac{l-1}{N-1}T_0,s\right) \phi_k(s)
-K\left(\frac{l-1}{N-1}T_0,\frac{i-1}{N-1}T_0
\right) \phi_k \left(\frac{i-1}{N-1}T_0 \right) \right| ds \nonumber \\
& \hspace{0.1in}+\frac{T_0}{N-1} \left |K \left(\frac{l-1}{N-1}T_0, T_0 \right) \right| \left| \phi_k(T_0) \right| \\
& \leq B_2 T_0^{1+\beta}
\frac{\left(k+B_1\right)^\tau}{(N-1)^\beta} +\frac{T_0
\bar{K}(T_0) \left| \phi_k(T_0) \right|}{N-1} \label{lasthow} \\
& \leq B_2 T_0^{1+\beta}
\frac{\left(k+B_1\right)^\tau}{(N-1)^\beta}
+\frac{T_0 \bar{K}(T_0) \left( B_3(k+B_4)^\tau T_0^\gamma+1 \right)}{N-1} \label{Dora1} \\
& \leq B_2 T_0^{1+\beta}
\frac{\left(k+B_1\right)^\tau}{(N-1)^\beta}
+\frac{T_0 \bar{K}(T_0) \left( B_3(k+B_4)^\tau T_0^\gamma+1 \right)}{(N-1)^\beta} \label{Dora2}\\
& \leq \frac{\left(B_2 T_0^{1+\beta}+T_0^{1+\gamma} \bar{K}(T_0)
B_3\right) \left(k+\max (B_1, B_4)\right)^\tau
+T_0 \bar{K}(T_0)}{(N-1)^\beta} \\
& \leq \frac{\max \left(1,\left(B_2 T_0^{1+\beta}+T_0^{1+\gamma}
\bar{K}(T_0) B_3\right)\right)\left( \left(k+\max (B_1,
B_4)\right)^\tau
+ \left(\left( T_0 \bar{K}(T_0)\right)^{1/\tau}\right)^\tau\right)}{(N-1)^\beta}\\
& \leq \frac{\max\left(1, 2^{1-\tau}\right) \max \left(1,\left(B_2
T_0^{1+\beta}+T_0^{1+\gamma} \bar{K}(T_0) B_3\right)\right)
\left(k+\max (B_1, B_4) +\left( T_0
\bar{K}(T_0)\right)^{1/\tau}\right)^\tau
}{(N-1)^\beta} \label{runrunrun}\\
& \leq B_6 \frac{(k+B_7')^\tau}{(N-1)^\beta} \label{Dora3}
\end{align}
where (\ref{lasthow}) follows because the random process satisfies condition 3 in Section \ref{defineA}, and because $\bar{K}(T_0)$ is defined as
\begin{align}
\max_{1 \leq l \leq N} \left |K \left(\frac{l-1}{N-1}T_0, T_0
\right) \right|
\end{align}
and is a finite nonnegative number since $K(t,s)$ satisfies
condition 2 in Section \ref{defineA} and thus, is continuous, (\ref{Dora1}) follows from (\ref{Dora4}), and
(\ref{Dora2}) follows because
$\beta \leq 1$ from condition 3 in Section \ref{defineA}, and
(\ref{runrunrun}) follows because for $\forall u,v>0$
\begin{align}
u^{\tau}+v^{\tau} &\leq (u+v)^{\tau}, \quad \tau \geq 1 \label{mymy2}\\
\frac{u^{\tau}+v^{\tau}}{2} & \leq
\left(\frac{u+v}{2}\right)^{\tau}, \quad 0 \leq  \tau <1 \label{mymy3}
\end{align}
(\ref{Dora3}) comes because we define the variables $B_6$ and
$B_7'$, which are both independent of $k$ and $N$, as
\begin{align}
B_6 & = \max\left(1, 2^{1-\tau}\right) \max \left(1,\left(B_2 T_0^{1+\beta}+T_0^{1+\gamma} \bar{K}(T_0) B_3\right)\right)\\
B_7' & = \left\lceil\max (B_1, B_4) +\left( T_0
\bar{K}(T_0)\right)^{1/\tau} \right\rceil
\end{align}
Note that $B_7'$ is a positive integer. Finally, we calculate the
norm of vector $\mathbf{a}_k^{(N)}$ as
\begin{align}
\left| \left| \mathbf{a}_k^{(N)} \right| \right|&=\sqrt{\sum_{l=1}^N
\left(\sqrt{\frac{T_0}{N-1}} \epsilon_k^N \left(\frac{l-1}{N-1} T_0
 \right) \right)^2} \\
 & \leq \sqrt{\frac{N T_0}{N-1}} B_6 \frac{(k+B_7')^\tau}{(N-1)^\beta} \\
 & \leq 2B_6 \sqrt{T_0} \frac{(k+B_7')^\tau}{(N-1)^\beta} \label{largeNagain}
\end{align}
where (\ref{largeNagain}) follows when $N$ is large enough, more specifically, there exists an interger $N_1$ such that when $N > N_1$,
we have $\sqrt{\frac{N}{N-1}} \leq 2$.

Now, we will calculate the norm of vector $\mathbf{b}_k^{(N)}$. We write
\begin{align}
1=\int_0^{T_0} \phi^2_k(s) ds =\sum_{i=1}^{N} \frac{T_0}{N-1}
\phi_k^2\left(\frac{i-1}{N-1}T_0 \right)+\delta_N^k \label{Dora7}
\end{align}
where $\delta_N^k$ is defined as
\begin{align}
\delta_N^k= \sum_{i=1}^{N-1}
\int_{\frac{i-1}{N-1}T_0}^{\frac{i}{N-1}T_0} \left( \phi^2_k(s)-
\phi_k^2\left(\frac{i-1}{N-1}T_0 \right) \right) ds -
\frac{T_0}{N-1} \phi^2_k(T_0)
\end{align}
Using (\ref{worried4}), we have for any $s_1,s_2 \in [0,T_0]$,
\begin{align}
\left|\phi_k^2(s_1)-\phi_k^2(s_2) \right| & = \left|\phi_k(s_1)+\phi_k(s_2) \right|\left|\phi_k(s_1)-\phi_k(s_2) \right| \\
& \leq 2 \max_{s \in [0,T_0]} \left| \phi_k(s) \right| B_3(k+B_4)^\tau \left| s_1-s_2 \right|^\gamma \\
& \leq 2 \left(B_3(k+B_4)^\tau T_0^\gamma+1 \right) B_3(k+B_4)^\tau \left| s_1-s_2 \right|^\gamma \label{worried5}
\end{align}
where (\ref{worried5}) follows from (\ref{Dora4}).
The approximation error, $\delta_N^k$ satisfies
\begin{align}
|\delta_N^k| & \leq \sum_{i=1}^{N-1}
\int_{\frac{i-1}{N-1}T_0}^{\frac{i}{N-1}T_0} \left|\phi^2_k(s)-
\phi_k^2\left(\frac{i-1}{N-1}T_0 \right) \right| ds +\frac{T_0 \phi^2_k(T_0)}{N-1}  \\
& \leq T_0^{1+\gamma}
\frac{2 \left(B_3(k+B_4)^\tau T_0^\gamma+1 \right) B_3(k+B_4)^\tau}{(N-1)^\gamma} +\frac{T_0 \phi^2_k(T_0)}{N-1} \\
& \leq T_0^{1+\gamma}
\frac{2 \left(B_3(k+B_4)^\tau T_0^\gamma+1 \right) B_3(k+B_4)^\tau}{(N-1)^\gamma}
 +\frac{T_0 \left(B_3(k+B_4)^\tau T_0^\gamma+1\right)^2}{N-1} \label{Dora5} \\
& \leq T_0^{1+\gamma}
\frac{2 \left(B_3(k+B_4)^\tau T_0^\gamma+1 \right) B_3(k+B_4)^\tau}{(N-1)^\gamma}
 +\frac{T_0 \left(B_3(k+B_4)^\tau T_0^\gamma+1\right)^2}{\left(N-1\right)^\gamma} \label{Dora6} \\
& \leq  \frac{3 T_0^{1+2 \gamma} B_3^2 (k+B_4)^{2\tau}+4 T_0^{1+\gamma} B_3 (k+B_4)^\tau}{(N-1)^\gamma} +
\frac{T_0}{(N-1)^\gamma}
\end{align}
where (\ref{Dora5}) follows from (\ref{Dora4}), and (\ref{Dora6})
follows from the condition of $\gamma \leq 1$ in condition 3 in
Section \ref{defineA}. Due to the fact that $K_1(N)$
satisfies (\ref{conwor1}) and (\ref{conwor2}),
for a fixed constant $B_5$ that satisfies $0< B_5 <1$,
Then, there exists an integer $N_0>0$, such
that for $N \geq N_0$,
\begin{align}
\frac{3 T_0^{1+2 \gamma} B_3^2 (K_1(N)+B_4)^{2\tau}+4 T_0^{1+\gamma} B_3 (K_1(N)+B_4)^\tau}{(N-1)^\gamma} +
\frac{T_0}{(N-1)^\gamma} & \leq B_5
\end{align}
Hence, for any $k \leq K_1(N)$ and $N \geq N_0$, we have
\begin{align}
|\delta_N^k|
& \leq \frac{3 T_0^{1+2 \gamma} B_3^2 (k+B_4)^{2\tau}+4 T_0^{1+\gamma} B_3 (k+B_4)^\tau}{(N-1)^\gamma} +
\frac{T_0}{(N-1)^\gamma}\\
& \leq \frac{3 T_0^{1+2 \gamma} B_3^2 (K_1(N)+B_4)^{2\tau}+4 T_0^{1+\gamma} B_3 (K_1(N)+B_4)^\tau}{(N-1)^\gamma} +
\frac{T_0}{(N-1)^\gamma}\\
& \leq B_5
\end{align}
Finally, by the definition of $\mathbf{b}_k^{(N)}$, we have
\begin{align}
\left| \left| \mathbf{b}_k^{(N)} \right| \right|& = \sqrt{\sum_{i=1}^{N} \frac{T_0}{N-1}
\phi_k^2\left(\frac{i-1}{N-1}T_0 \right)}\\
& = \sqrt{1-\delta_N^k} \label{Dora8} \\
& \geq \sqrt{1-\left|\delta_N^k \right| }\\
& \geq \sqrt{1-B_5}
\end{align}
where (\ref{Dora8}) follows from (\ref{Dora7}). Similarly, we have
\begin{align}
\left| \left| \mathbf{b}_k^{(N)} \right| \right| \leq \sqrt{1+B_5} \label{constantl}
\end{align}

Next, we show that based on the orthogonality of the eigenfunctions of $\phi_k(t)$, the sampled version $\mathbf{b}_k^{(N)}$s are
almost orthogonal. Using (\ref{worried4}), we have
\begin{align}
&\left|\phi_m(s_1) \phi_l(s_1)-\phi_m(s_2) \phi_l(s_2) \right| \\
& = \left|\phi_m(s_1) \phi_l(s_1)-\phi_m(s_1) \phi_l(s_2)
+\phi_m(s_1) \phi_l(s_2) -\phi_m(s_2) \phi_l(s_2)  \right| \\
& \leq \left|\phi_m(s_1) \phi_l(s_1)-\phi_m(s_1) \phi_l(s_2) \right|+\left|\phi_m(s_1) \phi_l(s_2) -\phi_m(s_2) \phi_l(s_2)  \right| \\
& \leq \max_{s_1 \in [0,T_0]} \left|\phi_m(s_1) \right| \left| \phi_l(s_1)-\phi_l(s_2)\right|+ \max_{s_2 \in [0,T_0]}
\left|\phi_l(s_2) \right| \left| \phi_m(s_1)-\phi_m(s_2)\right| \\
& \leq \left(B_3(m+B_4)^\tau T_0^\gamma+1\right) B_3(l+B_4)^\tau \left|s_1-s_2 \right|^\gamma
+\left(B_3(l+B_4)^\tau T_0^\gamma+1\right) B_3(m+B_4)^\tau \left|s_1-s_2 \right|^\gamma \label{conwor3} \\
&= \left(2 B_3^2 (m+B_4)^\tau (l+B_4)^\tau T_0^\gamma+B_3(l+B_4)^\tau+B_3(m+B_4)^\tau \right)\left|s_1-s_2 \right|^\gamma
\end{align}
where (\ref{conwor3}) follows from (\ref{Dora4}).
Let $m$ and $l$ be two different integers, that belong to $\{1,2,\cdots,N\}$. Then, we have
\begin{align}
0=\int_0^{T_0} \phi_m(t) \phi_l(t) dt=\sum_{i=1}^N \frac{T_0}{N-1} \phi_m \left(\frac{i-1}{N-1}T_0 \right)\phi_l \left(\frac{i-1}{N-1}T_0 \right)
+ \varepsilon_N^{m,l}
\end{align}
Then, we have
\begin{align}
\left| \varepsilon_N^{m,l}\right|
= & \left| \int_0^{T_0} \phi_m(t) \phi_l(t) dt-
\sum_{i=1}^N \frac{T_0}{N-1} \phi_m \left(\frac{i-1}{N-1}T_0 \right)\phi_l \left(\frac{i-1}{N-1}T_0 \right) \right| \\
 \leq &\sum_{i=1}^{N-1} \int_{\frac{i-1}{N-1}T_0}^{\frac{i}{N-1}T_0} \left|\phi_m(t) \phi_l(t)-
 \phi_m \left(\frac{i-1}{N-1}T_0 \right)\phi_l \left(\frac{i-1}{N-1}T_0 \right)\right| dt \nonumber \\
 &+ \frac{T_0}{N-1} \left|\phi_m(T_0)\right|
 \left| \phi_l(T_0) \right| \\
  \leq & T_0^{1+\gamma} \frac{2 B_3^2 (m+B_4)^\tau (l+B_4)^\tau T_0^\gamma+B_3(l+B_4)^\tau+B_3(m+B_4)^\tau}{(N-1)^\gamma} \\
 &+T_0 \frac{\left(B_3(m+B_4)^\tau T_0^\gamma+1 \right)\left(B_3(l+B_4)^\tau T_0^\gamma+1 \right)}{N-1} \\
 \leq &T_0^{1+\gamma} \frac{2 B_3^2 (m+B_4)^\tau (l+B_4)^\tau T_0^\gamma+B_3(l+B_4)^\tau+B_3(m+B_4)^\tau}{(N-1)^\gamma} \\
 &+T_0 \frac{\left(B_3(m+B_4)^\tau T_0^\gamma+1 \right)\left(B_3(l+B_4)^\tau T_0^\gamma+1 \right)}{(N-1)^\gamma} \\
 =&  \frac{3 B_3^2 (m+B_4)^\tau (l+B_4)^\tau T_0^{1+2\gamma}+2B_3(l+B_4)^\tau T_0^{1+\gamma}+2B_3(m+B_4)^\tau T_0^{1+\gamma}+T_0}{(N-1)^\gamma}
\end{align}
For $m,l \leq K_1(N)$, we have
\begin{align}
\left| \varepsilon_N^{m,l}\right|& \leq
\frac{3 B_3^2 (K_1(N)+B_4)^{2\tau}  T_0^{1+2\gamma}+4B_3(K_1(N)+B_4)^\tau T_0^{1+\gamma}+T_0}{(N-1)^\gamma} \\
& \leq \frac{4 B_3^2 (K_1(N)+B_4)^{2\tau}  T_0^{1+2\gamma}}{(N-1)^\gamma} \label{conwor5}
\end{align}
where (\ref{conwor5}) follows when $N$ is large enough due to the fact that $K_1(N)$ satisfies (\ref{conwor1}), i.e., there exists an integer $N_2$
such that when $N>N_2$, (\ref{conwor5}) is true. The right hand side of (\ref{conwor5})
converges to zero as $N$ goes to infinity due to the fact that $K_1(N)$ satisfies (\ref{conwor2}). We have
\begin{align}
\left|\mathbf{b}_m^{(N)^T} \mathbf{b}_l^{(N)}\right|&=
\left|\sum_{i=1}^N \frac{T_0}{N-1} \phi_m \left(\frac{i-1}{N-1}T_0 \right)\phi_l \left(\frac{i-1}{N-1}T_0 \right) \right|
=\left| \varepsilon_N^{m,l}\right| \\
& \leq \frac{4 B_3^2 (K_1(N)+B_4)^{2\tau}  T_0^{1+2\gamma}}{(N-1)^\gamma} \label{GG6}
\end{align}
which means that vectors $\mathbf{b}_m^{(N)}$ and $\mathbf{b}_l^{(N)}$ become more
orthogonal as $N$ gets larger.

Now, we are ready to establish the link between the eigenvalues of $\Sigma_N'$ and $\lambda_k$. From (\ref{worried3}), we have
\begin{align}
\mathbf{b}_k^{(N)} = \left(\Sigma_N'-\lambda_k I
\right)^{-1}\left(-\mathbf{a}_k^{(N)}\right)
\end{align}
Hence,
\begin{align}
\left|\left| \mathbf{b}_k^{(N)}  \right| \right| & \leq  \left|
\left|\left(\Sigma_N'-\lambda_k I \right)^{-1}
 \right| \right|_2
\left|\left| \mathbf{a}_k^{(N)}  \right| \right| \\
& = \left( \min_{0 \leq m \leq N-1} \left(\mu_m^{(N)'} -\lambda_k
\right) \right)^{-1} \left|\left| \mathbf{a}_k^{(N)}  \right| \right| \label{conwor9}
\end{align}
Thus, we have
\begin{align}
\min_{0 \leq m \leq N-1} \left(\mu_m^{(N)'} -\lambda_k \right)
&\leq \frac{\left|\left| \mathbf{a}_k^{(N)}  \right| \right|}
{\left|\left| \mathbf{b}_k^{(N)}  \right| \right|} \label{Dora9} \\
& \leq \frac{2 B_6 \sqrt{T_0}\frac{(k+B_7')^\tau}{(N-1)^\beta}}{\sqrt{1-B_5 }}\\
& \leq d_0 \frac{\left(k+B_7'\right)^\tau}{(N-1)^\beta}
\label{runrunrunrun}
\end{align}
where (\ref{runrunrunrun}) follows by defining $d_0$ as
\begin{align}
d_0=\frac{2 B_6 \sqrt{T_0}}{\sqrt{1-B_5}}
\end{align}

Hence, for $k=0, 1,2,\cdots,K_1(N)$, there exists an eigenvalue
$\mu'^{(N)}$ of $\Sigma_N'$ such that
\begin{align}
\left| \mu'^{(N)} -\lambda_k \right|  & \leq d_0
\frac{\left(k+B_7'\right)^\tau}{(N-1)^\beta} \label{worried}
\end{align}
when $N$ is large enough, more specifically, when $N \geq
\max(N_0, N_1, N_2)$.

For $k=0, 1,2,\cdots,K_1(N)$, if we label the $\mu'^{(N)}$ that satisfies (\ref{worried}) to be $\mu_k^{(N)'}$, then
when $\lambda_k$ for different $k$s are sufficiently close, more specifically,
\begin{align}
\left|\lambda_m-\lambda_l \right| \leq 2d_0
\frac{\left(K_1(N)+B_7'\right)^\tau}{(N-1)^\beta}, \quad m, l \leq K_1(N), m \neq l
\end{align}
$\mu_m^{(N)'}$ and $\mu_l^{(N)'}$ might correspond to the same eigenvalue of
$\Sigma_N'$, which is undesirable. If we relax the minimum distance of $d_0
\frac{\left(k+B_7'\right)^\tau}{(N-1)^\beta} $, we will be able to eliminate this problem.
Thus, we will next show that for $k=0, 1,2,\cdots,K_1(N)$, there exists an eigenvalue
 $\mu'^{(N)}$ of $\Sigma_N'$, different for each $k$,  such that
 \begin{align}
\left| \mu'^{(N)} -\lambda_k \right|  & \leq (2 \bar{\chi}+1)\sqrt{d_2} d_0
\frac{\left(k+\bar{\chi}+B_7'\right)^\tau}{(N-1)^\beta}
\end{align}
when $N$ is large enough, where
we define $\bar{\chi}\overset{\triangle}{=}\max
(K_0+1+c_u+c_l, 2c_u+2c_l+1)$ and constant $d_2$ as the
 largest root of the following second-order equation
\begin{align}
(1-B_5)d_2^2-2\left((1-B_5)+3 \bar{\chi}(1+B_5) \right) d_2+ (1-B_5)+ 2 \bar{\chi}(1+B_5) = 0 \label{defeqn}
\end{align}
It can be checked that both roots of the above equation are real, and the largest root is a positive constant,
strictly larger than $\frac{2 \bar{\chi}(1+B_5)}{1-B_5}+1$,
that is a function of $\bar{\chi}$
and $B_5$.

First, let us define a cluster of $\lambda$s. We say that $\chi$ $\lambda$s are a
cluster, where with no loss of generality, we may label these
$\lambda$s $\lambda_k, \lambda_{k+1}, \cdots, \lambda_{k+\chi-1}$, if
\begin{align}
\lambda_{k+l}-\lambda_{k+l+1} \leq 2 \sqrt{d_2} d_0
\frac{\left(k+\bar{\chi}+B_7'\right)^\tau}{(N-1)^\beta}, \quad l=0,1,\cdots, \chi-1 \label{sufficient}
\end{align}
Note here that whether the $\lambda$s are in a cluster depends on $N$.

Next, we prove that the number of $\lambda$s within a cluster is upper bounded by $\bar{\chi}$ when $N$ is large enough.
For $k>K_0$, we have
\begin{align}
\frac{d}{(k+c_l)^x} &\leq \lambda_k \leq \frac{d}{(k-c_u)^x} \\
\frac{d}{(k+2c_l+c_u+1)^x} &\leq \lambda_{k+c_u+c_l+1} \leq \frac{d}{(k+c_l+1)^x}
\end{align}
Hence, for every $k \geq K_0$, the distance between $\lambda_k$ and $\lambda_{k+c_u+c_l+1}$ satisfies
\begin{align}
\lambda_k-\lambda_{k+c_u+c_l+1} \geq \frac{d}{(k+c_l)^x} - \frac{d}{(k+c_l+1)^x}
\end{align}
which is a non-increasing function of $k$.
Thus, for all $K_0 < k \leq K_1(N)$, the distance between $\lambda_k$ and $\lambda_{k+c_u+c_l+1}$ satisfies
\begin{align}
\lambda_k-\lambda_{k+c_u+c_l+1} & \geq \frac{d}{(K_1(N)+c_l)^x} - \frac{d}{(K_1(N)+c_l+1)^x} \\
&=\frac{d}{(K_1(N)+c_l)^x} \left(1-\left(1-\frac{1}{K_1(N)+c_l+1}\right)^x \right) \\
& \geq \frac{d}{(K_1(N)+c_l)^x} \left(x \frac{1}{K_1(N)+c_l+1}-\frac{x(x-1)}{2}\frac{1}{(K_1(N)+c_l+1)^2} \right)\\
& = \frac{xd}{(K_1(N)+c_l)^{x+1}}-\frac{x(x-1)d}{2(K_1(N)+c_l)^{x+2}}\\
& \geq \frac{xd}{2(K_1(N)+c_l)^{x+1}} \label{conwor6} \\
& > 2 \sqrt{d_2} d_0
\frac{\left(K_1(N)+\bar{\chi}+B_7'\right)^\tau}{(N-1)^\beta} \label{conwor7}
\end{align}
where (\ref{conwor6}) is true when $N$ is large enough due to the fact that $K_1(N)$ satisfies (\ref{conwor1}), i.e.,
there exists an integer $N_3$, such that when $N>N_3$, (\ref{conwor6}) is true, and (\ref{conwor7}) is true when $N$ is large enough,
due to the fact that $K_1(N)$ satisfies (\ref{worried2}), i.e., there exists an integer $N_4$, such that when $N>N_4$, (\ref{conwor7}) is true.

Hence, for all $K_0 < k \leq K_1(N)$, when $N$ is large enough, more specifically, when $N > \max(N_3,N_4)$, due to the sufficient distance
between $\lambda_k$ and $\lambda_{k+c_u+c_l+1}$, shown in
(\ref{conwor7}), they cannot be in the same cluster. Hence, we may conclude that
for large enough $N$, the size of a cluster is
at most $\bar{\chi}$, which is a finite number.

Following from (\ref{worried3}), we have
\begin{align}
\left(\lambda_k I-\Sigma_N'\right) \mathbf{b}_k^{(N)}=\mathbf{a}_k^{(N)}
\end{align}
Let the eigenvalues and the corresponding eigenvectors of $\Sigma_N'$ be $\mu_i^{(N)'}$ and $\mathbf{u}_i^{(N)}$, $i=1,2,\cdots,N$,
with arbitrary labelling of the eigenvalues and
eigenvectors.
Then we have
\begin{align}
\sum_{i=1}^N \left(\lambda_k-\mu_i^{(N)'}\right) \mathbf{u}_i^{(N)} \mathbf{u}_i^{(N)^T} \mathbf{b}_k^{(N)}
=\mathbf{a}_k^{(N)}
\end{align}
We take the norm squared on both sides, and due to the orthogonality of eigenvectors $\mathbf{u}_i^{(N)}$, we have
\begin{align}
\sum_{i=1}^N \left(\lambda_k-\mu_i^{(N)'}\right)^2 \left(\mathbf{u}_i^{(N)^T} \mathbf{b}_k^{(N)}\right)^2
=\left|\left|\mathbf{a}_k^{(N)}\right|\right|^2, \quad k=0,1,2,\cdots \label{GG1}
\end{align}
and we also have
\begin{align}
\sum_{i=1}^N \left(\mathbf{u}_i^{(N)^T} \mathbf{b}_k^{(N)}\right)^2 & = \left|\left|\mathbf{b}_k^{(N)}\right|\right|^2, \quad
k=0,1,2,\cdots \label{GG2}
\end{align}

Let $\lambda_k, \lambda_{k+1}, \cdots, \lambda_{k+\chi-1}$ be a cluster, and from previous arguments, we know
$\chi \leq \bar{\chi}$. Furthermore, we are only interested in the first $K_1(N)+1$ eigenvalues, and therefore
$k+\chi-1 \leq K_1(N)$. We will prove by contradiction.
Suppose that only $\varsigma$ number of $\mu_i^{(N)'}$s are within distance
\begin{align}
\sqrt{d_2} d_0 \frac{(k+\bar{\chi}+B_7')^\tau}{(N-1)^\beta} \label{conwor8}
\end{align}
from
any of the $\lambda_k, \lambda_{k+1}, \cdots, \lambda_{k+\chi-1}$, with $1 \leq \varsigma < \chi$, we will show that there is a contradiction, and therefore,
we can conclude that our assumption that $\varsigma< \chi$
number of $\mu_i^{(N)'}$s are within distance (\ref{conwor8}) from any of the
$\lambda_k, \lambda_{k+1}, \cdots, \lambda_{k+\chi}$ is not correct. Note that $\varsigma \geq 1$ because we have already proved (\ref{worried}).
Based on (\ref{runrunrunrun}), the distance in (\ref{conwor8}) satisfies
\begin{align}
\sqrt{d_2} d_0 \frac{(k+\bar{\chi}+B_7')^\tau}{(N-1)^\beta} \geq
\frac{\sqrt{d_2} ||\mathbf{a}_{k+l}^{(N)}||}{||\mathbf{b}_{k+l}^{(N)}||}, \quad l=0,1,\cdots, \chi-1
\end{align}
Then, based on (\ref{GG1}), we have
\begin{align}
\left(\frac{\sqrt{d_2}||\mathbf{a}_{k+l}^{(N)}||}{||\mathbf{b}_{k+l}^{(N)}||}\right)^2
\sum_{i=\varsigma+1}^N \left(\mathbf{u}_i^{(N)^T} \mathbf{b}_{k+l}^{(N)}\right)^2
& \leq \left|\left|\mathbf{a}_{k+l}^{(N)}\right|\right|^2, \quad l=0,1,\cdots,\chi-1
\end{align}
where we have labelled the $\mu'^{(N)}$ that are within distance
(\ref{conwor8}) from
any of the $\lambda_k, \lambda_{k+1}, \cdots, \lambda_{k+\chi}$ $\mu_1^{(N)'}, \mu_2^{(N)'}, \cdots,
\mu_\varsigma^{(N)'}$.
Hence, we have
\begin{align}
\sum_{i=\varsigma+1}^N \left(\mathbf{u}_i^{(N)^T} \mathbf{b}_{k+l}^{(N)}\right)^2 \leq \frac{||\mathbf{b}_{k+l}^{(N)}||^2}{d_2}, \quad
l=0,1,\cdots, \chi-1
\end{align}
Together with (\ref{GG2}), we have
\begin{align}
\sum_{i=1}^\varsigma \left(\mathbf{u}_i^{(N)^T} \mathbf{b}_{k+l}^{(N)}\right)^2 \geq \frac{(d_2-1)||\mathbf{b}_{k+l}^{(N)}||^2}{d_2}, \quad
l=0,1,\cdots, \chi-1 \label{largeportion}
\end{align}
Since the $\mathbf{u}_i^{(N)}$ form a complete set of basis in $\mathbb{R}^N$, we can write $\mathbf{b}_{k+l}^{(N)}$ as
\begin{align}
\mathbf{b}_{k+l}^{(N)}=\sum_{i=1}^\varsigma \alpha_{k+l,i} \mathbf{u}_i^{(N)}+\mathbf{v}_{k+l}^{(N)}, \quad
l=0,1,\cdots, \chi-1 \label{expansion}
\end{align}
where $\mathbf{v}_{k+l}^{(N)}$ is orthogonal to $\mathbf{u}_i^{(N)}$, for $i=1,2,\cdots,\varsigma$. If we
take the expression of $\mathbf{b}_{k+l}^{(N)}$ in (\ref{expansion}) and plug it in (\ref{largeportion}), we get
\begin{align}
\sum_{i=1}^\varsigma \left(\alpha_{k+l,i}\right)^2 \geq \frac{(d_2-1)||\mathbf{b}_{k+l}^{(N)}||^2}{d_2}, \quad
l=0,1,2,\cdots, \chi-1 \label{GG7}
\end{align}
From (\ref{expansion}), we get
\begin{align}
||\mathbf{b}_{k+l}^{(N)}||^2=\sum_{i=1}^\varsigma \left(\alpha_{k+l,i}\right)^2 +||\mathbf{v}_{k+l}^{(N)}||^2,
\quad l=0,1,\cdots, \varsigma-1
\end{align}
Hence, we conclude that
\begin{align}
||\mathbf{v}_{k+l}^{(N)}||^2 \leq \frac{||\mathbf{b}_{k+l}^{(N)}||^2}{d_2},
\quad l=0,1,\cdots, \varsigma-1 \label{GG4}
\end{align}
Furthermore, from (\ref{expansion}), we have
\begin{align}
\mathbf{b}_{k+m}^{(N)^T} \mathbf{b}_{k+l}^{(N)}=\sum_{i=1}^\varsigma \alpha_{k+m,i} \alpha_{k+l,i}
+ \mathbf{v}_{k+m}^{(N)^T} \mathbf{v}_{k+l}^{(N)}, \quad m,l=0,1,\cdots, \varsigma-1, m \neq l
\end{align}
Hence, we have
\begin{align}
\sum_{i=1}^\varsigma \alpha_{k+m,i} \alpha_{k+l,i}=\mathbf{b}_{k+m}^{(N)^T} \mathbf{b}_{k+l}^{(N)}
-\mathbf{v}_{k+m}^{(N)^T} \mathbf{v}_{k+l}^{(N)},  \quad m,l=0,1,\cdots, \varsigma-1, m \neq l
\end{align}
and for $m,l=0,1,\cdots, \varsigma-1, m \neq l$, we have
\begin{align}
\left|\sum_{i=1}^\varsigma \alpha_{k+m,i} \alpha_{k+l,i}\right| & \leq
\left|\mathbf{b}_{k+m}^{(N)^T} \mathbf{b}_{k+l}^{(N)}\right|
+\left|\mathbf{v}_{k+m}^{(N)^T} \mathbf{v}_{k+l}^{(N)} \right| \\
& \leq \frac{4 B_3^2 (K_1(N)+B_4)^{2\tau}  T_0^{1+2\gamma}}{(N-1)^\gamma}+
\left| \left|\mathbf{v}_{k+m}^{(N)^T} \right| \right| \left| \left|\mathbf{v}_{k+l}^{(N)}  \right| \right| \label{whatnow1}\\
& \leq \frac{4 B_3^2 (K_1(N)+B_4)^{2\tau}  T_0^{1+2\gamma}}{(N-1)^\gamma}+
\frac{\left| \left|\mathbf{b}_{k+m}^{(N)^T} \right| \right| \left| \left|\mathbf{b}_{k+l}^{(N)}  \right| \right|}{d_2} \label{whatnow2} \\
& \leq \frac{4 B_3^2 (K_1(N)+B_4)^{2\tau}  T_0^{1+2\gamma}}{(N-1)^\gamma}+
\frac{1+B_5}{d_2} \label{whatnow3} \\
& \leq \frac{2(1+B_5)}{d_2} \label{GG9}
\end{align}
where (\ref{whatnow1}) follows from (\ref{GG6}) when $N>N_2$, (\ref{whatnow2}) follows from (\ref{GG4}), (\ref{whatnow3})
follows from (\ref{constantl}) when $N>N_0$, and
 (\ref{GG9}) follows when $N$ is large enough, due to the fact that $K_1(N)$ satisfies (\ref{conwor2}), i.e.,
there exists an integer $N_6$, when $N > N_6$, (\ref{GG9}) is true.
Let us define matrix $A$ to be
\begin{align}
A= \begin{bmatrix}\alpha_{k,1} & \alpha_{k,2} & \cdots & \alpha_{k,\varsigma} \\
 \alpha_{k+1,1} & \alpha_{k+1,2} & \cdots & \alpha_{k+1, \varsigma} \\
 \vdots & \vdots & \ddots & \vdots \\
 \alpha_{k+\varsigma-1,1} & \alpha_{k+\varsigma-1,2} & \cdots & \alpha_{k+\varsigma-1,\varsigma}  \end{bmatrix}
 \label{defineAha}
\end{align}
and define vectors $\mathbf{b}$, $\mathbf{v}$, $\mathbf{u}$ to be
\begin{align}
\mathbf{b}=\begin{bmatrix}\mathbf{b}_{k}^{(N)^T} \mathbf{b}_{k+\varsigma}^{(N)} \\
\mathbf{b}_{k+1}^{(N)^T} \mathbf{b}_{k+\varsigma}^{(N)}\\
\vdots\\
 \mathbf{b}_{k+\varsigma-1}^{(N)^T} \mathbf{b}_{k+\varsigma}^{(N)}\end{bmatrix}, \quad
 \mathbf{v}=  \begin{bmatrix} \mathbf{v}_{k}^{(N)^T}\mathbf{b}_{k+\varsigma}^{(N)} \\
 \mathbf{v}_{k+1}^{(N)^T}\mathbf{b}_{k+\varsigma}^{(N)}  \\
 \vdots \\
 \mathbf{v}_{k+\varsigma-1}^{(N)^T}\mathbf{b}_{k+\varsigma}^{(N)}  \end{bmatrix}, \quad
 \mathbf{u}=\begin{bmatrix} \mathbf{u}_1^{(N)^T}\mathbf{b}_{k+\varsigma}^{(N)} \\
 \mathbf{u}_2^{(N)^T}\mathbf{b}_{k+\varsigma}^{(N)}  \\
 \vdots \\
 \mathbf{u}_{\varsigma}^{(N)^T}\mathbf{b}_{k+\varsigma}^{(N)}  \end{bmatrix}
\end{align}
Then, by (\ref{expansion}), we have
\begin{align}
\mathbf{b}=
 A \mathbf{u}
 +\mathbf{v}
\end{align}
In other words,
\begin{align}
\mathbf{u}=A^{-1}\left( \mathbf{b}-\mathbf{v} \right)
\end{align}
thus, we have
\begin{align}
\left| \left| \mathbf{u} \right| \right|^2 = \left| \left| A^{-1} \right| \right|_2^2 \left(\left| \left| \mathbf{b} \right| \right|+
\left| \left| \mathbf{v} \right| \right| \right)^2 \label{hard}
\end{align}
We start by evaluating $\left| \left| A^{-1} \right| \right|_2^2$, which is equal to the inverse of the
smallest eigenvalue of $A^TA$. From the definition of matrix $A$ in (\ref{defineAha}), we have
\begin{align}
A^TA= D+E
\end{align}
where $D$ is an $\varsigma \times \varsigma$ diagonal matrix with the $l$-th diagonal element being
$ \sum_{i=1}^\varsigma \left(\alpha_{k+l-1,i}\right)^2$, and $E$ is an $\varsigma \times \varsigma$ matrix
with zero diagonals and $(m,l)$-th element being $\sum_{i=1}^\varsigma \alpha_{k+m-1,i} \alpha_{k+l-1,i}$,
when $m \neq l$.
The absolute difference between the smallest eigenvalue of $A^TA$ and $D$ is upper
bounded by $\left| \left| E \right| \right|_2$ \cite{Stewart:1993}. The smallest eigenvalue of $D$ is
\begin{align}
\min_{l \in \{0,1,\cdots, \varsigma-1\}} \sum_{i=1}^\varsigma \left(\alpha_{k+l,i}\right)^2 & \geq \min_{l}
\frac{(d_2-1)||\mathbf{b}_{k+l}^{(N)}||^2}{d_2} \label{GG8} \\
& \geq \frac{(d_2-1)(1-B_5)}{d_2} \label{whatnow5}
\end{align}
where (\ref{GG8}) follows from (\ref{GG7}), and (\ref{whatnow5}) follows from (\ref{constantl}) when $N>N_0$ since $k+\varsigma-1 \leq K_1(N)$.
 We can upper bound the spectral norm of matrix $E$,
i.e., $\left| \left| E \right| \right|_2$, by the Frobenius norm of $E$, i.e,
\begin{align}
\left| \left| E \right| \right|_2^2 & = \sum_{m \neq l}
\left(\sum_{i=1}^\varsigma \alpha_{k+m-1,i} \alpha_{k+l-1,i} \right)^2 \\
& \leq \varsigma^2 \left(\frac{2(1+B_5)}{d_2} \right)^2 \label{GG10}\\
& < \chi^2 \frac{4 (1+B_5)^2}{d_2^2} \\
& \leq \bar{\chi}^2 \frac{4 (1+B_5)^2}{d_2^2}
\end{align}
where (\ref{GG10}) follows from (\ref{GG9}).
Hence, we may conclude that
\begin{align}
\left| \left| A^{-1} \right| \right|_2^2 <
\left(\frac{(d_2-1)(1-B_5)}{d_2}-\frac{2 \bar{\chi}(1+B_5)}{d_2} \right)^{-1} \label{tm3}
\end{align}
where the right hand side is a positive number, by the definition of $d_2$.
Next, we evaluate $\left| \left| \mathbf{v} \right| \right|^2$.
\begin{align}
\left| \left| \mathbf{v} \right| \right|^2 & = \sum_{i=0}^{\varsigma-1}
\left(\mathbf{v}_{k+i}^{(N)^T} \mathbf{b}_{k+\varsigma}^{(N)} \right)^2\\
& \leq  \sum_{i=0}^{\varsigma-1} \left| \left|\mathbf{v}_{k+i}^{(N)^T} \right| \right|^2 \left| \left| \mathbf{b}_{k+\varsigma}^{(N)} \right| \right|^2 \\
& \leq \frac{\left| \left| \mathbf{b}_{k+\varsigma}^{(N)} \right| \right|^2 }{d_2} \sum_{i=0}^{\varsigma-1}
\left| \left| \mathbf{b}_{k+i}^{(N)} \right| \right|^2 \label{GG3}\\
& \leq \frac{\varsigma (1+B_5)}{d_2}\left| \left| \mathbf{b}_{k+\varsigma}^{(N)} \right| \right|^2 \label{whatnow4}\\
& < \frac{\chi (1+B_5)}{d_2}\left| \left| \mathbf{b}_{k+\varsigma}^{(N)} \right| \right|^2 \\
& \leq \frac{\bar{\chi} (1+B_5)}{d_2}\left| \left| \mathbf{b}_{k+\varsigma}^{(N)} \right| \right|^2 \label{tm1}
\end{align}
where (\ref{GG3}) follows from (\ref{GG4}), and (\ref{whatnow4}) follows from (\ref{constantl}) when $N>N_0$ since $k+\varsigma-1 \leq K_1(N)$.
Finally, we evaluate $\left| \left| \mathbf{b} \right| \right|^2$.
\begin{align}
\left| \left| \mathbf{b} \right| \right|^2 & = \sum_{i=0}^{\varsigma-1}
\left(\mathbf{b}_{k+i}^{(N)^T} \mathbf{b}_{k+\varsigma}^{(N)} \right)^2\\
& \leq \varsigma \left(\frac{4 B_3^2 (K_1(N)+B_4)^{2\tau}  T_0^{1+2\gamma}}{(N-1)^\gamma} \right)^2 \label{GG5}\\
& < \chi \left(\frac{4 B_3^2 (K_1(N)+B_4)^{2\tau}  T_0^{1+2\gamma}}{(N-1)^\gamma} \right)^2 \\
& \leq \bar{\chi} \left(\frac{4 B_3^2 (K_1(N)+B_4)^{2\tau}  T_0^{1+2\gamma}}{(N-1)^\gamma} \right)^2 \label{tm2}
\end{align}
where (\ref{GG5}) follows from (\ref{GG6}) when $N>N_2$.

Following from (\ref{hard}), using (\ref{tm1}), (\ref{tm2}) and (\ref{tm3}), we have
\begin{align}
\left| \left| \mathbf{u} \right| \right|^2 &= \left| \left| A^{-1} \right| \right|_2^2 \left(\left| \left| \mathbf{b} \right| \right|+
\left| \left| \mathbf{v} \right| \right| \right)^2 \\
& < \left(\frac{(d_2-1) (1-B_5)}{d_2}-
 \frac{ 2 \bar{\chi}(1+B_5)}{d_2} \right)^{-1} \nonumber \\
& \quad \left(
\sqrt{\frac{\bar{\chi} (1+B_5)}{d_2}}\left| \left| \mathbf{b}_{k+\varsigma}^{(N)} \right| \right|+
 \sqrt{\bar{\chi}} \left(\frac{4 B_3^2 (K_1(N)+B_4)^{2\tau}  T_0^{1+2\gamma}}{(N-1)^\gamma} \right)\right)^2 \\
 & \leq \left(\frac{(d_2-1) (1-B_5)}{d_2}-
\frac{ 2 \bar{\chi}(1+B_5)}{d_2}\right)^{-1}
\quad \left(2
\sqrt{\frac{\bar{\chi} (1+B_5)}{d_2}}\left| \left| \mathbf{b}_{k+\varsigma}^{(N)} \right| \right|\right)^2 \label{hard2}\\
& = \frac{d_2-1}{d_2} \left| \left| \mathbf{b}_{k+\varsigma}^{(N)} \right| \right|^2 \label{hard3}
\end{align}
where (\ref{hard2}) follows when $N$ is large enough, due to the fact that $K_1(N)$ satisfies (\ref{conwor2}), i.e, there exists an integer
$N_5$, such that when $N> N_5$,
\begin{align}
\sqrt{\bar{\chi}} \left(\frac{4 B_3^2 (K_1(N)+B_4)^{2\tau}  T_0^{1+2\gamma}}{(N-1)^\gamma} \right)
\leq \frac{\sqrt{\bar{\chi} (1+B_5)}}{d_2} \sqrt{1-B_5}
\end{align}
and (\ref{hard2}) is true, and (\ref{hard3}) follows from the definition of $d_2$ by (\ref{defeqn}). Hence,
when $N$ is large enough, more specifically, when $N > \max(N_0, N_2, N_3, N_4,N_5, N_6)$, we have a contradiction with (\ref{largeportion}). Therefore, we conclude that
there must be at least $\chi$ eigenvalues of $\Sigma_N'$ within distance (\ref{conwor8}) away from any of the clustered $\lambda$s, furthermore,
from the definition of a cluster in (\ref{sufficient}), there must be
at least $\chi$ eigenvalues within distance
\begin{align}
\left(2\chi+1  \right) \sqrt{d_2} d_0 \frac{(k+ \bar{\chi}+B_7')^\tau}{(N-1)^\beta} \leq
\left(2\bar{\chi}+1  \right) \sqrt{d_2} d_0 \frac{(k+\bar{\chi}+B_7')^\tau}{(N-1)^\beta} \label{much}
\end{align}
away from all of the clustered $\lambda$s. We pick $\chi$ eigenvalues of $\Sigma_N'$
which are within distance (\ref{much}) and arbitrarily pair each clustered $\lambda$ with one of the eigenvalues.
These eigenvalues will not be paired with any other $\lambda$ because all other clusters of $\lambda$s are at least distance $2 \sqrt{d_2} d_0
\frac{\left(k+\bar{\chi}+B_7'\right)^\tau}{(N-1)^\beta}$ apart from this cluster.

Finally, by letting
\begin{align}
d_1 &= \left(2\bar{\chi}+1  \right) \sqrt{d_2} d_0, \quad B_7 =\bar{\chi}+B_7'
\end{align}
we have the desired results when $N$ is large enough, i.e., $N > \max (N_0,N_1,N_2,N_3,N_4,N_5,N_6)$.

\subsubsection{Proof of Lemma \ref{divergence}}
In the proof of Lemma \ref{divergence}, we will need results from
Lemma \ref{lambdak2} and \ref{appro}. Thus, we will first prove
that under the condition of Lemma \ref{divergence}, the results of
Lemma \ref{lambdak2} and \ref{appro} apply. Since
\begin{align}
\lim_{N \rightarrow \infty} \vartheta_U^N=0 \label{mymy1}
\end{align}
for any $0 < \kappa <1$, when $N$ is large enough, $\theta' <
\vartheta_U^N$ is small enough, which means that the result of
Lemma \ref{lambdak2} is valid. Now we show that the result of
Lemma \ref{appro} is also true. Let $K_1(N)=
\left(\frac{d}{\vartheta_L^N} \right)^{\frac{1}{x}}+c_u$. Because
of
\begin{align}
\lim_{N \rightarrow \infty} \frac{1}{\vartheta_L^N N^{\frac{x
\gamma}{2\tau}}} =0, \quad \lim_{N \rightarrow \infty} \frac{1}{\vartheta_L^N N^{\frac{x
\beta}{x+1+\tau}}} =0
\end{align}
we have (\ref{conwor2}) and (\ref{worried2}).
Because of (\ref{mymy1}) and the fact that
$
\vartheta_L^{N} \leq \vartheta_U^N
$,
we have (\ref{conwor1}).

Hence, for any $0 \leq k \leq \left\lfloor \left(\frac{d}{\theta'}
\right)^{\frac{1}{x}}+c_u \right\rfloor$, result of Lemma
\ref{appro} applies because
\begin{align}
k \leq \left\lfloor \left(\frac{d}{\vartheta_L^N}
\right)^{\frac{1}{x}}+c_u \right\rfloor \leq K_1(N)
\end{align}
and $N$ is large enough.

Now, we will use the result of Lemma \ref{lambdak2} and
\ref{appro} to prove Lemma \ref{divergence}. From the properties
of the Karhunen-Loeve expansion, we know that
\begin{align}
\sum_{k=0}^\infty \lambda_k=\int_{0}^{T_0} K(t,t) dt < \infty
\end{align}
Thus, for any constant $0 < \kappa < 1$, we have
\begin{align}
\sum_{k=0}^{\left\lfloor \left(\frac{d}{\theta'}
\right)^{\frac{1}{x}}+c_u \right\rfloor} \lambda_k &=
\sum_{k=0}^\infty \lambda_k-\sum_{k=\left\lfloor
\left(\frac{d}{\theta'} \right)^{\frac{1}{x}}+c_u
\right\rfloor+1}^\infty
 \lambda_k\\
&  \geq \int_{0}^{T_0} K(t,t)dt-\sum_{k=\left\lfloor
\left(\frac{d}{\theta'} \right)^{\frac{1}{x}}+c_u
\right\rfloor+1}^\infty
 \lambda_k'' \\
 & \geq \int_{0}^{T_0} K(t,t)dt -\frac{ d^\frac{1}{x}}{(x-1) \kappa} \theta'^{1-\frac{1}{x}}
\label{nolaststep}
\end{align}
where we have used (\ref{needupper}) in Lemma \ref{lambdak2} to
obtain (\ref{nolaststep}).

From the definition of matrix $\Sigma_N$, we have
\begin{align}
\sum_{k=0}^{N-1} \mu_k^{(N)'}= \frac{T_0}{N-1}\text{tr}
\left(\Sigma_N \right)=\frac{T_0}{N-1} \sum_{i=0}^{N-1}
K\left(\frac{i}{N-1}T_0, \frac{i}{N-1}T_0 \right)
\end{align}
Thus,
\begin{align}
\sum_{k=\left\lfloor \left(\frac{d}{\theta'}
\right)^{\frac{1}{x}}+c_u \right\rfloor +1}^{N-1} \mu_k^{(N)'} &=
\sum_{k=0}^{N-1} \mu_k^{(N)'}-
\sum_{k=0}^{\left\lfloor \left(\frac{d}{\theta'} \right)^{\frac{1}{x}}+c_u \right\rfloor} \mu_k^{(N)'} \\
&\leq \frac{T_0}{N-1} \sum_{i=0}^{N-1} K\left(\frac{i}{N-1}T_0,
\frac{i}{N-1}T_0 \right) -\sum_{k=0}^{\left\lfloor
\left(\frac{d}{\theta'} \right)^{\frac{1}{x}}+c_u \right\rfloor}
 \left( \lambda_k-d_1 \frac{\left(k+B_7\right)^\tau}{(N-1)^\beta} \right) \label{sumjensen}\\
& \leq \frac{T_0}{N-1} \sum_{i=0}^{N-1} K\left(\frac{i}{N-1}T_0,
\frac{i}{N-1}T_0 \right)
- \int_{0}^{T_0} K(t,t)dt \nonumber \\
& \hspace{0.1in}+\frac{d^\frac{1}{x}}{(x-1) \kappa} \theta'^{1-\frac{1}{x}}+
\frac{d_1}{(N-1)^\beta}
\sum_{k=0}^{\left\lfloor \left(\frac{d}{\theta'} \right)^{\frac{1}{x}}+c_u \right\rfloor} (k+B_7)^\tau \label{singgy3}\\
& \leq \frac{B T_0^{1+\alpha}
2^{\frac{\alpha}{2}}}{(N-1)^\alpha}+\frac{T_0 K(0,0)}{N-1}+\frac{
d^\frac{1}{x}}{(x-1) \kappa} \theta'^{1-\frac{1}{x}}
 + \frac{d_1}{(N-1)^\beta}
\sum_{k=0}^{\left\lfloor \left(\frac{d}{\theta'}
\right)^{\frac{1}{x}}+c_u \right\rfloor+B_7} k^\tau
\label{nolaststep2}\\
& \leq \frac{B T_0^{1+\alpha} 2^{\frac{\alpha}{2}}+T_0
K(0,0)}{(N-1)^\alpha}+\frac{ d^\frac{1}{x}}{(x-1) \kappa}
\theta'^{1-\frac{1}{x}} \nonumber \\
& \hspace{0.1in} + \frac{d_1}{(\tau+1)(N-1)^\beta} \left(\left\lfloor
\left(\frac{d}{\theta'} \right)^{\frac{1}{x}}+c_u
\right\rfloor+B_7\right)^{\tau+1}
\label{usesumtau}\\
& \leq \frac{B T_0^{1+\alpha} 2^{\frac{\alpha}{2}}+T_0
K(0,0)}{(N-1)^\alpha}+\frac{ d^\frac{1}{x}}{(x-1) \kappa}
\theta'^{1-\frac{1}{x}}
 + \frac{2 d_1 d^{\frac{\tau+1}{x}}}{(\tau+1) \theta'^{\frac{\tau+1}{x}}(N-1)^\beta} \label{thetasmall47} \\
& \leq \frac{ d^\frac{1}{x}}{(x-1) \kappa^2}
\theta'^{1-\frac{1}{x}}\label{niao}
\end{align}
where (\ref{sumjensen}) follows by Lemma \ref{appro}. We have used
(\ref{nolaststep}) to obtain (\ref{singgy3}), and condition 2 in
Section \ref{defineA} to obtain (\ref{nolaststep2}),
(\ref{usesumtau}) follows from the fact that $\alpha \leq 1$ and
that
\begin{align}
\sum_{k=0}^n k^\tau \leq \int_0^n y^\tau dy =\frac{1}{\tau+1}
n^{\tau+1} \label{sumtau}
\end{align}
(\ref{thetasmall47}) follows because (\ref{mymy1})
and when $N$ is large enough, we have
\begin{align}
c_u+B_7 < \frac{d}{\vartheta_U^N} \leq \frac{d}{\theta'}
\end{align}
(\ref{niao}) follows because
\begin{align}
\lim_{N \rightarrow \infty} \frac{1}{\vartheta_L^N N^{\frac{\alpha
x}{x-1}}}=0 \label{vartheta2}
\end{align}
and
\begin{align}
\lim_{N \rightarrow \infty} \frac{1}{\vartheta_L^N N^{\frac{\beta
x}{1+x+\tau}}}=0 \Rightarrow \lim_{N \rightarrow \infty} \frac{1}{\vartheta_L^N N^{\frac{\beta
x}{x+\tau}}}=0 \label{vartheta3}
\end{align}
and when $N$ large enough, i.e., there exists a $N_5(\kappa)>0$
such that when $N> N_5(\kappa)$, we have
\begin{align}
\frac{\frac{B T_0^{1+\alpha} 2^{\frac{\alpha}{2}}+T_0
K(0,0)}{(N-1)^\alpha}}{\frac{ d^\frac{1}{x}}{(x-1)}
\theta'^{1-\frac{1}{x}}} \leq \frac{\frac{B T_0^{1+\alpha}
2^{\frac{\alpha}{2}}+T_0 K(0,0)}{(N-1)^\alpha}}{\frac{
d^\frac{1}{x}}{(x-1)}
\left(\vartheta_L^N\right)^{(1-\frac{1}{x})}} & \leq \frac{1}{2} \left(\frac{1}{\kappa^2}-\frac{1}{\kappa} \right) \\
\frac{\frac{2 d_1 d^{\frac{\tau+1}{x}}}{(\tau+1)
\theta'^{\frac{\tau+1}{x}}(N-1)^\beta}} {\frac{
d^\frac{1}{x}}{(x-1)} \theta'^{1-\frac{1}{x}}} \leq \frac{\frac{2
d_1 d^{\frac{\tau+1}{x}}}{(\tau+1) \left(\vartheta_L^N\right)
^{\frac{\tau+1}{x}}(N-1)^\beta}} {\frac{ d^\frac{1}{x}}{(x-1)}
\left(\vartheta_L^N\right)^{1-\frac{1}{x}}} & \leq \frac{1}{2}
\left(\frac{1}{\kappa^2}-\frac{1}{\kappa} \right)
\end{align}
Therefore, for any $0 < \kappa < 1$, (\ref{nomorename}) holds for
$\theta' \in [\vartheta_L^N, \vartheta_U^N]$ when $N$ is large
enough.

\subsection{Proof of Lemma \ref{nothingleft}} \label{Rlowerupper}
Since the condition of Lemma \ref{nothingleft} is the same as
Lemma \ref{divergence}, the results of Lemma \ref{lambdak2},
\ref{appro} and \ref{divergence} hold. By the same argument as
Lemma \ref{lambdak2}, Lemma \ref{lambdak1} holds as well.

We first prove (\ref{singapore1}). Since $\vartheta_L^N$ satisfies
\begin{align}
\lim_{N \rightarrow \infty} \frac{1}{\vartheta_L^N N^{ \frac{\beta
x}{x+\tau+1} }} =0 \Rightarrow
\lim_{N \rightarrow \infty} \frac{1}{\vartheta_L^N N^x}=0 \label{runrepeat5}
\end{align}
when $N$ is large enough such that
\begin{align}
\left\lfloor \left(\frac{d}{\theta'} \right)^{\frac{1}{x}}+c_u
\right\rfloor+1 \leq \left\lfloor \left(\frac{d}{\vartheta_L^N}
\right)^{\frac{1}{x}}+c_u \right\rfloor+1 < N-1
\end{align}
we can provide an upper bound on $R_a^N(\theta')$ by splitting the
sum of $N$ variables into two parts,
\begin{align}
R_a^N(\theta')&= \sum_{k=0}^{\left\lfloor \left(\frac{d}{\theta'}
\right)^{\frac{1}{x}}+c_u \right\rfloor} \frac{1}{2} \log
\left(1+\frac{\mu_k^{(N)'}}{\theta'} \right) +
\sum^{N-1}_{k=\left\lfloor \left(\frac{d}{\theta'}
\right)^{\frac{1}{x}}+c_u \right\rfloor+1}
 \frac{1}{2} \log \left(1+\frac{\mu_k^{(N)'}}{\theta'} \right)
\label{rateterm2}
\end{align}
For any $0<\kappa<1$, we start with the first term in
(\ref{rateterm2}).
\begin{align}
& \sum_{k=0}^{\left\lfloor \left(\frac{d}{\theta'}
\right)^{\frac{1}{x}}+c_u \right\rfloor}
 \frac{1}{2} \log \left(1+\frac{\mu_k^{(N)'}}{\theta'} \right) \nonumber \\
&  \leq \sum_{k=0}^{\left\lfloor \left(\frac{d}{\theta'}
\right)^{\frac{1}{x}}+c_u \right\rfloor} \frac{1}{2} \log
\left(1+ \frac{\lambda_k}{\theta'}+d_1\frac{(k+B_7)^\tau}{\theta'(N-1)^\beta} \right) \label{needreason1}\\
& \leq \sum_{k=0}^{\left\lfloor \left(\frac{d}{\theta'}
\right)^{\frac{1}{x}}+c_u \right\rfloor} \frac{1}{2} \log
\left(1+\frac{\lambda_k}{\theta'} \right)
 +\frac{d_1}{2\theta' (N-1)^\beta} \sum_{k=0}^{\left\lfloor \left(\frac{d}{\theta'} \right)^{\frac{1}{x}}
 +c_u \right\rfloor}
(k+B_7)^\tau \label{derivative3}\\
& \leq \sum_{k=0}^{\left\lfloor \left(\frac{d}{\theta'}
\right)^{\frac{1}{x}}+c_u \right\rfloor} \frac{1}{2} \log
\left(1+\frac{\lambda_k''}{\theta'} \right)+\frac{d_1}{2\theta'
(N-1)^\beta} \sum_{k=0}^{\left\lfloor \left(\frac{d}{\theta'}
\right)^{\frac{1}{x}}
 +c_u\right\rfloor+B_7}
k^\tau \label{introducelambda2prime}\\
& \leq  \left(\frac{\log 2+x}{2 \kappa}  \right) d^{\frac{1}{x}}
\theta'^{-\frac{1}{x}} +\frac{d_1}{2(\tau+1)\theta' (N-1)^\beta}
\left(\left\lfloor \left(\frac{d}{\theta'}
\right)^{\frac{1}{x}} \right\rfloor +c_u+B_7\right)^{\tau+1}\label{fromlower3}\\
& \leq \left(\frac{\log 2+x}{2 \kappa}  \right) d^{\frac{1}{x}}
\theta'^{-\frac{1}{x}}
+\frac{d_1 d^{\frac{\tau+1}{x}}}{(\tau+1)\theta'^{\frac{\tau+1+x}{x}} (N-1)^\beta} \label{smalltheta37}\\
& \leq \left(\frac{\log 2+x}{2 \kappa^2}  \right) d^{\frac{1}{x}}
\theta'^{-\frac{1}{x}} \label{changemind}
\end{align}
where (\ref{needreason1}) follows from Lemma \ref{appro}.
(\ref{derivative3}) follows because the derivative of the function
$\frac{1}{2} \log(1+x)$ is bounded by $\frac{1}{2}$ for $x \geq
0$, (\ref{introducelambda2prime})  follows from the definition of
the sequence $\lambda_k''$ in (\ref{definelambdak2}) and the
observation in (\ref{orderorder}), (\ref{fromlower3}) follows
because of (\ref{needupper2}) in Lemma \ref{lambdak2}, and the
observation in (\ref{sumtau}). (\ref{smalltheta37}) follows
because of the same reason as (\ref{thetasmall47}), and
(\ref{changemind}) follows because of (\ref{vartheta3}),
and when $N$ is large enough, more specifically, there exists an
$N_6(\kappa)>0$ such that when $N > N_6(\kappa)$, we have
\begin{align}
\frac{\frac{d_1
d^{\frac{\tau+1}{x}}}{(\tau+1)\theta'^{\frac{\tau+1+x}{x}}
(N-1)^\beta}} {\left(\frac{\log 2+x}{2 }  \right) d^{\frac{1}{x}}
\theta'^{-\frac{1}{x}}} \leq \frac{\frac{d_1
d^{\frac{\tau+1}{x}}}{(\tau+1)
\left(\vartheta_L^N\right)^{\frac{\tau+1+x}{x}} (N-1)^\beta}}
{\left(\frac{\log 2+x}{2 }  \right) d^{\frac{1}{x}}
\left(\vartheta_L^N\right)^{-\frac{1}{x}}} <
\left(\frac{1}{\kappa^2}-\frac{1}{\kappa} \right)
\end{align}

Now, we will study the second term of (\ref{rateterm2}). Using
Jensen's inequality \cite{Cover:book}, the second term of
(\ref{rateterm2}) is bounded by
\begin{align}
& \sum^{N-1}_{k=\left\lfloor \left(\frac{d}{\theta'}
\right)^{\frac{1}{x}}+c_u \right\rfloor+1} \frac{1}{2} \log
\left(1+\frac{\mu_k^{(N)'}}{\theta'} \right) \\
&\leq \frac{N-\left\lfloor \left(\frac{d}{\theta'}
\right)^{\frac{1}{x}}+c_u \right\rfloor-1}{2} \log
\left(1+\frac{1}{\theta'} \frac{1}{N-\left\lfloor
\left(\frac{d}{\theta'} \right)^{\frac{1}{x}}+c_u \right\rfloor-1}
\sum^{N-1}_{k=\left\lfloor \left(\frac{d}{\theta'} \right)^{\frac{1}{x}}+c_u \right\rfloor +1} \mu_k^{(N)'} \right) \label{todiscuss1}\\
& \leq \frac{ d^{\frac{1}{x}}}{2(x-1) \kappa^2}
\theta'^{-\frac{1}{x}} \label{changemind2}
\end{align}
where in obtaining (\ref{changemind2}),
 we have used (\ref{nomorename}) in Lemma \ref{divergence} and the fact that $\log(1+x)\leq x$.

We combine the results of (\ref{changemind}) and
(\ref{changemind2}) and obtain
\begin{align}
R_a^N(\theta') \leq \frac{ d^{\frac{1}{x}}\left(x^2-(1-\log
2)x+(1-\log 2)\right)}{2(x-1) \kappa^2}
\theta'^{-\frac{1}{x}}\label{rateupperbound}
\end{align}
Using similar methods, we may also lower bound $R_a^N(\theta')$.
We write
\begin{align}
R_a^N(\theta')&= \sum_{k=0}^{\left\lfloor \left(\frac{d}{\theta'}
\right)^{\frac{1}{x}}-c_l \right\rfloor} \frac{1}{2} \log
\left(1+\frac{\mu_k^{(N)'}}{\theta'} \right) +
\sum^{N-1}_{k=\left\lfloor \left(\frac{d}{\theta'}
\right)^{\frac{1}{x}}-c_l \right\rfloor+1}
 \frac{1}{2} \log \left(1+\frac{\mu_k^{(N)'}}{\theta'} \right)
\label{rateterm}
\end{align}
We start with the first term of (\ref{rateterm}),
\begin{align}
 \sum_{k=0}^{\left\lfloor \left(\frac{d}{\theta'} \right)^{\frac{1}{x}}-c_l \right\rfloor}
 \frac{1}{2} \log \left(1+\frac{\mu_k^{(N)'}}{\theta'} \right)
&  \geq \sum_{k=0}^{\left\lfloor \left(\frac{d}{\theta'}
\right)^{\frac{1}{x}}-c_l \right\rfloor} \frac{1}{2} \log
\left(1+ \frac{\lambda_k}{\theta'}-d_1 \frac{(k+B_7)^\tau}{\theta'(N-1)^\beta} \right)\label{smalltheta57}\\
& \geq \sum_{k=0}^{\left\lfloor \left(\frac{d}{\theta'}
\right)^{\frac{1}{x}}-c_l \right\rfloor} \frac{1}{2} \log
\left(1+\frac{\lambda_k}{\theta'} \right) -\frac{d_1}{2 \theta'
(N-1)^\beta } \sum_{k=0}^{\left\lfloor \left(\frac{d}{\theta'}
\right)^{\frac{1}{x}}-c_l \right\rfloor}
(k+B_7)^\tau \label{derivative1}\\
& \geq \sum_{k=0}^{\left\lfloor \left(\frac{d}{\theta'}
\right)^{\frac{1}{x}}-c_l \right\rfloor} \frac{1}{2} \log
\left(\frac{\lambda_k'}{\theta'} \right)
 -\frac{d_1}{2 \theta' (N-1)^\beta }
\sum_{k=0}^{\left\lfloor \left(\frac{d}{\theta'}
\right)^{\frac{1}{x}}-c_l \right\rfloor+B_7}
k^\tau\\
& \geq  \frac{\kappa x d^{\frac{1}{x}}}{2} \theta'^{-\frac{1}{x}} -
\frac{d_1}{2(\tau+1) \theta' (N-1)^\beta } \left(\left\lfloor
\left(\frac{d}{\theta'}
\right)^{\frac{1}{x}}-c_l \right\rfloor+B_7 \right)^{\tau+1} \label{fromlower} \\
& \geq \frac{\kappa x d^{\frac{1}{x}}}{2} \theta'^{-\frac{1}{x}}-
\frac{d_1 d^{\frac{\tau+1}{x}}}
{(\tau+1) \theta'^{\frac{\tau+1+x}{x}} (N-1)^\beta } \label{GradyLittle} \\
 & \geq \frac{\kappa x d^{\frac{1}{x}}}{4} \theta'^{-\frac{1}{x}} \label{kangwei}
\end{align}
where (\ref{smalltheta57}) follows when applying the result of
Lemma \ref{appro}. (\ref{derivative1}) follows because the
function $\frac{1}{2} \log (1+x)$ has derivative bounded by
$\frac{1}{2}$, when $x \geq 0$. The first term in
(\ref{fromlower}) follows because of (\ref{needlower2}) in Lemma
\ref{lambdak1}. The second term follows because of (\ref{sumtau}),
(\ref{GradyLittle}) follows because of similar reasons as
(\ref{thetasmall47}), and (\ref{kangwei}) follows because of (\ref{vartheta3}),
and when $N$ is large enough, we have
\begin{align}
\frac{\frac{d_1 d^{\frac{\tau+1}{x}}} {(\tau+1)
\theta'^{\frac{\tau+1+x}{x}} (N-1)^\beta } }{\frac{x
d^{\frac{1}{x}}}{4} \theta'^{-\frac{1}{x}}} \leq \frac{\frac{d_1
d^{\frac{\tau+1}{x}}} {(\tau+1) \left(\vartheta_L^N
\right)^{\frac{\tau+1+x}{x}} (N-1)^\beta } } {\frac{x
d^{\frac{1}{x}}}{4} \left(\vartheta_L^N \right)^{-\frac{1}{x}}}
\leq \frac{1}{8}
\end{align}

A lower bound on the second term of (\ref{rateterm}) is zero.
Hence, we can conclude that
\begin{align}
R_a^N(\theta') \geq \frac{\kappa x d^{\frac{1}{x}}}{4}
\theta'^{-\frac{1}{x}}
\end{align}


Now we evaluate $D_b^N(\theta')$ for large enough $N$ and $\theta'
\in [\vartheta_L^N, \vartheta_U^N]$, and prove
(\ref{nuoconclude1}).
\begin{align}
D_b^N(\theta') &= T_0^{-1} \sum_{k=0}^{N-1}
\left(\frac{1}{\theta'}+
\frac{1}{\mu_k^{(N)'}} \right)^{-1}\\
&= T_0^{-1} \sum_{k=0}^{k=\left\lfloor \left(\frac{d}{\theta'}
\right)^{\frac{1}{x}}+c_u \right\rfloor }
 \left(\frac{1}{\theta'}+
\frac{1}{\mu_k^{(N)'}}
\right)^{-1}+T_0^{-1}\sum^{N-1}_{k=\left\lfloor
\left(\frac{d}{\theta'} \right)^{\frac{1}{x}}+c_u \right\rfloor
+1}
 \left(\frac{1}{\theta'}+
\frac{1}{\mu_k^{(N)'}} \right)^{-1} \label{finalterm}
\end{align}
where (\ref{finalterm}) follows because of the same reason as
(\ref{rateterm2}). The first term of (\ref{finalterm}) can be
bounded as
\begin{align}
& T_0^{-1} \sum_{k=0}^{\left\lfloor \left(\frac{d}{\theta'}
\right)^{\frac{1}{x}}+c_u \right\rfloor}
 \left(\frac{1}{\theta'}+
\frac{1}{\mu_k^{(N)'}} \right)^{-1} \nonumber \\
& \leq T_0^{-1} \sum_{k=0}^{\left\lfloor \left(\frac{d}{\theta'}
\right)^{\frac{1}{x}}+c_u \right\rfloor}
 \left(\frac{1}{\theta'}+
\frac{1}{\lambda_k+ d_1\frac{(k+B_7)^\tau}{(N-1)^\beta}} \right)^{-1} \label{thetasmall67}\\
& \leq T_0^{-1} \sum_{k=0}^{\left\lfloor \left(\frac{d}{\theta'}
\right)^{\frac{1}{x}}+c_u \right\rfloor} \left(\frac{1}{\theta'}+
\frac{1}{\lambda_k''+ d_1\frac{(k+B_7)^\tau}{(N-1)^\beta}} \right)^{-1}\\
& \leq T_0^{-1} \sum_{k=0}^{\left\lfloor \left(\frac{d}{\theta'}
\right)^{\frac{1}{x}}+c_u \right\rfloor}
 \left(\frac{1}{\theta'}+
\frac{1}{\lambda_k'' } \right)^{-1}  + \frac{d_1}{T_0 (N-1)^\beta}
\sum_{k=0}^{\left\lfloor \left(\frac{d}{\theta'}
\right)^{\frac{1}{x}}+c_u \right\rfloor}(k+B_7)^\tau
\label{derivative2}\\
& \leq T_0^{-1} \sum_{k=0}^{\left\lfloor \left(\frac{d}{\theta'}
\right)^{\frac{1}{x}}+c_u \right\rfloor}
 \min \left( \theta', \lambda_k''
\right)  + \frac{d_1}{T_0 (N-1)^\beta} \sum_{k=0}^{\left\lfloor
\left(\frac{d}{\theta'}
\right)^{\frac{1}{x}}+c_u \right\rfloor+B_7} k^\tau \label{usetaoyan} \\
& = T_0^{-1} \left\lfloor \left(\frac{d}{\theta'}
\right)^{\frac{1}{x}}+c_u+1 \right\rfloor \theta'
  + \frac{d_1}{(\tau+1)T_0 (N-1)^\beta}
\left(\left\lfloor \left(\frac{d}{\theta'}
\right)^{\frac{1}{x}}+c_u \right\rfloor+B_1\right)^{\tau+1} \label{whichK}\\
& \leq \frac{ d^{\frac{1}{x}}}{T_0} \theta'^{1-\frac{1}{x}}+
\frac{2 d_1 d^{\frac{\tau+1}{x}}}{(\tau+1) T_0 (N-1)^\beta
\theta'^{\frac{\tau+1}{x}}}
+\frac{c_u+1}{T_0} \theta' \label{smalltheta67}\\
& \leq \frac{ d^{\frac{1}{x}} }{T_0 \kappa}
\theta'^{1-\frac{1}{x}} \label{conclude1}
\end{align}
where (\ref{thetasmall67}) is true because of Lemma \ref{appro}.
(\ref{derivative2}) follows because the derivative of the function
$\left(\frac{1}{\theta'}+\frac{1}{x} \right)^{-1}$ is bounded by
1, and (\ref{usetaoyan}) follows from the fact that for $a,b \geq
0$, $\left(\frac{1}{a}+\frac{1}{b} \right)^{-1} \leq \min (a,b)$.
(\ref{whichK}) follows because of (\ref{sumtau}) and the fact that
when $K = \left\lfloor \left(\frac{d}{\theta'}
\right)^{\frac{1}{x}}+c_u \right\rfloor$,
\begin{align}
\lambda_{K+1}'' \leq \theta' \leq \lambda_K''
\end{align}
(\ref{smalltheta67}) follows because of the same reason as
(\ref{thetasmall47}), and finally (\ref{conclude1}) follows
because (\ref{vartheta3})
and (\ref{mymy1})
and when $N$ is large enough, i.e., for any $0 < \kappa < 1$,
there exists $N_7(\kappa)>0$ such that when $N > N_7(\kappa)$, we
have
\begin{align}
\frac{\frac{2 d_1 d^{\frac{\tau+1}{x}}}{(\tau+1) T_0 (N-1)^\beta
\theta'^{\frac{\tau+1}{x}}}} {\frac{ d^{\frac{1}{x}}}{T_0}
\theta'^{1-\frac{1}{x}}} \leq \frac{\frac{2 d_1
d^{\frac{\tau+1}{x}}}{(\tau+1) T_0 (N-1)^\beta \left(\vartheta_L^N
\right)^{\frac{\tau+1}{x}}}} {\frac{ d^{\frac{1}{x}}}{T_0}
\left(\vartheta_L^N \right)^{1-\frac{1}{x}}} &\leq \frac{1}{2}
\left(\frac{1}{\kappa}
-1 \right) \\
\frac{\frac{c_u+1}{T_0} \theta'}{\frac{ d^{\frac{1}{x}}}{T_0}
\theta'^{1-\frac{1}{x}}} \leq \frac{\frac{c_u+1}{T_0}
\vartheta_U^N}{\frac{ d^{\frac{1}{x}}}{T_0}
\left(\vartheta_U^N\right)^{1-\frac{1}{x}}} &\leq \frac{1}{2}
\left(\frac{1}{\kappa} -1 \right)
\end{align}

The second term of (\ref{finalterm}) can be bounded by using
Jensen's inequality,
\begin{align}
&T_0^{-1}\sum^{N-1}_{\left\lfloor \left(\frac{d}{\theta'}
\right)^{\frac{1}{x}}+c_u \right\rfloor+1}
 \left(\frac{1}{\theta'}+
\frac{1}{\mu_k^{(N)'}} \right)^{-1}  \nonumber \\
& \leq \frac{N-\left\lfloor \left(\frac{d}{\theta'}
\right)^{\frac{1}{x}}+c_u \right\rfloor-1}{T_0}
 \left(\frac{1}{\theta'}+
\frac{1}{\frac{1}{N-\left\lfloor \left(\frac{d}{\theta'}
\right)^{\frac{1}{x}}+c_u \right\rfloor-1}
 \sum^{N-1}_{k=\left\lfloor \left(\frac{d}{\theta'} \right)^{\frac{1}{x}}+c_u \right\rfloor+1}
 \mu_k^{(N)'}} \right)^{-1}\\
& \leq \frac{N}{T_0}
 \left(\frac{1}{\theta'}+
\frac{1}{\frac{1}{\kappa N} \sum^{N-1}_{k=\left\lfloor
\left(\frac{d}{\theta'} \right)^{\frac{1}{x}}+c_u \right\rfloor+1}
\mu_k^{(N)'}} \right)^{-1}\label{largeN}\\
& \leq \frac{N}{T_0} \min \left(\theta', \frac{
d^{\frac{1}{x}}}{(x-1) \kappa^3} \frac{\theta'^{1-\frac{1}{x}}}{N}
\right)
\label{nuoexplain1}\\
& \leq \frac{ d^{\frac{1}{x}}}{(x-1) T_0 \kappa^3}
\theta'^{1-\frac{1}{x}} \label{powerloose}
\end{align}
where (\ref{largeN}) follows because (\ref{runrepeat5})
and when $N$ is large enough, i.e., there exists an
$N_8(\kappa)>0$, such that when $N > N_8(\kappa)$, we have
\begin{align}
\frac{\left\lfloor \left(\frac{d}{\theta'}
\right)^{\frac{1}{x}}+c_u \right\rfloor+1}{N} \leq
\frac{\left\lfloor \left(\frac{d}{\vartheta_L^N}
\right)^{\frac{1}{x}}+c_u \right\rfloor+1}{N} \leq 1-\kappa
\end{align}
(\ref{nuoexplain1}) follows from (\ref{nomorename}) in Lemma
\ref{divergence} and (\ref{powerloose}) follows because of
(\ref{runrepeat5}) and when $N$ is large enough, we have
\begin{align}
\frac{\frac{ d^{\frac{1}{x}}}{(x-1) \kappa^3}
\frac{\theta'^{1-\frac{1}{x}}}{N}}{\theta'} \leq \frac{\frac{
d^{\frac{1}{x}}}{(x-1) \kappa^3} \frac{\left(\vartheta_L^N
\right)^{1-\frac{1}{x}}}{N}}{\left(\vartheta_L^N \right)} \leq 1
\end{align}

Thus, combining (\ref{conclude1}) and (\ref{powerloose}), we have
\begin{align}
D_b^N(\theta') \leq \frac{ d^{\frac{1}{x}} \left(1+\kappa^2(x-1)
\right)}{\kappa^3 (x-1) T_0} \theta'^{1-\frac{1}{x}}
\end{align}
Therefore, for any $0 < \kappa <1$, (\ref{singapore1}) and
(\ref{nuoconclude1}) are true for $\theta' \in [\vartheta_L^N,
\vartheta_U^N]$ when $N$ is large enough.

\subsection{Proof of Theorem \ref{singhope}} \label{hopehopehope}

Note that (\ref{singapore1}) implies that
\begin{align}
\frac{\kappa^x x^x d}{4^x R^x} \leq \theta_a^N(R) \leq \left(\frac{
d^{\frac{1}{x}}\left(x^2-(1-\log 2)x+(1-\log 2)\right)}{2(x-1)
\kappa^2}\right)^x R^{-x} \label{partial}
\end{align}
for large enough $N$ and $R$ in the interval of
\begin{align}
\left[\frac{ d^{\frac{1}{x}}\left(x^2-(1-\log 2)x+(1-\log
2)\right)}{2(x-1) \kappa^2}
 \left(\vartheta_U^N \right)^{-\frac{1}{x}}, \frac{\kappa x
d^{\frac{1}{x}}}{4} \left(\vartheta_L^N
\right)^{-\frac{1}{x}}\right] \label{partialinterval1}
\end{align}
From the definition of $D_a(R)$ in (\ref{lab}), we have
\begin{align}
D_a(R) &=D_a^N(\theta_a^N(R))\\
& \leq 2A^{(N)}+B^{(N)}+D_b^N(\theta_a^N(R)) \label{bbhoujia}\\
& \leq 2A^{(N)}+B^{(N)}+ \frac{ d^{\frac{1}{x}}
\left(1+\kappa^2(x-1) \right)}{\kappa^3 (x-1) T_0}
\left(\theta_a^N(R)\right)^{1-\frac{1}{x}}
\label{bbnuoconclude1} \\
& \leq O \left(N^{-\alpha} \right)+ O \left(N^{1/2-\alpha}
\right)+\frac{ d(1+\kappa^2(x-1))\left( x^2-(1-\log 2) x + (1-\log
2)\right)^{x-1}}{T_0 \kappa^{2x+1}2^{x-1}(x-1)^x} R^{1-x}
\label{bb}
\end{align}
where (\ref{bbhoujia}) follows from (\ref{houjia}), (\ref{bbnuoconclude1}) follows because of
(\ref{nuoconclude1}), (\ref{bb}) follows from (\ref{AA}),
(\ref{BB}), (\ref{partial}) and the fact that $R$ in
(\ref{partialinterval}) implies that $R$ is in
(\ref{partialinterval1}), and when $R$ is in
(\ref{partialinterval1}), $\theta_a^N(R)$ is in $[\vartheta_L^N,
\vartheta_U^N]$. When $R$ is in (\ref{partialinterval}), we have
that the third term in (\ref{bb}) is much larger than the sum of
the first and second terms when $N$ is large enough due to the fact that
\begin{align}
\lim_{N \rightarrow \infty} \frac{1}{\vartheta_{LL}^N N^{\frac{(\alpha-1/2) x}{x-1}
}} =0 \Rightarrow
\lim_{N \rightarrow \infty} \frac{1}{\vartheta_{LL}^N N^{\frac{\alpha x}{x-1}
}} =0
\end{align}
i.e., there
exists an $N_{9}(\kappa)>0$ such that when $N > N_{9}(\kappa)$, we
have
\begin{align}
& \frac{O \left(N^{-\alpha} \right)+ O \left(N^{1/2-\alpha}
\right)}{\frac{ d(1+\kappa^2(x-1))\left( x^2-(1-\log 2) x +
(1-\log 2)\right)^{x-1}}{T_0
2^{x-1}(x-1)^x} R^{1-x} } \nonumber \\
&\leq \frac{O \left(N^{-\alpha} \right)+ O \left(N^{1/2-\alpha}
\right)} {\frac{ d(1+\kappa^2(x-1))\left( x^2-(1-\log 2) x +
(1-\log 2)\right)^{x-1}}{T_0 2^{x-1}(x-1)^x} \left(\frac{x
d^{\frac{1}{x}}}{8} \left(\vartheta_{LL}^N
\right)^{-\frac{1}{x}}\right)^{1-x} } \\
& \leq \frac{1}{\kappa^{2x+2}}-\frac{1}{\kappa^{2x+1}}
\end{align}
Therefore, for $0<\kappa<1$, (\ref{siggy1}) is true for $R$ in the
interval of (\ref{partialinterval}) when $N$ is large enough.

\subsection{Proof of Theorem \ref{bestluck}} \label{bestluck1}
Pick the sequences $\vartheta_{LL}^N$ and $\vartheta_U^N$ as
\begin{align}
\vartheta_{LL}^N  = \left(\frac{\nu}{\frac{x
d^{\frac{1}{x}}}{8}} \log N P(N) \right)^{-x}, \quad
\vartheta_U^N  = \left( \frac{\nu}{\frac{
d^{\frac{1}{x}}\left(x^2-(1-\log 2)x+(1-\log 2)\right)}{2(x-1)
\kappa^2}} \log NP(N) \right)^{-x/2}
\end{align}
Then, because $P(N)$ satisfies (\ref{powersad1}) and
(\ref{powerfinalhope}),
 $\vartheta_{LL}^N$  satisfies (\ref{bestbestluck}) and $\vartheta_U^N$ satisfies (\ref{increasing}).
According to (\ref{goli}), we have the achievable rate in the
interval of (\ref{partialinterval}), and thus, when $N$ is large
enough, Theorem \ref{singhope} applies. Hence, an upper bound on
the minimum achievable expected distortion, or equivalently, the
achievable rate in the separation-based scheme is
\begin{align}
D_u^N &= D_a\left(C_a^N\right) \\
&\leq \frac{ d(1+\kappa^2(x-1))\left( x^2-(1-\log 2) x + (1-\log
2)\right)^{x-1}}{T_0 \kappa^{3x+1}2^{x-1}(x-1)^x \nu^{x-1}}
\left(\frac{1}{\log (NP(N))}\right)^{x-1}
\end{align}
Therefore, when $P(N)$ satisfies (\ref{powersad1}) and
(\ref{powerfinalhope}), for any $0 < \kappa <1$,  (\ref{singgy5})
holds when $N$ is large enough.

\bibliographystyle{unsrt}
\bibliography{ref}

\end{document}